\documentclass[aps,prx,showpacs,preprintnumbers,twocolumn,superscriptaddress,nofootinbib]{revtex4-2}
\usepackage{amsmath,amssymb}
\usepackage{bm}
\usepackage{tipa}
\usepackage{upgreek}
\usepackage{mathrsfs}
\usepackage{graphicx}
\usepackage{lipsum}
\usepackage{braket}
\usepackage{enumerate}
\usepackage{mathbbol}
\usepackage{booktabs}
\usepackage{gensymb}
\usepackage[normalem]{ulem}

\usepackage{color}

\usepackage[colorlinks=true,linkcolor=blue,allcolors=blue]{hyperref}

\def\equationautorefname~#1\null{Eq. (#1)\null}

\newcommand{\appref}[1]{\hyperref[#1]{App.~\ref*{#1}}}

\definecolor{purple}{rgb}{0.55,0.15,0.55}
\definecolor{nicegreen}{rgb}{0.28,0.75,0.19}

\usepackage[final]{changes}
\definechangesauthor[name={Round 1}, color=blue]{R1}
\definechangesauthor[name={Round 2}, color=orange]{R2}
\newcommand{\addedtwo}[1]{\added[id=R2]{#1}}
\newcommand{\deletedtwo}[1]{\deleted[id=R2]{#1}}
\newcommand{\replacedtwo}[2]{\replaced[id=R2]{#1}{#2}}
\usepackage{etoolbox}
\robustify\ref
\robustify\cite

\usepackage{bm}
\newcommand\intsec{\mathfrak{int}_t}

\begin{document}

\title{Solvable Quantum Circuits in Tree+1 Dimensions}% Force line breaks with \\

\author{Oliver Breach}
\email{oliver.breach@physics.ox.ac.uk}
\affiliation{Rudolf Peierls Centre for Theoretical Physics, Parks Road, Oxford, OX1 3PU, UK}

\author{Benedikt Placke}
\affiliation{Rudolf Peierls Centre for Theoretical Physics, Parks Road, Oxford, OX1 3PU, UK}

\author{Pieter W. Claeys}
\affiliation{Max Planck Institute for the Physics of Complex Systems, N\"othnitzer Str. 38, Dresden 01187, Germany}

\author{S.A. Parameswaran}
\affiliation{Rudolf Peierls Centre for Theoretical Physics, Parks Road, Oxford, OX1 3PU, UK}

\date{\today}

\begin{abstract}
We devise tractable models of unitary quantum many-body dynamics on tree graphs, as a first step towards a deeper understanding of dynamics in non-Euclidean spaces. To this end, we first demonstrate how to construct strictly local quantum circuits that preserve the symmetries of trees, such that their dynamical light cones grow isotropically. \replacedtwo{F}{We show that, f}or trees with coordination number $z$, such circuits can be built from $z$-site gates. We then introduce a family of gates for which the dynamics is exactly solvable; these satisfy a set of constraints that we term `tree-unitarity'. Notably, tree-unitarity reduces to the previously-established notion of dual-unitarity for $z=2$, when the tree reduces to a line. Among the unexpected features of tree-unitarity is a trade-off between `maximum \added[id=R1]{butterfly} velocity' dynamics of out-of-time-order correlators and the existence of non-vanishing correlation functions in multiple directions, a tension  absent in one-dimensional dual-unitary models and their Euclidean generalizations. 
\added[id=R1]{We connect the existence of (a wide class of) solvable dynamics with non-maximal butterfly velocity directly to a property of the underlying circuit geometry called $\delta$-hyperbolicity, and argue that such dynamics can only arise in non-Euclidean geometries.}
We give various examples of tree-unitary gates, discuss dynamical correlations, out-of-time-order correlators, and  entanglement growth, and show that the kicked Ising model on a tree is a physically-motivated example of maximum-velocity tree-unitary dynamics.

\end{abstract}

\maketitle

\section{Introduction}

The  emergence and ongoing development of experimental platforms with precise control over quantum degrees of freedom has paved the way for detailed investigation into out-of-equilibrium quantum systems. These platforms include both analog devices such as ultracold atom simulators~\cite{schfer_tools_2020}, Rydberg arrays~\cite{ebadi_quantum_2021}, and trapped ion devices~\cite{blatt_quantum_2012}, as well as digital platforms such as superconducting qubit arrays~\cite{arute_quantum_2019, kim_evidence_2023} where discrete operations may be performed on sets of qubits.
Such digital platforms raise new fundamental questions about the nature of many-body quantum dynamics under discrete local gates~\cite{nielsen2010quantum}, both regarding novel universal phenomena that may appear, and their utility in describing more general out-of-equilibrium dynamics. 
Discrete time evolution in the form of `brickwork quantum circuits' has gained intense attention as a minimally structured model for unitary dynamics generated by local interactions~\cite{von_keyserlingk_operator_2018, nahum_operator_2018,khemani_operator_2018,fisher_random_2023}.
These brickwork models can be used to gain theoretical insight into universal aspects of many-body dynamics, and are tailor-made for quantum simulation, being natively realized in digital quantum computing setups. 

Analytically studying the properties of such models is challenging. Currently, two main approaches have allowed theoretical progress to be made by simplifying this problem. In the first, randomness is introduced and the focus is shifted to ensemble-averaged dynamics rather than any specific realization; such averaging often allows for an effective description in terms of classical degrees of freedom~\cite{von_keyserlingk_operator_2018, nahum_operator_2018,khemani_operator_2018}. This approach has proven to be highly successful in understanding the dynamics of operator scrambling and entanglement spreading as well as  phase transitions in these phenomena induced by repeated measurements (for a recent review, see Ref.~\onlinecite{fisher_random_2023}).
A second approach is to identify classes of models with special properties that enable analytic treatment. Dual-unitary circuits are a prime example~\cite{gopalakrishnan_unitary_2019,bertini_exact_2019}, where the discreteness of both space and time can be used to realize dynamics that possesses unitarity in both the space and time directions. Dual-unitary brickwork dynamics is both provably quantum chaotic and exactly solvable, and through the study of these models many of the insights from random circuits could be demonstrated to extend to more structured settings. For instance, dual-unitarity has enabled exact calculation of the dynamics of correlation functions~\cite{bertini_exact_2019,claeys_ergodic_2021}, entanglement ~\cite{bertini_entanglement_2019,gopalakrishnan_unitary_2019, zhou_maximal_2022}, out-of-time-order correlators~\cite{claeys_maximum_2020,bertini2020scrambling}, deep thermalization~\cite{ho_exact_2022,claeys_emergent_2022,ippoliti_dynamical_2023,stephen_universal_2024}, and Hilbert space delocalization~\cite{claeys_fock-space_2024}.
Dual-unitary circuits have also been realized experimentally in various quantum computing platforms~\cite{chertkov_holographic_2022, mi_information_2021,fischer_dynamical_2024}. Since the discovery of dual-unitarity, further examples of solvable dynamics have been found, such as triunitarity and round-a-face circuits involving 3-site operators~\cite{jonay_triunitary_2021,prosen_many-body_2021,claeys_dual-unitary_2024}, in circuits with (2+1)d square lattices~\cite{suzuki_computational_2022,milbradt_ternary_2023}, and in (1+1)d nonsquare lattices with different space-time symmetries, leading to hierarchical extensions of dual-unitarity~\cite{yu_hierarchical_2024,rampp_entanglement_2023,rampp_geometric_2025,sommers_zero-temperature_2024,foligno_quantum_2024,liu_solvable_2025}. 

While most work in these directions has focused on standard Euclidean geometries, the versatility of quantum devices has now reached a level where this is not a necessary restriction~\cite{kollar2019hyperbolic,avikar2021programmable,ramette2022any_to_any,xu2024constant,bluvstein2024logical,grass_synthetic_2025}.
In fact, implementing non-Euclidean geometries may be highly beneficial. For example, they potentially allow the implementation of  quantum error correcting codes with far lower overhead~\cite{breuckmann2016hyperbolic,breuckmann2017hyperbolic,fahimniya2023hyperbolic_floquet,higgott2024hyperbolic_floquet,breuckmann2021qldpc_review} and, from a more physical perspective~\cite{rakovszky2023ldpc1,rakovszky2023ldpc2}, have been shown to host unique dynamical and thermodynamic phases of matter~\cite{laumann2017localization,chao2024localization,deRoeck2024qldpc_stability,chao2024qldpc_stability,hong2025self_correction,placke2024tqsg}. 
Quantum states defined on hyperbolic geometries and constructed out of perfect tensors \deletedtwo{(a generalization of dual-unitary gates)} have also been proposed as minimal models for holography~\cite{pastawski_holographic_2015,evenbly_hyper-invariant_2017,harlow_tasi_2018,bistron_bulk-boundary_2025}. 
Among non-Euclidean lattice geometries, one of the simplest is that of the Cayley tree. The absence of loops has made trees a fruitful source of analytically solvable statistical mechanics models~\cite{gujrati_bethe_1995, sommers_dynamically_2025}, which can be strikingly different from their Euclidean counterparts~\cite{laumann_absence_2009}.

Non-Euclidean lattice geometries explicitly break the symmetry between space and time, such that it is not clear if and how they support a notion of space-time duality. Additionally, the exact solvability of dual-unitary circuits is underpinned by the notion of an operator propagating in a unique `maximum velocity' direction~\cite{claeys_maximum_2020}, where the more complicated geometry of non-Euclidean lattices typically gives rise to an exponentially growing number of such directions. The corresponding difficulty of simulating the quantum dynamics due to this rapid growth further motivates the study of exactly solvable models of many-body dynamics on tree geometries, the focus of the present work. Before proceeding to a summary of our results, a word of clarification is in order: many existing results on dynamics on trees \cite{feng_measurement-induced_2023,sommers_dynamically_2025,ravindranath_entanglement_2025,nahum_measurement_2021,feng_charge_2025} invoke a tree-like tensor network structure of {\it spacetime}. In contrast, here we exclusively consider the setting where only the spacelike direction is a tree: in other words, the tensor network for the time evolution over discrete time $t$ consists of $t$ tree-like time slices, with local unitary gates acting between time slices [see, e.g. Fig.~\ref{fig:tree_brickwork}(b) or Fig.~\ref{fig:3site_coloring}(b)]. This more closely resembles the kind of  evolution we might imagine implementing on a quantum simulator with flexible spatial connectivity, and which would apply in the physical setting of expander-graph constructions of quantum codes. An intuitive and vivid description for our construction, reflected in our title, is that we work in ``tree+1'' dimensions, with the extra, non-tree-like dimension being time.\footnote{We thank Vedika Khemani for suggesting this terminology.} 
\added{Such a ``graph+1'' structure also appears when using a code on a given graph as a quantum memory, with the time direction arising from the need for repeated (noisy) syndrome measurements, although in this case the problem on the resulting geometry is a classical statistical mechanics model rather than a quantum dynamics one \cite{dennis_topological_2002,li_perturbative_2025}. }

Since brickwork structures have provided rich insights into local unitary dynamics in the Euclidean context,  we begin by extending the concept to the Cayley tree. A na\"ive generalization of the notion of brickwork circuits in terms of 2-site unitary gates gives rise to highly anisotropic dynamics (Sec.~\ref{sec:2site}). Instead, we show that for a tree with coordination number $z$, a brickwork structure in terms of $z$-site gates should be deployed in order to recover isotropic dynamics (Sec.~\ref{subsec:isotropic_zsite}). For the corresponding $z$-site gates, the notion of dual-unitarity can then be extended, \addedtwo{despite the lack of clear space-time duality,}  to a notion that we dub `tree-unitarity', where the tree geometry results in different space-time isometries (Sec.~\ref{subsec:def_treeunitary}). 
Dual-unitarity can be realized through kicked Ising dynamics on a one-dimensional lattice, and kicked Ising dynamics on the tree geometry realizes tree-unitarity (Sec.~\ref{subsec:kim}). More generally, for $z=2$, tree-unitary circuits reduce to the familiar dual-unitary brickwork circuits. In both dual-unitary and tree-unitary models, the dynamical correlation functions are restricted to the edge of a causal light cone (light tree), where they can be exactly calculated using a quantum channel approach (Sec.~\ref{subsec:correlations}). 
Out-of-time-order correlation functions (OTOCs) can be similarly characterized, although tree-unitary circuits no longer automatically exhibit the ballistic spreading with maximal butterfly velocity that is characteristic of dual-unitary circuits (Sec.~\ref{subsec:otoc}). Tree-unitarity does result in maximal entanglement growth, where for specific initial states and subregions, the subregion entanglement grows exponentially before saturating at the maximally mixed value (Sec.~\ref{subsec:entanglement}). 
\added{We then argue that the non-Euclidean geometry is crucial in obtaining solvable dynamics without maximal butterfly velocity (Sec.~\ref{subsec:tree_solvability}).}
A maximal butterfly velocity is recovered upon imposing additional constraints on top of tree-unitarity, which also imply that dynamical correlations along any non-maximal-velocity direction vanish identically (Sec.~\ref{sec:maxvel}). In this way, these models we introduce highlight the features of operator spreading that are particular to the tree geometry, including the proliferation of different paths, the exponential growth of subsystem entanglement, and the trade-off between maximal butterfly velocity and non-vanishing correlation functions. We conclude with a discussion and outlook (Sec.~\ref{sec:conclusion}).

\section{Motivating the $z$-site unitary circuit construction}
\label{sec:2site}

\begin{figure*}[t!]
\centering
\includegraphics{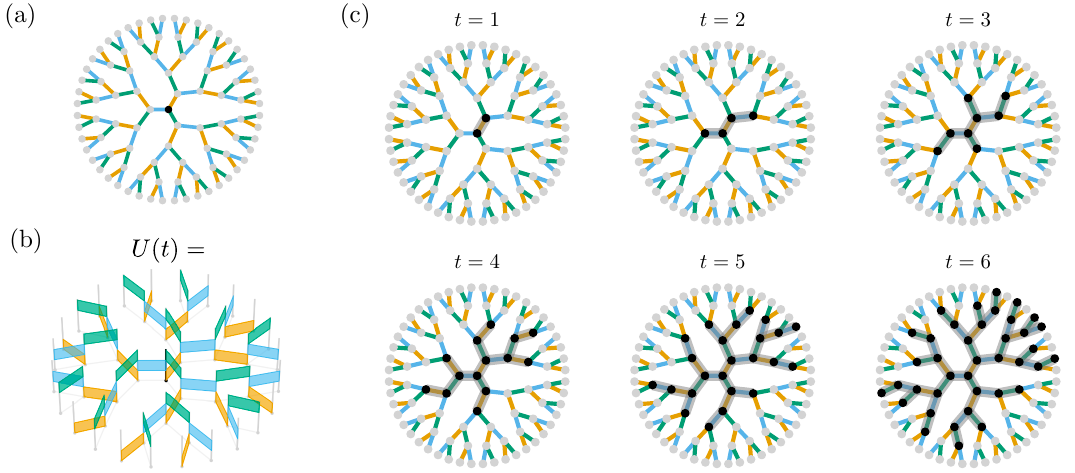}
\caption{ (a)~A coloring of the $z=3$ tree with 3 colors. A given circuit realization may apply unitaries in the ordering orange, blue, green, for example. (b) The corresponding time-evolution operator for 6 layers (2 full periods). The colored rectangles correspond to the two-site unitary gates, which need not all be identical. (c) The light cone for a circuit constructed by $z$ layers of two-site unitary gates is strongly asymmetric. Here we highlight how an operator initially at the root will spread upon application of layers in the order orange, blue, green. Along a given path, the motion appears as a sequence of steps which may be either `fast' (occur immediately) or `slow' (occur after an additional time step).}
\label{fig:tree_brickwork}
\end{figure*}

In this section we will sketch an (ultimately unsatisfactory) attempt to construct unitary circuits on the Cayley tree by using 2-site unitary gates in a `brickwork' construction. This initially may seem the most natural way to lift the ideas of dual-unitarity to the tree setting. However, as we show below, any unitary circuit built from two-site gates on a tree will necessarily involve several pathologies. Ultimately, these are linked to the necessity of choosing a sequence in which to apply the $z$ possible 2-site gates at each node, which breaks an extensive number of tree symmetries. The origins of the issue are already visible in the (1+1)d brickwork setting, where the 2-site decomposition necessarily breaks the symmetry between even and odd sites on the lattice. This is exponentially magnified in the tree setting, since there is a sequence choice to be made at {\it each} node. The end result is a highly anisotropic `light cone' on the tree: the time taken for correlations to spread to different sites at a fixed depth from a given site depends strongly on the specific sequence of steps taken to reach that site, and the resulting order of application of unitaries, violating the (large) set of symmetries of the tree.

To see this more concretely, consider such a  2-site unitary 
 construction applied to the $z=3$  Cayley tree in Fig.~\ref{fig:tree_brickwork}(a). We color the bonds of the tree such that no bonds of the same color share a vertex, which prescribes a particular sequence of application of 2-site gates. 
One such coloring is shown in  Fig.~\ref{fig:tree_brickwork}(a), in which we apply 2-site gates to the orange, blue, and green bonds in sequence, as shown in Fig.~\ref{fig:tree_brickwork}(b).
 
 Under such a brickwork unitary evolution,  an operator initially localized to a single site spreads as shown in Fig.~\ref{fig:tree_brickwork}(c). To simplify the picture, we only draw the tree and the gates, and highlight by dark shading the possible support of the operator as it evolves.
Since $z=3$, for any step except for the first, there are two  `forward' directions in which the operator could spread (for the first step, there are three choices). Since  different forward directions by construction have different colors, the forward propagation under the unitary evolution occurs first in one direction, then the other.

Since such choice occurs at {\it each} vertex, any path between two vertices is characterized by a sequence of `fast' or `slow' steps. In a fast step the support grows immediately along the path, while in a slow step the front waits for one time step before progressing. 
Consequently, a brickwork construction necessarily produces a highly asymmetric ``light cone'' (i.e. the boundary of operator support at any time that is consistent with unitarity of all gates), and in particular the existence of a unique ``fastest'' path. The fastest path is defined by the sequence of sites in the operator support that are farthest from the origin at any given time $t$.

For brickwork circuits in one dimension, the light cone spreads with the same speed in both directions, with a small anistropy based on the fact that in the very first step an operator either propagates to the left or right depending on whether it is associated with an even or an odd site. 
For higher but still finite dimensions, the situation is slightly more involved, but the light cone nevertheless typically retains the discrete symmetry of the lattice, or a subset thereof. For example, 2-site brickwork unitary evolution on the square lattice entails distinguishing `fast' paths between the original site and the corners of the `light pyramid' and slower paths to its perimeter, but preserves the four-fold rotational and two-fold mirror symmetries of the lattice. 
This should be contrasted with the tree setting: if we consider the `root' of the tree to be at the position of the operator at $t=0$, then demanding isotropy under the tree symmetries imposes the much more stringent constraint of a uniform operator front at a given depth (possibly up to small differences based on the starting site). Evidently, this is violated by the 2-site brickwork evolution, where the anisotropy occurs {\it on the light cone}.

While a fundamental property of any unitary brickwork construction involving 2-site gates, the consequences of this anisotropy are particularly severe for dual-unitary gates\deletedtwo{, where (in finite dimension) all the nontrivial correlations lie {\it on} the light cone}. This na\"ive generalization of dual-unitarity to the tree setting exhibits various somewhat pathological features. The resulting dynamics may be summarized as follows (details are provided in App.~\ref{app:2site_dynamics} for the interested reader):
\begin{enumerate}[(i)]
    \item two-point correlation functions vanish everywhere except on the light cone in the direction of the unique fastest path;
    \item in the fastest direction, operators spread at the maximum velocity allowed by unitarity;
    \item in other directions, operators do not spread at the maximum velocity allowed by unitarity; and
    \item entanglement dynamics can be calculated exactly for special initial states; in such states, the entanglement entropy of a subregion of `radius' $r$ grows as $S(t)=(z-1)^r+\frac{2(z-1)^r}{z-2}\left[1-(z-1)^{-t}\right]$, saturating at time $t=r$.
\end{enumerate}
While possibly interesting in their own right, these results  highlight that \replacedtwo{such an}{the na\"ive} extension of dual-unitarity is not particularly appealing, especially from the symmetry perspective. Therefore, it is clearly desirable to identify a different generalization of the brickwork construction and of dual-unitarity that preserves more of the symmetries of the tree. In the remainder of this paper, we address this question by introducing the distinct notion of `tree-unitarity', and exploring its consequences.

\section{$z$-site Unitary Circuits and `Tree-Unitarity'}
\label{sec:zsite}

From the results of the previous section, it is clear that the coordination number of the tree should be directly reflected in the choice of unitary gates. In this section, we introduce a different approach to unitary dynamics on trees based on groupings of $z$-site operators, which presents a more natural generalization of circuit dynamics (both unitary and dual-unitary) to the tree setting. In other words, the `natural' Trotterization of unitary dynamics on trees involves 2-colorings of the tree into $z$-site gates, rather than $z$-colorings into 2-site gates.  Extensions of dual-unitarity can be formulated for these circuits in a direct way, and we will identify the necessary set of conditions leading to the analogous property, which we dub `tree-unitarity'. Imposing these conditions results in solvable dynamics of correlation functions, out-of-time-order correlation functions, and entanglement, while avoiding the pathological behavior highlighted in the preceding section.

\subsection{Isotropic unitary dynamics on trees}
\label{subsec:isotropic_zsite}

\begin{figure}[t]
\centering
\includegraphics[width=1.0\columnwidth]{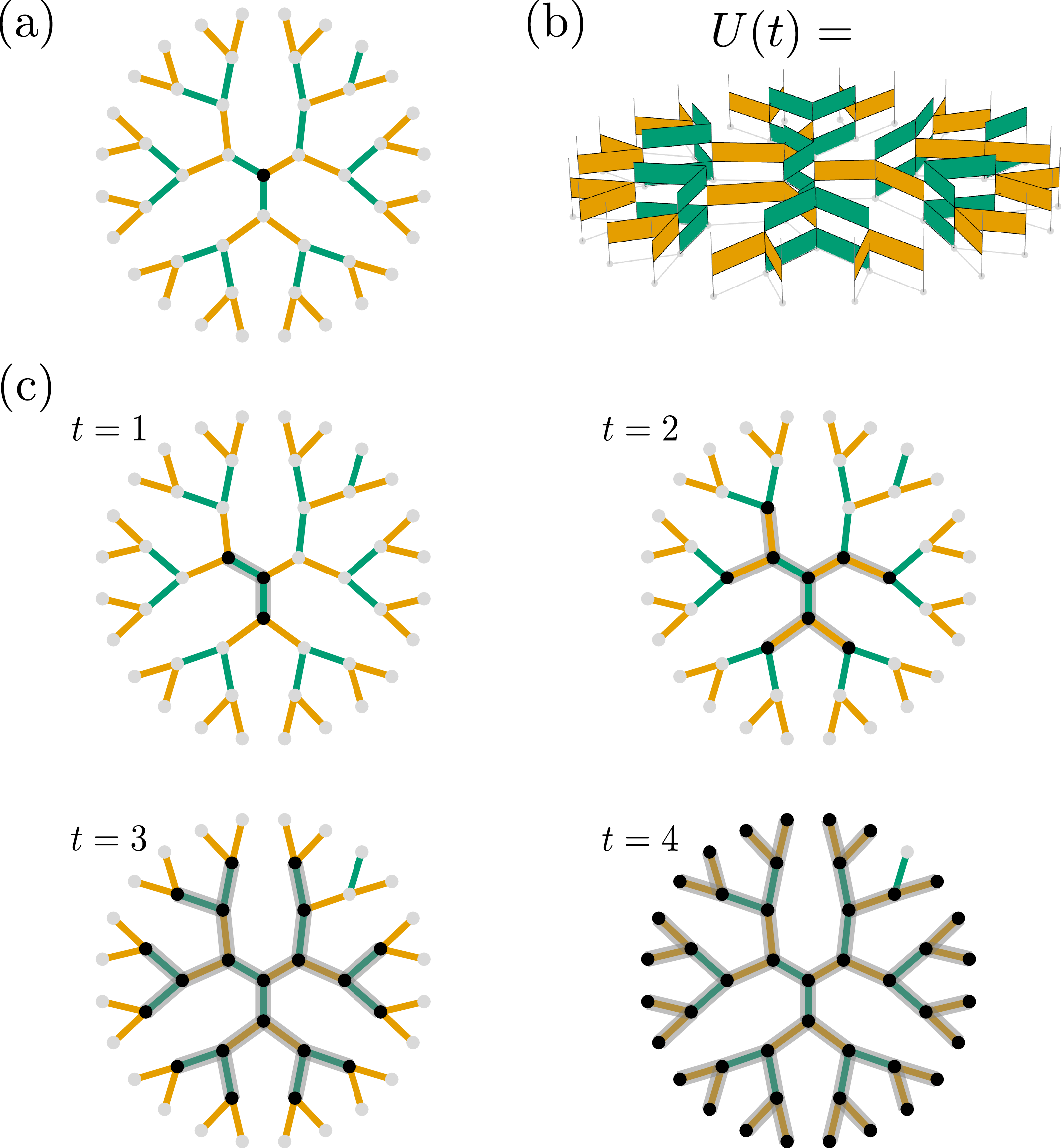}
\caption{(a) A coloring of the $z=3$ Cayley tree into two sets of 3-site unitaries. (b) The corresponding time-evolution operator for 3 layers ($1.5$ full periods). (c) The light cone for a circuit constructed by 2 layers of $z$-site gates generates a symmetric light cone, here highlighted in grey. 
}
\label{fig:3site_coloring}
\end{figure} 

As a first step, recall that the conventional brickwork Trotter decomposition in one dimension separates the unitary evolution into gates acting on even and odd bonds.  In the same spirit, for dynamics on a tree we can identify two `layers' of the time evolution by decomposing the bonds of the tree into two sets\footnote{Such a decomposition is also used for algorithms such as infinite time-evolving block decimation on a tree~\cite{nagaj_quantum_2008}, where taking $z$-site gates allows as much symmetry as possible to be preserved when Trotterizing the evolution.}. 

To do so, we iteratively `grow' a 2-coloring of the edges of the tree into green and orange sets (Fig.~\ref{fig:3site_coloring}(a)) by alternating between odd and even steps as follows. In an odd step, we consider each vertex on the boundary of the $2-$colored region, and ensure that exactly one of its $z$ neighbours is green and the remaining $z-1$ are orange. In an even step, we do the converse, i.e. ensure that exactly one of the neighbours is orange and the remaining $z-1$ are green. 
Iterating this procedure generates a $2$-coloring of the tree edges that partitions it into clusters of $z$ vertices connected by edges of the same color, such that every vertex (in the finite case, every vertex that is not a leaf of the tree) belongs to exactly two clusters of opposite color. We can then apply $z$-site unitary gates to all the clusters of a given color in one time step, and then apply them to clusters of the opposite color in the next time step, thereby directly generalizing the two-layer unitary circuits from one dimension to the tree setting.

We now show that this decomposition into $z$-site unitaries leads to an isotropic light cone, with the same velocity in any direction, up to a possible correction of a single time step (similar to the odd-even anisotropy in one dimension). Specifically, we prove that the number of time steps $t_{i\to j}$ for an operator initially localized at site $i$ to spread to site $j$ satisfies the inequality
\begin{equation}
r_{i,j}-1\leq t_{i\to j} \leq r_{i,j}+1,\label{eq:timebound}
\end{equation}
where $r_{i,j}$ is the graph distance between $i$ and $j$.

To see this, first observe that there is a unique path that connects $i$ and $j$ on the tree; the $2$-coloring of the graph edges then induces an edge coloring of this path. Given our layered unitary construction, at any time step the operator front can advance from its current site to any site linked to it by (one or more) edges of the same color. Thus, depending on whether the first gate applied is the same or opposite color as the first edge of the path, we see that $t_{i,j} = n_c$ or $t_{i,j} = n_c+1$, where $n_c$ is the number of edge color changes in the path, and the offset in the second case captures the need to wait for a single time step at the outset before the front starts moving. 
To determine $n_c$, we can take the two sites to be part of some rooted tree; this can always be done without loss of generality assuming a sufficiently large system. The $2$-coloring then guarantees that each $z$-site cluster has one site at depth $d$ and $z-1$ sites at depth $d+1$ relative to this root. Now, any path between two sites $i$ and $j$ at a fixed distance $r_{i,j}=r$ from each other involves $r_+$ steps `upwards' towards the root and $r_-$ steps `downwards' towards the leaves, with $r_++r_- =r$. Evidently, paths with $r_+=0$ or $r_-=0$ never `turn away' from the root, and it is straightforward to see that the edges along such paths strictly alternate between colors, so that $n_c = r$. The only other possibility is that the path turns exactly once; in this case, the two edges adjacent to the turning vertex are of the same color, and the rest strictly alternate as before, so that $n_c=r-1$.  
Combining the results for $n_c$ with those for $t_{i\to j}$, we arrive at the inequality in \eqref{eq:timebound}.

\begin{figure}[t]
\centering
\includegraphics[width=0.45\columnwidth]{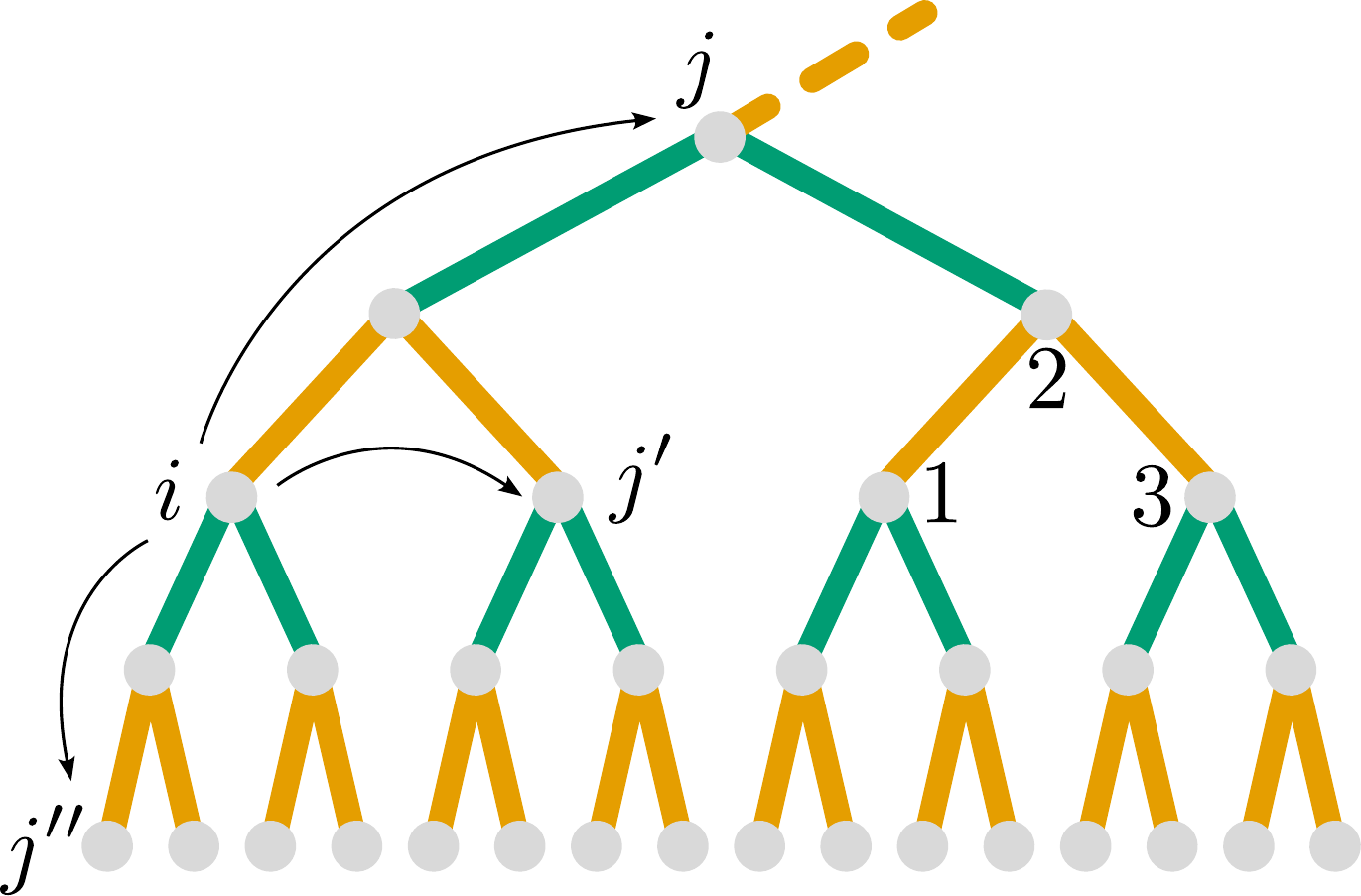}
\caption{For a pair of nodes $i,j$ separated by a fixed graph distance, the path joining them can turn away from the root at most once. The time taken for an operator to spread from $i$ to any possible $j$ at fixed distance can therefore differ by at most $2$ time steps. To the right side of the tree, we also show the labeling convention for the legs of a $z=3$ gate on the tree.}
\label{fig:rooted_tree}
\end{figure} 

\subsection{Tree-unitarity}
\label{subsec:def_treeunitary}
For brickwork circuits, imposing additional unitarity constraints such as dual-unitarity has given rise to a wealth of solvable models. Here, we extend these notions to dynamics on trees. Focusing again on $z=3$ for concreteness, where extensions to arbitrary $z$ are direct, we consider unitary three-site gates which we graphically represent as
\begin{equation}
U=\vcenter{\hbox{\includegraphics[width=0.1\columnwidth]{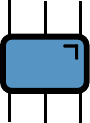}}}\,,\quad
U^{\dagger}=\vcenter{\hbox{\includegraphics[width=0.1\columnwidth]{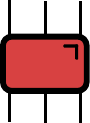}}}\,.
\end{equation}
Introducing a `folded' notation (the graphical equivalent of the operator-to-state mapping), we can write the superoperator that implements unitary conjugation by $U$ and $U^\dagger$ as
\begin{equation}
\vcenter{\hbox{\includegraphics[width=0.12\columnwidth]{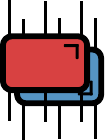}}}=\vcenter{\hbox{\includegraphics[width=0.1\columnwidth]{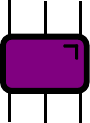}}}.
\end{equation}
Unitarity can then be graphically represented via
\begin{equation}
UU^{\dagger}=\vcenter{\hbox{\includegraphics[width=0.11\columnwidth]{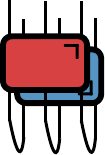}}}=
\vcenter{\hbox{\includegraphics[width=0.1\columnwidth]{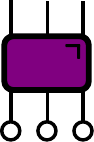}}}=
\vcenter{\hbox{\includegraphics[width=0.1\columnwidth]{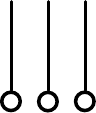}}}=
\vcenter{\hbox{\includegraphics[width=0.1\columnwidth]{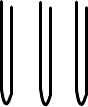}}}
\end{equation}
and
\begin{equation}
    U^{\dagger}U=\vcenter{\hbox{\includegraphics[width=0.1\columnwidth]{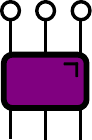}}}=
    \vcenter{\hbox{\includegraphics[width=0.1\columnwidth]{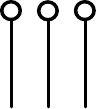}}}.
\end{equation}
This folded notation will prove to be convenient when describing operator dynamics.

In order to extend dual-unitarity to the Cayley tree and obtain solvable dynamics, we consider gates that satisfy the additional properties
\begin{equation}\label{eq:tree_unitarity_conditions}
\vcenter{\hbox{\includegraphics[width=0.1\columnwidth]{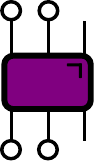}}}=
\vcenter{\hbox{\includegraphics[width=0.1\columnwidth]{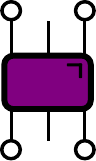}}}=
\vcenter{\hbox{\includegraphics[width=0.1\columnwidth]{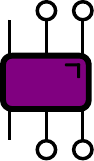}}}=q^{z-2}~
\vcenter{\hbox{\includegraphics[width=0.025\columnwidth]{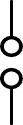}}}\,,
\end{equation}
with $q$ the dimension of the local Hilbert space and $z=3$ in this example. These constraints can be extended to more general $z$, in which we trace over $z-1$ pairs of sites and require that we obtain a term proportional to the identity. We denote this set of conditions \emph{tree-unitarity}.

Expressed in the components of $U$, these conditions read 
\begin{align}
    \sum_{ijlm} U^{ijk}_{lmn} (U^{\dagger})^{lmc}_{ijf} &= \sum_{ijlm} U^{jki}_{mnl} (U^{\dagger})^{mcl}_{jfi}\nonumber\\
    &=\sum_{ijlm}U^{kij}_{nlm} (U^{\dagger})^{clm}_{fij}
    =\addedtwo{q^{z-2}}\delta^k_f \delta^c_n.
\end{align}
As will be shown in the next section, tree-unitarity results in dynamics that exhibits the same qualitative behavior as for dual-unitarity, in that, strikingly, all dynamical correlation functions vanish everywhere except on the edge of the causal light cone. 

The conditions are qualitatively similar to dual-unitarity, but the tree structure results in some crucial differences. The conditions from Eq.~\eqref{eq:tree_unitarity_conditions} impose that the unitary gates are isometries from the composite Hilbert spaces corresponding to any pair of opposite legs to the remaining $2(z-1)$ legs.\footnote{Experts may recognize a \deletedtwo{family} resemblance to $2z$-leg {\it perfect tensors}, which are isometries from {\it any} set of $z'\leq z$ legs to the remaining $2z-z'$ legs. This is equivalent, when all the legs are identical, to {\it unitarity} for any {\it balanced} bipartition where $z'=z$. The tree-unitary constraints are weaker than this, since they only require isometry for bipartitions with $z'=2$ and legs on opposite sides. Perfect tensors are thus a subset of tree-unitary gates, with special features discussed in Sec.~\ref{sec:maxvel}.} 
This is to be contrasted with the dual-unitarity condition for 2-site gates where the single additional condition is that of {\it unitarity}; for $z$-site gates with $z >2$, the $z$ additional conditions are necessarily modified to isometries. 
Dual-unitarity is explicitly recovered for $z=2$. \replaced{For $z>2$ tree-unitarity imposes $z$ independent isometric constraints, whereas for $z=2$ the two unitary constraints are equivalent.}{(the two possible ways of writing (\ref{eq:tree_unitarity_conditions}) for $z=2$ lead to equivalent conditions, which is not the case for $z>2$.).}

It is natural to ask if and how tree-unitary gates can be constructed for different coordination number $z$ and local Hilbert space dimensions $q$.
Numerically, tree-unitary gates can be constructed \deletedtwo{in a straightforward manner} through an extension of the algorithm proposed in Ref.~\onlinecite{rather_creating_2020} for the construction of dual-unitary gates, as detailed in App.~\ref{app:gen_tu}. A numerical estimation of the dimension of the manifold of tree-unitary gates using the approach of Ref.~\onlinecite{prosen_many-body_2021} suggests that, for 3-site gates with $q=2$, tree-unitary gates form a $37$-dimensional space. This dimension is substantially smaller than that of the space of 3-site unitary operators on qubits, which has dimension 64, but nevertheless is a large parameter space containing a range of interesting behavior. 

For $z=3$, the triunitary gates introduced in Ref.~\onlinecite{jonay_triunitary_2021} can be deformed into 3-site tree-unitary gates (see App.~\ref{app:triunitary}). However, gates constructed this way are non-generic (in a sense made precise in Sec.~\ref{sec:maxvel}), with tree-unitarity gates presenting a much broader class, and are furthermore not restricted to $z=3$.

\subsection{Kicked Ising Model}
\label{subsec:kim}

A particular tree-unitary circuit can be obtained by considering kicked Ising dynamics. The self-dual kicked Ising model (KIM) is a paradigmatic realization of dual-unitary dynamics~\cite{Gutkin,gopalakrishnan_unitary_2019,gutkin_exact_2020,ho_exact_2022,stephen_universal_2024,claeys_operator_2024} and can be directly extended to the Cayley tree. While the resulting gates do not exhaust the full space of tree-unitary gates, and are in some aspects nongeneric, this model serves to both highlight that tree-unitary gates can be constructed for any coordination number $z$, and introduces a circuit that is naturally realized in (Hamiltonian) Floquet dynamics.

We motivate this construction by first considering Floquet dynamics on a Cayley tree, with every vertex supporting a spin-$1/2$ degree of freedom. Dynamics under a classical Ising Hamiltonian $H_{\textrm{I}}$ is periodically alternated with a transverse kick $H_{\textrm{K}}$, where
\begin{align}
    H_{\textrm{I}} = J \sum_{\langle i,j \rangle} \sigma^z_{i} \sigma^z_{j} +\sum_{i} h_j \sigma^z_j, \quad
    H_{\textrm{K}} =  b \sum_{j} \sigma_j^y.
\end{align}
Here the Ising interaction acts on neighboring sites $\langle i,j \rangle$,  $J$ and $b$ are the Ising interaction strength and the transverse kick strength, respectively, and $\sigma_i^{\alpha}$ with $\alpha \in \{x,y,z\}$ are the Pauli matrices. $\{h_j\}$ describes a (possibly inhomogeneous) longitudinal field. The dynamics after $t$ periods can be written as
\begin{align}
\label{eq:kim_ut}
    U(t) = \left(e^{-i H_\textrm{K}}e^{-i H_\textrm{I}}\right)^t,
\end{align}
where we have absorbed all time scales into the couplings in $H_I$ and $H_K$. When restricted to a one-dimensional lattice, this dynamics can be recast as a dual-unitary circuit at the parameter values $J = b = \pi/4$~\cite{bertini_exact_2018}. On the Cayley tree, it is now possible to recast this dynamics as a tree-unitary circuit. Introducing 2-site Ising gates and 1-site kicked gates as
\begin{align}
    \mathcal{I}_{12} = e^{-iJ \sigma^z_{1} \sigma^z_2 -i (h_1 \sigma^z_1+h_2 \sigma^z_2)/2}, \qquad \mathcal{K} = e^{-i b \sigma^y},
\end{align}
we can identify the dynamics with a unitary circuit generated by arranging $z$-site gates according to the geometry shown in Fig.~\ref{fig:3site_coloring}. The constituting gates are of the form
\begin{align}\label{eq:KIM_gate}
    U = \prod_{\langle i,j\rangle}^z\mathcal{I}_{ij} \prod_{j=1}^z \mathcal{K}_j \prod_{\langle i,j\rangle}^z\mathcal{I}_{ij} ,
\end{align}
where the Ising gates act on all neighboring \addedtwo{lattice} sites inside the cluster of $z$ sites. These gates satisfy the tree-unitary property [Eq.~\eqref{eq:tree_unitarity_conditions}] for the dual-unitary parameters $|J|=|b|=\pi/4$. 
Note that $U$ reduces to a Clifford gate if $h_j = \mathbb{Z} \pi/8, \forall j$; this fact allows for numerical simulations of tree-unitary dynamics for large system sizes.

It is easiest to see that \deletedtwo{that} gates of this form satisfy tree-unitarity using the graphical tensor network language of Refs.~\onlinecite{ho_exact_2022,stephen_universal_2024}. By introducing the building blocks
\begin{align}
\label{eq:hadamard_gate}
\vcenter{\hbox{\includegraphics[height=0.12\columnwidth]{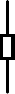}}}=\frac{1}{\sqrt{2}}
\begin{pmatrix}
    1 & -i \\
    -i & 1
\end{pmatrix}=H , \quad \vcenter{\hbox{\includegraphics[height=0.12\columnwidth]{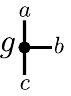}}}=\delta_{abc} e^{-ig(1-2a)},
\end{align}
 we can write the Ising and kick terms as
 \begin{align}
 \label{eq:ising_kick}
\mathcal{I}_{12} = \added{\sqrt{2}~}\,\vcenter{\hbox{\includegraphics[height=0.12\columnwidth]{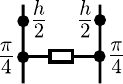}}}\,, \quad \mathcal{K} = \vcenter{\hbox{\includegraphics[height=0.12\columnwidth]{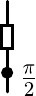}}}.
 \end{align}
In terms of these, the $z=3$ gate~\eqref{eq:KIM_gate} can be graphically represented as 
\begin{align}
\label{eq:u_kim_tensor}
       U = \added{4~}\,\vcenter{\hbox{\includegraphics[height=0.17\columnwidth]{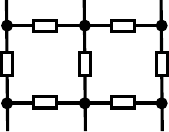}}}\,,
\end{align}
where we have suppressed writing the phases in the $\delta$-tensors. 
Similarly, the $z=4$ gate is
\begin{align}
       U = \added{8~}\, \vcenter{\hbox{\includegraphics[height=0.24\columnwidth]{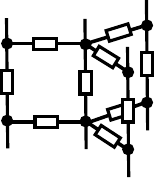}}}\,.
\end{align}

In proving identities such as tree-unitarity, it is useful to note that in the folded representation, 
\begin{gather}
\label{eq:had_identities}
\vcenter{\hbox{\includegraphics[height=0.15\columnwidth]{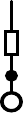}}}=\vcenter{\hbox{\includegraphics[height=0.15\columnwidth]{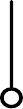}}}, \qquad
\vcenter{\hbox{\includegraphics[height=0.17\columnwidth]{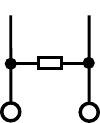}}}=\added{\frac{1}{2}~}\,\vcenter{\hbox{\includegraphics[height=0.17\columnwidth]{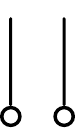}}} \,.
\end{gather}

Using these, we can show that the KIM satisfies tree-unitarity as follows% \added{(with the $4^2$ arising from the $4$ in Eq.~\ref{eq:u_kim_tensor} and its conjugate folded behind)}
\begin{align}
\label{eqn:kim_tu}\added{4^2~}
    \vcenter{\hbox{\includegraphics[height=0.2\columnwidth]{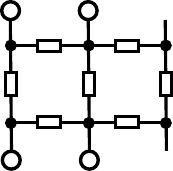}}}= \added{8~}\,
    \vcenter{\hbox{\includegraphics[height=0.2\columnwidth]{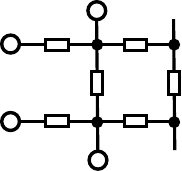}}}= \added{4~}\,
    \vcenter{\hbox{\includegraphics[height=0.2\columnwidth]{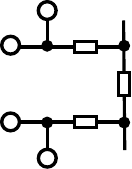}}} \nonumber\\
    =\added{4~}\,\vcenter{\hbox{\includegraphics[height=0.17\columnwidth]{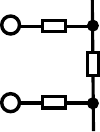}}}=\added{2~}\,
    \vcenter{\hbox{\includegraphics[height=0.15\columnwidth]{figures_final/double_identity.pdf}}}.
\end{align}
This derivation directly extends to the remaining tree-unitary conditions and arbitrary $z$.

\begin{figure}
\centering
\includegraphics[width=7.5cm]{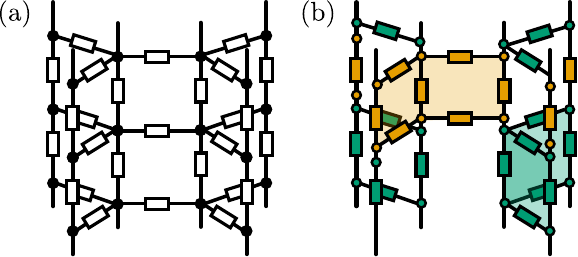}
\caption{\added{(a) Kicked Ising model (KIM) on the $z=3$ tree in full tensor network form. (b) Up to a boundary layer of unitary gates, the KIM tensor network can be regrouped into identical $z$-site gates in a 2-layer circuit structure. Here, orange and green indicate the two layers of the circuit, corresponding to Fig.~\ref{fig:3site_coloring}. Two complete gates in this section of the network are shaded orange and green respectively. }}
\label{fig:kim_bricking}
\end{figure}

\added{In Fig.~\ref{fig:kim_bricking} we explicitly show how one passes from a tensor network written in terms of kicks and Ising phases to a brickwork structure written in terms of $z$-site gates (up to a unitary transformation at the initial and final time, which does not change the bulk dynamics). A key point that enables the grouping of kicks and Ising gates into compound $z$-site gates is that during a single period each site experiences a kick followed by Ising interactions with all $z$ neighbors. Since the latter commute, these can be realized in any order. This rewriting of the KIM as a $z$-site brickwork circuit offers further motivation for the considered circuit structure. Making the transition to $z$-site gates allows us to identify how the solvability of the KIM in this geometry arises from the tree-unitary properties of the individual gates; these constraints in turn allow us to move beyond the single example of kicked Ising dynamics and consider a much broader class of models.}

\added{This grouping is not unique. It is also possible to identify the kicked Ising dynamics with 2-site gates, with layers forming a $z$-colouring of the circuit and using two different types of gate (either Ising phases or Ising phases with kicks). The important difference relative to the grouping in terms of $z$-site gates is that the resulting circuit has in a \textit{physical} light cone that, while isotropic, only grows at the rate of the `slow' steps, and which therefore lags the  \textit{geometric} light cone identified in Sec.~\ref{sec:2site}. This difference is due to the fact that different $2$-site Ising gates in this decomposition commute.
%the , with the \textit{physical} light cone lagging behind and growing isotropically at the rate of the `slow' steps, due to the specific form of 2-site gates. 
In other words, while it is possible at a technical level to write the KIM in terms of 2-site gates with correlations spreading isotropically, the resulting dynamics is unnatural as it leads to a disconnect between the physical and geometric light cones. The isotropy then also relies on fine-tuned properties of the 2-site gates for the KIM: for generic 2-site gates, the light cone would be anisotropic as shown in Sec.~\ref{sec:2site}. This observation further underscores the importance of our $z$-site gate construction to obtaining isotropic light cones for {\it generic} choices of unitary gates in a tree+1 circuit. }

\section{Dynamics of Tree-Unitary Circuits}
\label{sec:treeunitarydynamics}
Having introduced the basic structure and established the isotropy of the light cone, we now consider the dynamics of circuits on the tree. We focus on two-point correlation functions, out-of-time-order correlation functions, and entanglement growth following a quench. All the results of this section apply to {\it any} tree-unitary circuit; in the following Sec.~\ref{sec:maxvel}, we discuss special cases which constrain the dynamics further. Note that we will generally consider dynamics on the infinite Bethe lattice (regular tree). However, due to the causal structure of the circuit, dynamics can be restricted to the Cayley tree with the leaves on the edge of the causal light cone without loss of generality.

\subsection{Correlation Functions}
\label{subsec:correlations}

We first consider two-point correlation functions of the form
\begin{equation}\label{eq:corr_function}
    c_{\alpha \beta} (i,j;t)= \langle \sigma_{\beta}(i,0) \sigma_{\alpha}(j,t)\rangle,
\end{equation}
at infinite temperature, with $\langle \bullet \rangle \equiv \mathrm{tr}(\bullet)/\mathrm{tr}(\mathbb{1})$. Here $\sigma_{\alpha,\beta}$ are (possibly generalized) Pauli matrices satisfying $\mathrm{tr}(\sigma_{\alpha} \sigma_{\beta})/q = \delta_{\alpha \beta}$ for local Hilbert space dimension $q$. Identifying $\sigma_{0} = \mathbb{1}$, all other Pauli matrices are fixed to be traceless and chosen to be Hermitian, and we will focus on $\alpha,\beta \neq 0$ in what follows. We take $\sigma_{\alpha}(j)$ to act as $\sigma_{\alpha}$ on site $j$ and as the identity everywhere else and define the dynamics as $\sigma_{\alpha}(j,t) = U(t)^{\dagger} \sigma_{\alpha}(j) U(t)$. 

The correlation function \eqref{eq:corr_function} vanishes due to unitarity whenever one Pauli operator lies outside the light cone of the other. In (1+1)d, while correlation functions inside this light cone are generally exponentially costly to compute, correlation functions \textit{on} the 
light cone admit an efficient calculation in terms of quantum channels~\cite{bertini_exact_2019,claeys_maximum_2020}. For dual-unitary circuits, correlation functions inside the causal light cone additionally vanish due to the space-time duality, such that the only nonzero correlations are those on the edge of the causal light cone.

On the tree, we again have vanishing correlation functions outside the light cone, and correlation functions inside the light cone are exponentially costly to compute in general. On the edge of the light cone, we can calculate the correlation functions using quantum channels $\mathcal{M}_{e\tilde{e}}(\sigma)$ with $e,\tilde{e} \in \{1,2,\dots,z\}$. 
The action of these channels is defined by conjugating the operator on incoming leg $e$ with $U$, then tracing over all legs apart from $\tilde{e}\neq e$:
\begin{align}
\label{eq:M_def}
    \mathcal{M}_{e\tilde{e}}(\sigma) =  \overline{\mathrm{tr}}_{{\tilde{e}}}\left[U^{\dagger}\sigma(e) U\right]/q^{z-1},
\end{align}
where  $\overline{\mathrm{tr}}_{{\tilde{e}}}(\ldots)$ denotes a trace over all except  
the `outgoing' leg $\tilde{e}$ and $\sigma(e) = \mathbb{1}^{\otimes e-1} \otimes \sigma \otimes \mathbb{1}^{\otimes z-e}$ is fixed by the `incoming' leg $e$.
Observe that the case $e=\tilde{e}$ can never appear on the light cone, since this would correspond to the operator moving inside the light cone. 
We will also write $\mathcal{M}_{e\tilde{e}}(\sigma) \equiv \mathcal{M}_{e\tilde{e}}|\sigma)$, since these quantum channels can be interpreted as [super]operators acting on an [operator] Hilbert space spanned by the $\sigma_{\alpha}$. We define an inner product in this operator Hilbert space as $(\sigma_{\alpha}|\sigma_{\beta}) \equiv \mathrm{tr}(\sigma_{\alpha}^\dagger \sigma_{\beta})/q$, which will help to simplify the following expressions.

For fixed $z$, there are $z(z-1)$ choices of such quantum channels. Graphically, these can be represented as, for example (again fixing $z=3$ for concreteness), 
\begin{equation}
    \mathcal{M}_{23}(\sigma)=\frac{1}{q^2}\mathrm{tr_{12}}[U^{\dagger}(\mathbb{1} \otimes \sigma \otimes \mathbb{1} )U] =  \vcenter{\hbox{\includegraphics[width=0.15\columnwidth]{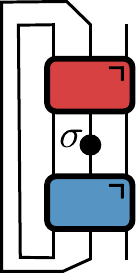}}}
\end{equation}
or in the folded representation, with all other $z=3$ channels,

\begin{align}
    \mathcal{M}_{23}|\sigma)=\vcenter{\hbox{\includegraphics[height=0.2\columnwidth]{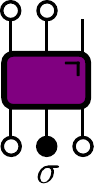}}}\,,~
    \mathcal{M}_{21}|\sigma)=\vcenter{\hbox{\includegraphics[height=0.2\columnwidth]{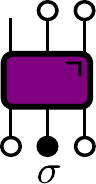}}}\,,~
    \mathcal{M}_{12}|\sigma)=\vcenter{\hbox{\includegraphics[height=0.2\columnwidth]{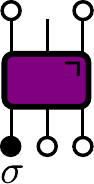}}}\,,~\nonumber \\
    \mathcal{M}_{13}|\sigma)=\vcenter{\hbox{\includegraphics[height=0.2\columnwidth]{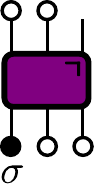}}}\,,~
    \mathcal{M}_{31}|\sigma)=\vcenter{\hbox{\includegraphics[height=0.2\columnwidth]{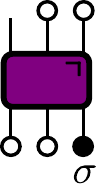}}}\,,~
    \mathcal{M}_{32}|\sigma)=\vcenter{\hbox{\includegraphics[height=0.2\columnwidth]{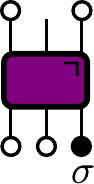}}}\,.~
\end{align}
\added{When $z=2$, these reduce to the usual quantum channels for the calculation of light-cone correlation functions in (1+1)d \cite{bertini_exact_2019},}
\begin{equation}
\mathcal{M}_{12}|\sigma)=\vcenter{\hbox{\includegraphics[height=0.2\columnwidth]{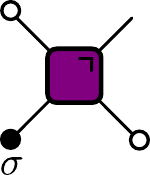}}}\,,~
    \mathcal{M}_{21}|\sigma)=\vcenter{\hbox{\includegraphics[height=0.2\columnwidth]{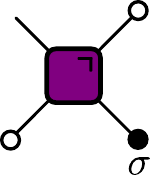}}}\,.
\end{equation}

Light-cone correlation functions between two operators can be calculated through the successive application of such quantum channels, where the precise sequence depends on the path, as we explain below. The argument is identical in most respects to the quantum channel construction for (1+1)d brickwork circuits (see, e.g., Refs.~\onlinecite{bertini_exact_2019,claeys_maximum_2020}), so we here only quote the final result and comment on key distinctions.
We consider the path for an operator initially localized at
site $i$ to spread to site $j$ as outlined in Sec.~\ref{subsec:isotropic_zsite}, and fix the number of time steps $t = t_{i \to j}$. The path connecting $i$ and $j$ consists of a set of vertices $(v_1, v_2, \dots v_t)$, with $v_1=i$ and $v_t=j$. 
For the gate that spreads the operator from $v_{\tau}$ to $v_{\tau+1}$, we take $e_{\tau}$ to be the leg acting on $v_{\tau}$ and $\tilde{e}_{\tau}$ to be the leg acting on $v_{\tau+1}$.
The resulting correlation function reads
\begin{align}
    c_{\alpha \beta}(i,j;t) =  (\sigma_{\beta}|\mathcal{M}_{e_{t}\tilde{e}_{t}} \dots \mathcal{M}_{e_{2}\tilde{e}_{2}}\mathcal{M}_{e_{1}\tilde{e}_{1}} |\sigma_{\alpha})\,.\label{eq:channelcorrelator}
\end{align}
This expression can be efficiently evaluated for arbitrarily long times. One key difference from the (1+1)d case is already apparent in the freedom to choose a (nearly) arbitrary set of ingoing and outgoing legs of the $\mathcal{M}_{e\tilde{e}}$ in \eqref{eq:channelcorrelator}. This stems from the fact that there are many distinct paths and hence choices of $e, \tilde{e}$ that remain on the light cone, in contrast to the (1+1)d case where there is a unique choice (so that, for example, propagation along the left light cone is generated by repeated iterations of the channel $\mathcal{M}_{12}$, and along the right light cone by iterations of $\mathcal{M}_{21}$). For a tree {with the layout as in Fig.~\ref{fig:3site_coloring}}, one can clearly make several more choices depending on the precise choice of path, with one important constraint: the ``single-turn'' rule (that we used in our proof of isotropy in Sec.~\ref{subsec:isotropic_zsite}) means that there can be at most one appearance of a channel of the form $\mathcal{M}_{e\tilde{e}}$ for which $e$ and $\tilde{e}$ are {\it both} `leaf' legs.
Here, the terms `leaf' and `root' leg refer to the location of the legs of the gate, with respect to the rooted tree on which we are considering the light cone, as in our discussion in  Sec.~\ref{subsec:isotropic_zsite}; each gate acts on $z$ sites, the `root' of which is nearest-neighbour to all $z-1$ `leaf' sites.
The resulting exponentially growing choice of channels to apply is in accord with the exponentially growing light cone on the tree. 

These quantum channels are unital, i.e. they satisfy $\mathcal{M}_{ij}|\mathbb{1}) = |\mathbb{1})$, similar to the quantum channels appearing in dual-unitarity. As such, the identity is a trivial eigenoperator with eigenvalue one. For ergodic dynamics the remaining eigenvalues generally lie inside the unit circle, such that the correlation functions decay exponentially with the graph distance. 
To illustrate this construction, let us consider the correlation function down the right-most path in Fig.~\ref{fig:rooted_tree}. For each step, the ingoing leg is $e=2$ and the outgoing site is $\tilde{e}=3$, such that the correlation function is
\begin{equation}
    c_{\alpha \beta}(i,j;t)= (\sigma_{\beta}|\mathcal{M}_{23}^t|\sigma_{\alpha}) = 
    \mathrm{tr}\left[\sigma_{\beta}\mathcal{M}_{23}^t(\sigma_{\alpha})\right]/q.
\end{equation}
On the other hand, a path that departs from this at the very last leaf, for example, is characterized by
\begin{align}
    c_{\alpha \beta}(i,j;t)&= (\sigma_{\beta}|\mathcal{M}_{21}\mathcal{M}_{23}^{t-1}|\sigma_{\alpha})\nonumber\\ &= 
    \mathrm{tr}\left[\sigma_{\beta}\mathcal{M}_{21}\mathcal{M}_{23}^{t-1}(\sigma_{\alpha})\right]/q,
\end{align}
and so on.

\begin{figure*}[t!]
\centering
\includegraphics[width=2\columnwidth]{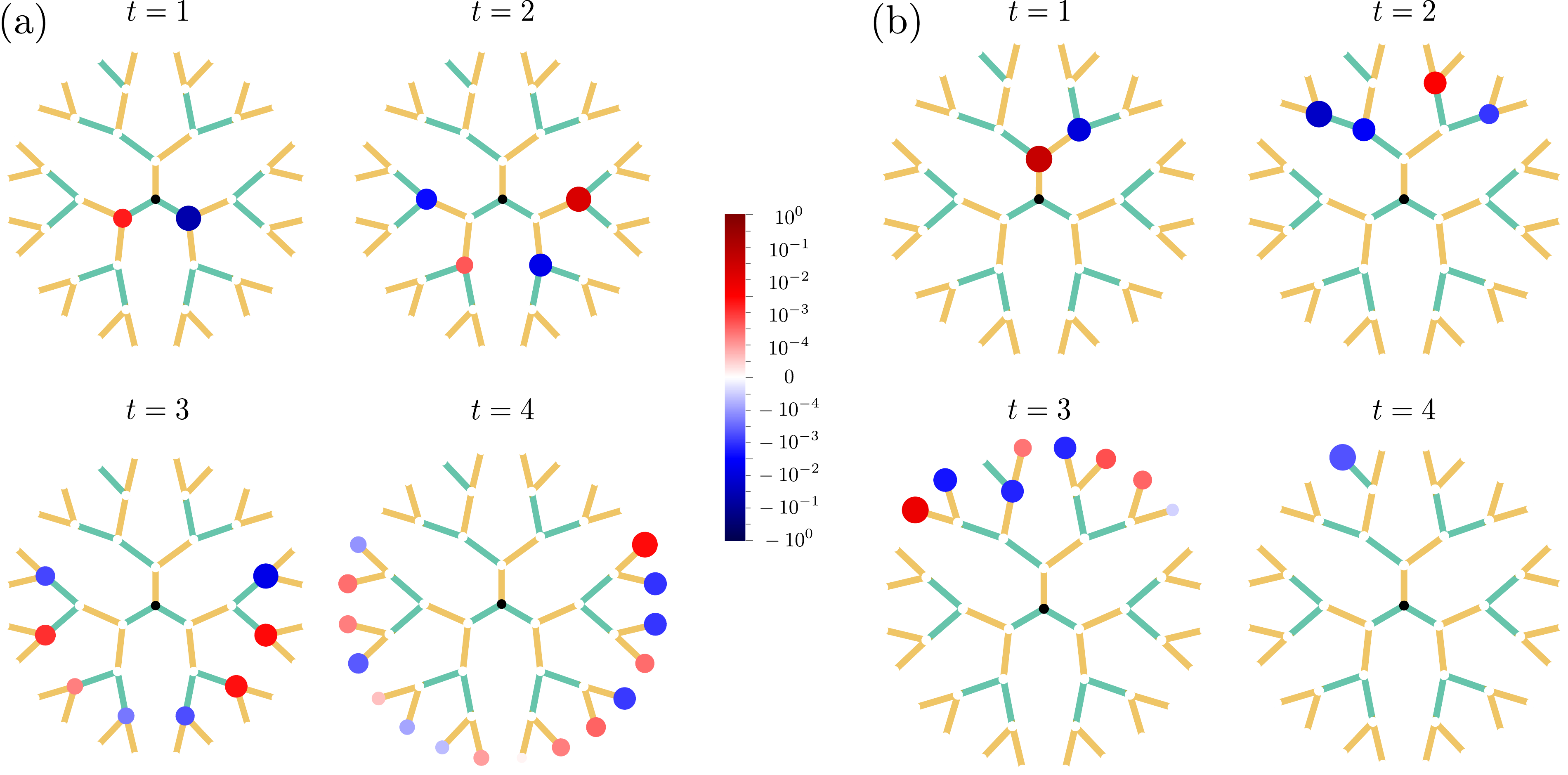}
\caption{A heatmap showing the values of correlation functions $c_{\alpha \beta}(x,t)$ with $\sigma_{\alpha} = \frac{1}{\sqrt{2}} (X+Z)$ and $\sigma_{\beta} = Z$, for a generic tree-unitary brickwork circuit, generated using the algorithm in App.~\ref{app:gen_tu}. The size of the points is proportional to the value of $\ln{|c_{\alpha \beta}(x,t)|}$. The origin (initial location for $\sigma_{\alpha}(0,0)$) is colored black for reference, but $c_{\alpha\beta}(0,t)=0$ for all times $t$ shown. The correlation functions are $0$ everywhere not on the light cone. In (a), the first layer is chosen to act on green bonds, leading to non-vanishing correlations around a fraction $(z-1)/z$ of the light cone. In (b), the first layer is chosen to act on orange bonds, leading to non-vanishing correlations around a fraction $1/z$ of the light cone. }
\label{fig:cf_map}
\end{figure*}

All we have used at this stage is that the evolution on the tree is generated by unitary $z$-site gates. In principle, we could develop expressions similar to \eqref{eq:channelcorrelator} for two-point correlation functions {\it anywhere}, but the dimensions of the corresponding superoperators, and hence the complexity of the computations, grow exponentially in the distance off the light cone. 
In the (1+1)d case, a crucial simplification is that imposing dual-unitarity results in two-point correlation functions that vanish  \textit{everywhere inside} the light cone~\cite{bertini_exact_2019}. As a result, we can calculate {\it all} two-point correlation functions efficiently, leading to the `solvability' of such circuits. This argument directly extends to our $z$-site construction on imposing  tree-unitarity. In order to show this, it is useful to reinterpret the tree-unitarity conditions [Eq.~\eqref{eq:tree_unitarity_conditions}] on the level of operator spreading. For a general $z$-site unitary gate acting on sites $1, 2 \dots z$, the unitary transformation of a single-site Pauli matrix generally returns a linear combination of products of Pauli matrices,
\begin{align}
\label{eq:operator_spreading}
    U^{\dagger} \sigma_\alpha(i) U = \sum_{\alpha_1, \alpha_2, \dots, \alpha_z} c^{(\alpha,i)}_{\alpha_1  \alpha_2 \dots \alpha_z} \sigma_{\alpha_1} \otimes \sigma_{\alpha_2}\otimes \dots  \sigma_{\alpha_z}\,.
\end{align}
Correlation functions probe the terms in the expansion where only a single Pauli matrix is nontrivial (different from the identity), since due to the trace-orthonormality of the Pauli matrices
\begin{align}
    \addedtwo{\frac{1}{q}}\mathrm{tr}[\sigma_\beta(j) U^{\dagger} \sigma_\alpha(i) U] = c^{(\alpha,i)}_{0 \dots 0 \,\alpha_j = \beta \, 0 \dots 0}\,.
\end{align}
Tree-unitarity now strongly constrains the terms that can appear in the right-hand side of \eqref{eq:operator_spreading}, enforcing that terms with a single nontrivial Pauli matrix can only appear when this Pauli matrix is located on the edge of the causal light cone, similar to how dual-unitarity enforces maximal operator spreading~\cite{claeys_maximum_2020}.

Again turning to $z=3$ tree-unitary gates for concreteness, we can use tree-unitarity to constrain e.g. the evolution of a Pauli matrix located on the second site, $\sigma_{\alpha}(2)$, since 
\begin{align}
    \mathrm{tr}_{13}\left[U^{\dagger}\sigma_{\alpha}(2)U \right] = \vcenter{\hbox{\includegraphics[width=0.1\columnwidth]{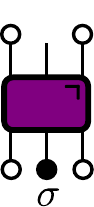}}} %\nonumber\\ 
    =
    \vcenter{\hbox{\includegraphics[height=0.21\columnwidth]{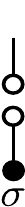}}} 
    \propto \mathrm{tr}(\sigma_\alpha) = 0,
\end{align}
where we have used that the Pauli matrices are traceless. 
As such, this dynamics does not generate any terms of the form $\mathbb{1} \otimes \sigma_{\beta} \otimes \mathbb{1}$, where the operator remains on the second site, since these would not vanish under the trace. The only possible terms either act nontrivially on two or more sites, and hence do not contribute to two-point functions, or the initial Pauli operator has `hopped' to one of the two other sites, i.e. $\mathbb{1} \otimes \sigma_{\alpha} \otimes \mathbb{1}\to\sigma_{\beta} \otimes \mathbb{1} \otimes \mathbb{1}$ or $\mathbb{1} \otimes \mathbb{1} \otimes \sigma_{\beta}$. These `hopping'  terms, however, give rise exactly to the correlation function on the light-cone, and thus all other correlation functions are identically zero. It is easy to see that the same reasoning applies when permuting the initial position of the operator. 
Similarly, for $z>3$, a tree-unitary can only generate superpositions of Pauli strings in which each string has at least one Pauli on one of the $z-1$ sites other than the input site.

For tree-unitary circuits, the single-site correlation functions hence vanish identically inside the light cone, but are generically non-zero on the edge of the light cone. This is demonstrated in Fig.~\ref{fig:cf_map} with a heatmap for values of correlation functions, both for the first layer being applied to green bonds (a) and orange bonds (b). Notice that in neither case are correlations non-zero around the entire light cone; the position of the first gate applied dictates that correlations can be non-zero either in the direction of $z-1$ \replacedtwo{leaves}{leafs} of the first gate (equivalent to moving `down' the rooted tree), or in the remaining direction. This is analogous to dual-unitary circuits, in which correlation functions will only be non-vanishing along the light cone in one direction, which depends on the positioning of the first layer of gates relative to the location of the operator at $t=0$. We emphasize that this is a special feature of the correlation functions in the tree-unitary case; for a generic unitary circuit, correlation functions would be non-vanishing all around the light cone, as well as inside the light cone.

\begin{figure}[t]
\centering
\includegraphics[width=8cm]{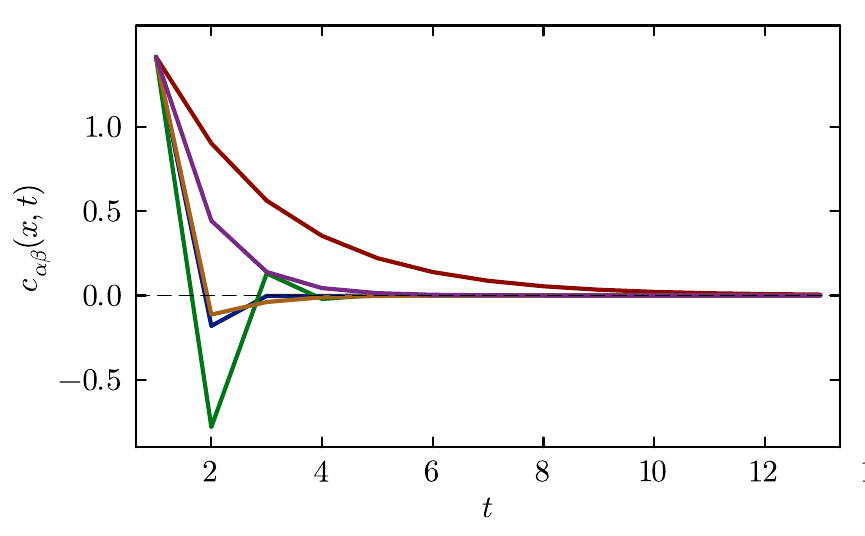}
\caption{Correlation function $c_{\alpha \beta}(x,t)$ on the light cone direction $2\to 3$ plotted with $\sigma_{\alpha} = \frac{1}{\sqrt{2}} (X+Z)$ and $\sigma_{\beta} = Z$, for three random tree-unitary gates generated using the method in App.~\ref{app:gen_tu}. }
\label{fig:cf_plot}
\end{figure}

This construction should be contrasted to previous extensions of dual-unitary circuits to higher dimensions~\cite{jonay_triunitary_2021, milbradt_ternary_2023}, for which correlation functions are non-vanishing only along particular rays. 
In Fig.~\ref{fig:cf_plot} we plot the light cone correlation function for five random tree-unitary gates. In each case the equilibrium value of $0$ is approached, with differing time scales depending on the eigenvalues of the channel $M_{23}$.

\subsection{Out-of-time-order correlation functions}
\label{subsec:otoc}

The exact solvability of the correlation functions for both tree-unitary and dual-unitary circuits is underpinned by the nature of operator spreading in these circuits. We now consider the out-of-time-order correlation function (OTOC) as a different probe of operator spreading, which will be useful in highlighting how moving from dual-unitarity to tree-unitarity induces qualitative changes in the dynamics. The OTOC is defined as 
\begin{equation}\label{eq:otoc_def}
    C_{\alpha \beta} (i,j;t) = \langle \sigma_{\beta}(i,0) \sigma_{\alpha}(j,t) \sigma_{\beta}(i,0) \sigma_{\alpha}(j,t) \rangle\,.
\end{equation}

Outside the light cone, the OTOC always takes the trivial value of $1$ because $\sigma_{\alpha}(0,t)$ and $\sigma_{\beta}(x,0)$ commute. 
In generic unitary circuits, the OTOC develops a front which propagates ballistically at the butterfly velocity $v_B<1$ and broadens diffusively~\cite{von_keyserlingk_operator_2018}. For (1+1)d circuits consisting of dual-unitary gates, the butterfly velocity takes its maximally allowed value $v_B=1$ consistent with the strict light cone, with such circuits referred to as `maximum velocity'~\cite{claeys_maximum_2020,rampp_dual_2023}. This maximum velocity has been experimentally observed in quantum simulators~\cite{mi_information_2021}. In recent years, various circuits exhibiting differing degrees of solvability have been identified, all of which exhibit such a maximum velocity~\cite{foligno_quantum_2024,rampp_entanglement_2023}, suggesting a strong connection between maximum butterfly velocity and solvability. Remarkably, in tree-unitary circuits this connection breaks down.

We restrict our analysis below to the OTOC on the edge of the causal light cone, and find that tree-unitarity alone does not lead to a nontrivial OTOC on the light cone, indicating a `butterfly velocity' that is not maximal. 
In (1+1)d circuits, the OTOC value on the light cone can be efficiently calculated using a quantum channel approach~\cite{claeys_maximum_2020}, and 
a similar result applies on the tree, where the results of the preceding section can be directly extended to obtain a quantum channel representation of the OTOC on the light cone. This can be expressed in terms of quantum channels involving two copies of the unitary gate and its conjugate (i.e. two replicas). We define
\begin{align}
    \vcenter{\hbox{\includegraphics[width=0.14\columnwidth]{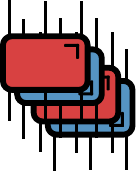}}}=
    \vcenter{\hbox{\includegraphics[width=0.1\columnwidth]{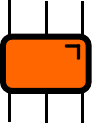}}}
\end{align}
along with the contractions 
\begin{align}
    \vcenter{\hbox{\includegraphics[height=0.125\columnwidth]{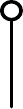}}}\,=\frac{1}{\sqrt{q}}\,
    \vcenter{\hbox{\includegraphics[width=0.1\columnwidth]{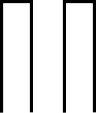}}}\,, \qquad
    \vcenter{\hbox{\includegraphics[width=0.025\columnwidth]{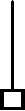}}}\,=\frac{1}{\sqrt{q}}\,
    \vcenter{\hbox{\includegraphics[width=0.1\columnwidth]{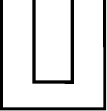}}}\,,
\end{align}
and superoperators
\begin{align}
    (\sigma|=\,\vcenter{\hbox{\includegraphics[height=0.16\columnwidth]{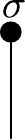}}}\,=\frac{1}{\sqrt{q}}\,
    \vcenter{\hbox{\includegraphics[height=0.16\columnwidth]{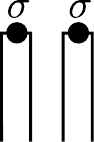}}} \,,\quad
    |\sigma)=\,\vcenter{\hbox{\includegraphics[height=0.16\columnwidth]{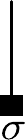}}}\,=\frac{1}{\sqrt{q}}\,
    \vcenter{\hbox{\includegraphics[height=0.16\columnwidth]{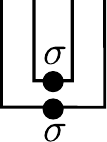}}}~.
\end{align}
The channel $\mathcal{T}_{e \tilde{e}}$ used to compute the OTOC can then be written graphically; for example,
\begin{align}\label{eq:Tee_otoc}
    \mathcal{T}_{23}|\sigma)=\vcenter{\hbox{\includegraphics[height=0.2\columnwidth]{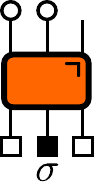}}}\,.
\end{align}

Using unitarity and the geometry of the circuit, the light-cone OTOC can be directly expressed as
\begin{align}
    C_{\alpha \beta}(i,j;t) =  (\sigma_{\beta}|\mathcal{T}_{e_{t}\tilde{e}_{t}} \dots \mathcal{T}_{e_{2}\tilde{e}_{2}}\mathcal{T}_{e_{1}\tilde{e}_{1}} |\sigma_{\alpha})\,.
\end{align}
The derivation is analogous to that for (1+1)d unitary circuits~\cite{claeys_maximum_2020}, so we do not give it here in full and only cite the final result. 

\begin{figure}[t]
\centering
\includegraphics[width=8cm]{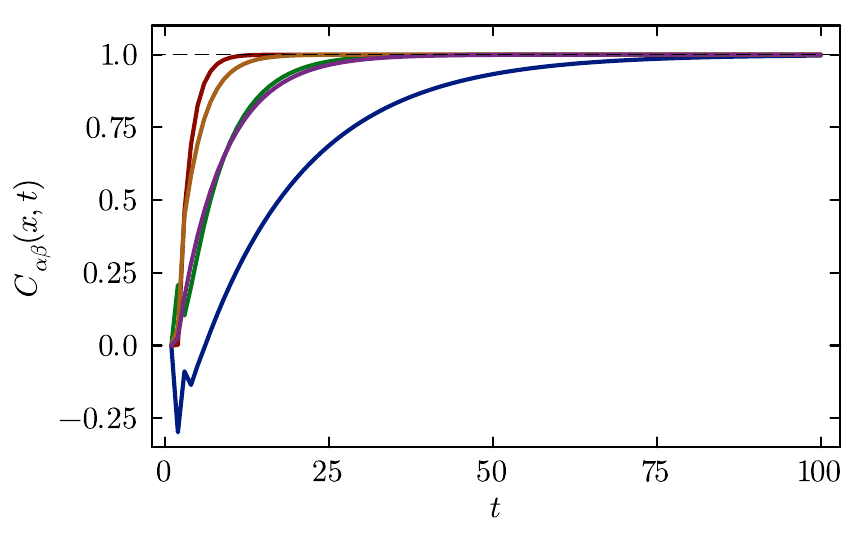}
\caption{OTOC $C_{\alpha \beta}(x,t)$ on the light cone direction $2\to 3$ with $\sigma_{\alpha} = \frac{1}{\sqrt{2}} (X+Z)$ and $\sigma_{\beta} = Z$, for the same random tree-unitary gates as in Fig.~\ref{fig:cf_plot}. The OTOC approaches the trivial value of $1$ on the light cone, indicating that operators do not spread at maximum velocity.}
\label{fig:otoc_plot}
\end{figure}

The maximal velocity property for dual-unitary circuits corresponds to showing that this OTOC decays to a nontrivial value. To do so, one can study the eigenspectrum of the relevant quantum channel, to characterize the nondecaying eigenoperators. Unitarity alone  
requires the existence of at least a single left and right eigenoperator with eigenvalue 1,
\begin{equation}
    |R)=\,\vcenter{\hbox{\includegraphics[height=0.125\columnwidth]{figures_final/squaretop.pdf}}}\,, \qquad (L|=\,\vcenter{\hbox{\includegraphics[height=0.125\columnwidth]{figures_final/circletop.pdf}}}\,.
\end{equation}
In isolation, however, this unitarity condition does not lead to a nontrivial behavior of the OTOC, which is sensible: maximal velocity is not a property of generic unitary circuits. If this is the only eigenoperator with unit-modulus eigenvalue, then the late-time behavior of the OTOC follows as a projection on these eigenoperators and
\begin{align}
    &\lim_{t \to \infty} C_{\alpha \beta}(i,j;t) = (\sigma_{\alpha}|R)(L|\sigma_{\beta})\nonumber\\
    &\quad=\vcenter{\hbox{\includegraphics[height=0.15\columnwidth]{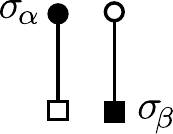}}}=\frac{1}{q^2}\mathrm{tr}[\sigma_{\alpha}^2]~\mathrm{tr}[\sigma_{\beta}^2]=1,
\end{align}
where both overlaps vanish because the Pauli matrices are traceless. Without any further structure imposed, there are generically no other unit-eigenvalue eigenoperators, so the OTOC approaches the trivial value of $1$, as shown in Fig.~\ref{fig:otoc_plot}. This indicates that the butterfly velocity $v_B<1$.

Imposing dual-unitarity in (1+1)d forces the existence of nontrivial non-decaying eigenoperators. Following the argument above, we then obtain a nontrivial value of the OTOC on the light cone, and hence a maximal velocity $v_B=1$. We will return to this argument and the explicit construction of such eigenoperators in Sec.~\ref{sec:maxvel}.

In striking contrast to this, while imposing tree-unitarity facilitates solvability (by forcing causal correlators to vanish off the light cone) just as in dual-unitary circuits, it \textit{does not} allow us to construct further unit-eigenvalue eigenoperators to the channels $\mathcal{T}_{e\tilde{e}}$ relevant to computing the OTOCs. Numerical inspection of the eigenspectrum of the light-cone channels for a `generic' tree-unitary circuit (from the construction in App.~\ref{app:gen_tu}) shows that indeed no additional such eigenoperators exist. Thus, we conclude that a generic tree-unitary circuit has butterfly velocity $v_B<1$ along any given path. However, in Sec.~\ref{sec:maxvel} we will see that  tree-unitary gates can also obey stronger additional constraints which {\it do} give $v_B=1$ in some directions, analogous to the situation in finite dimensions.

The reason for this difference relative to dual-unitary circuits is that while dual-unitarity  requires the operator front to grow with every time step, tree-unitarity does not. For dual-unitary 2-site gates $U^{\dagger}(\sigma \otimes \mathbb{1}) U$ consists of a linear combination of $\mathbb{1} \otimes \sigma$ and $\sigma \otimes \sigma$, such that the operator front always moves a single step to the right. However, tree-unitarity implies that e.g. $U^{\dagger}(\mathbb{1} \otimes \sigma \otimes \mathbb{1})U$ can have contributions of the form $\mathbb{1} \otimes \mathbb{1} \otimes \sigma$ and $\sigma \otimes \mathbb{1}  \otimes \mathbb{1}$ \addedtwo{for example}. While single-site Pauli matrices necessarily `hop' at every time step, this does not imply that the operator front also grows: for instance, the contribution  $\mathbb{1} \otimes \mathbb{1} \otimes \sigma$ does not grow the front along the `leftmost' direction. This freedom to `hop' single-site Paulis without growing the operator front in all directions is again due to the growing number of sites on the light cone on the tree, to be contrasted with the constant number of sites on the light cone in the (1+1)d brickwork circuit.

\subsection{Entanglement Dynamics}
\label{subsec:entanglement}

Both dual-unitarity and tree-unitarity can be used to obtain exact results for state dynamics in the Schr\"odinger picture, in addition to the operator dynamics in the Heisenberg picture considered so far. For dual-unitary circuits, such computations rely on the existence of special classes of `solvable' initial states that do not break spatial unitarity, and for which the dynamics of entanglement can be exactly characterized~\cite{piroli_exact_2020}. These states thermalize exactly after a fixed number of time steps, i.e. their reduced density matrix reduces to the maximally mixed state that maximizes the entropy. 

 The simplest example of a solvable state for a dual-unitary circuit is a product wave function consisting of Bell states on alternating bonds~\cite{piroli_exact_2020}. We now show that (generalized) GHZ states on $z$ sites furnish a natural extension of solvable states to tree-unitary dynamics. The resulting states also exhibit maximal entanglement growth, which is now exponential because of the geometry of the tree, to be contrasted with the linear growth exhibited in (1+1)d brickwork circuits. We consider GHZ states of the form
\begin{equation}\label{eq:GHZ}
  |\textrm{GHZ}\rangle = \frac{1}{\sqrt{q}} \sum_{a=1}^q |a, a, \dots ,a \rangle\,.
\end{equation}
These states can be interpreted as isometries in the folded picture. The GHZ states satisfy
\begin{align}
    \overline{\mathrm{tr}}_{e} ( |\textrm{GHZ}\rangle \langle \textrm{GHZ}| ) = \mathbb{1}_e,\label{eq:GHZcondition}
\end{align}
for all $e$, which can be compared with the tree-unitarity conditions [Eq.~\eqref{eq:tree_unitarity_conditions}]. 

\begin{figure}
\centering
\includegraphics[width=\columnwidth]{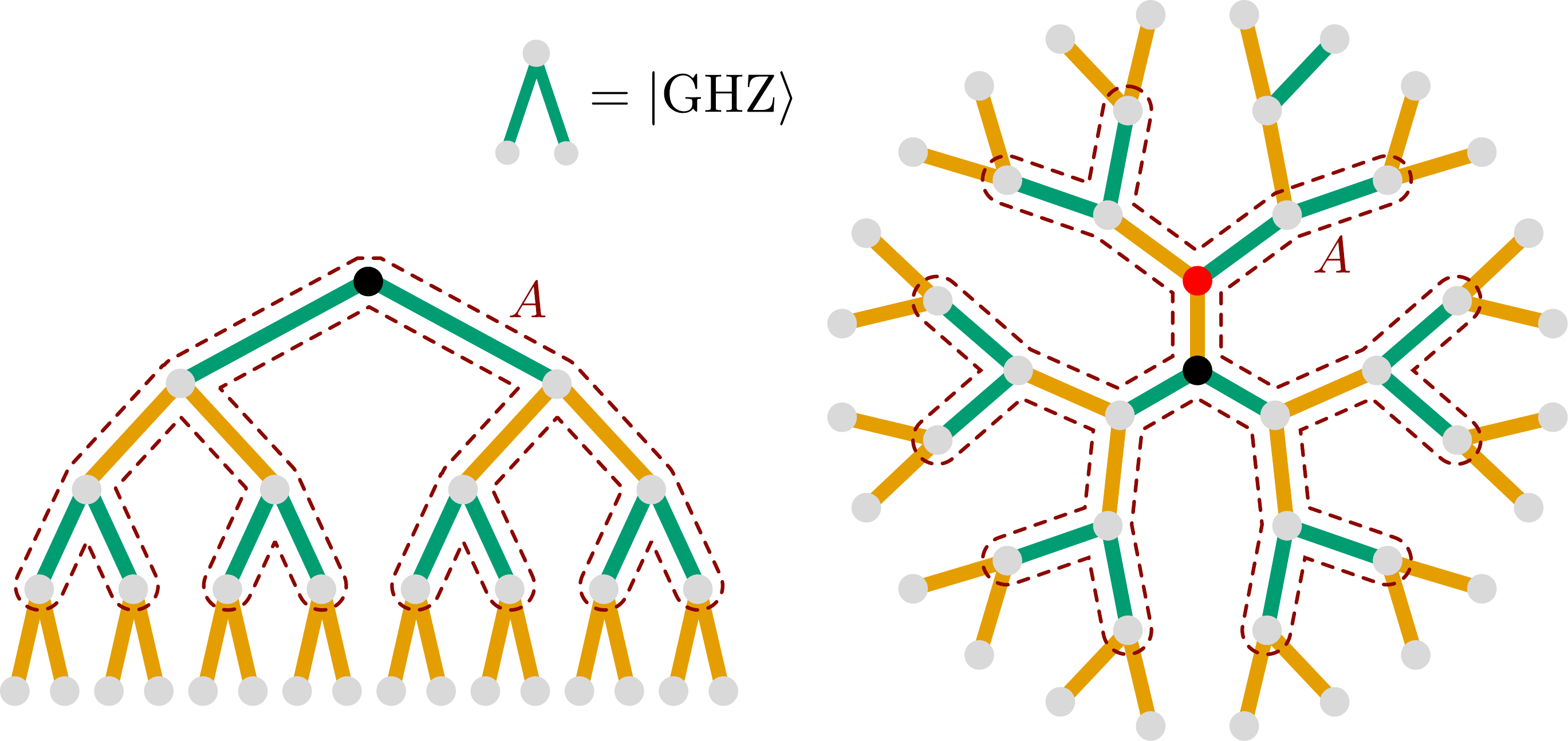}
\caption{Regions of the unrooted and rooted trees for which entanglement entropy can be computed exactly using tree-unitarity. Here, GHZ states [Eq.~\eqref{eq:GHZ}] are placed on green bonds, and the first layer of the circuit is applied to orange bonds. The `solvable regions' for which entanglement can be computed exactly correspond to interiors of light cones. In each case, we show a light cone generated by $t=3$ time steps from the black site. The red site provides a reference for Eq.~\eqref{eq:ee_unrooted_unitarity}.}
\label{fig:ee_regions}
\end{figure}

Considering again $z=3$ for concreteness, the GHZ state can be represented as
\begin{align}\label{eq:ghz_state}
    |\textrm{GHZ}\rangle =  \vcenter{\hbox{\includegraphics[width=0.1\columnwidth]{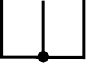}}} \,.
\end{align}
The identity \eqref{eq:GHZcondition} can be graphically represented for $e=2$ by using the folded picture, where we fold the ket $\ket{\rm GHZ}$ behind the bra $\bra{\rm GHZ}$ and indicate the trace with circles as
\begin{align}\label{eq:graphicalGHZcondition}
    \vcenter{\hbox{\includegraphics[width=0.12\columnwidth]{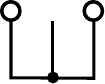}}}\,=\,
    \vcenter{\hbox{\includegraphics[height=0.1\columnwidth]{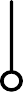}}}.
\end{align}
We take an initial state $\ket{\psi_0}$ that is a product state of GHZ states, arranged according to the 2-coloring, such that the states correspond to one color and the first layer of unitary gates corresponds to the second color, shown in Fig.~\ref{fig:ee_regions}. 
Using tree-unitarity, the entanglement entropy can be exactly computed for specific solvable subregions $A$. These subregions correspond to the interior of the light cone starting from any point on the lattice, such as any of the snapshots in Fig.~\ref{fig:3site_coloring}(c). 
For example, the regions shown in Fig.~\ref{fig:ee_regions} correspond to a light cone after three time steps. 
We consider the dynamics of the reduced density matrix
\begin{align}
\rho_A(t)=\mathrm{tr}_{\bar{A}} [U(t) \ket{\psi_0} \bra{\psi_0} U^{\dagger} (t)]. \label{eq:rhoAtdef}
\end{align}

Suppose we take a subregion $A$ which consists of a light cone grown from $r$ steps (for example, $r=3$ in Fig.~\ref{fig:ee_regions}).
The calculation of $\rho_A(t)$ for \textit{even} times proceeds as follows: 
\begin{enumerate}
\item At a fixed time $t$, we write $\rho_{A}$ as a tensor network [see below \eqref{eq:rho2}] and use unitarity as far as possible to contract the network outside of region $A$. This will allow us to fully remove gates and GHZ states only outside of a new boundary, extending the boundary of $A$ by $t$ steps. The new boundary will be called the `contraction front'; one can view this as evolving during the process of contracting the network.
\item We then use the GHZ contraction property (\ref{eq:graphicalGHZcondition}) and tree-unitarity to move the contraction front inwards. This can be performed $t$ times, such that the position of the contraction front at the final time $t$ is $t$ steps \textit{smaller} than region $A$. Sites which are inside region $A$, but lie outside the contraction front, are now fully thermalized, and contribute $\ln{q}$ to the entanglement entropy.
\item When $t=r$, after using unitarity and tree-unitarity, the contraction front has shrunk to its deepest possible position ($r=0$), and all states within $A$ have been thermalized. %
\end{enumerate}
The final contraction front (after using unitarity and tree-unitarity) at \textit{odd} times $t$ will be the same as at $t+1$ if the initial entanglement of $A$ was $0$, and the same as $t-1$ is the initial entanglement of $A$ was nonzero (i.e. GHZ states cross the boundary of $A$).

We now illustrate these steps by calculating the entanglement growth on the rooted tree first, because the manipulations involved are simpler to visualize. To do this, we introduce a new notation in terms of what we refer to as a `tree-folded' circuit. This will allow us to represent the state of a rooted tree conveniently in a plane, upon which the computation proceeds as a straightforward generalization of the (1+1)d calculation \cite{bertini_entanglement_2019}. The key step in tree-folding is to fold together the $z-1$ legs on the leaves of the $z$-site gate. For $z=3$, this is graphically depicted as
\begin{equation}
    \vcenter{\hbox{\includegraphics[width=0.14\columnwidth]{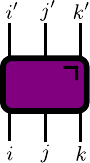}}}=
    \vcenter{\hbox{\includegraphics[width=0.12\columnwidth]{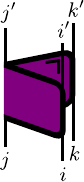}}}=
    \vcenter{\hbox{\includegraphics[width=0.12\columnwidth]{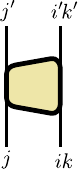}}}.
\end{equation}
In terms of the resulting folded gates, tree-unitarity is then expressed as
\begin{equation}
\label{eq:tu_folded}
    \vcenter{\hbox{\includegraphics[width=0.12\columnwidth]{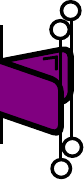}}}=~
    \vcenter{\hbox{\includegraphics[width=0.1\columnwidth]{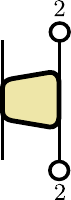}}}=~
    \vcenter{\hbox{\includegraphics[height=0.16\columnwidth]{figures_final/double_identity.pdf}}}
\end{equation}
and unitarity as
\begin{equation}
\label{eq:u_folded}
\vcenter{\hbox{\includegraphics[width=0.115\columnwidth]{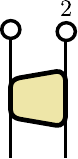}}}=
    \vcenter{\hbox{\includegraphics[height=0.18\columnwidth]{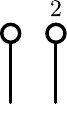}}}\,.
\end{equation}
Note that the legs on the right-hand side of each folded gate now have a larger Hilbert space compared to the legs on the left-hand side. This is in accord with the general move from unitarity to isometry in this work, which we make explicit in the graphical notation. We also introduce the tree-folded GHZ state
\begin{equation}
    \vcenter{\hbox{\includegraphics[height=0.12\columnwidth]{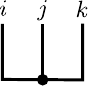}}}=
    \vcenter{\hbox{\includegraphics[height=0.16\columnwidth]{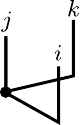}}}=
    \vcenter{\hbox{\includegraphics[height=0.13\columnwidth]{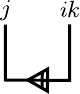}}},
\end{equation}
where the identity \eqref{eq:graphicalGHZcondition} with $n$ folded GHZ states can be represented as 
\begin{equation}
\label{eq:ghz_tr_folded}
    \vcenter{\hbox{\includegraphics[height=0.15\columnwidth]{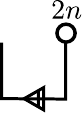}}}=
    \vcenter{\hbox{\includegraphics[height=0.12\columnwidth]{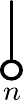}}}\,\,.
\end{equation}

Using this notation, a segment of two layers of the circuit acting on two GHZ states can be written
\begin{equation}
    \vcenter{\hbox{\includegraphics[height=0.35\columnwidth]{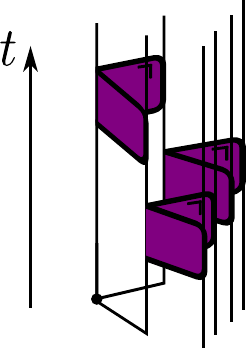}}}=
    \vcenter{\hbox{\includegraphics[height=0.35\columnwidth]{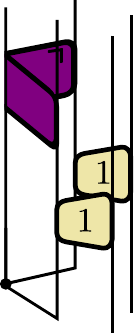}}}=
    \vcenter{\hbox{\includegraphics[height=0.35\columnwidth]{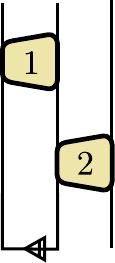}}}\,\,
\end{equation}
where for clarity we have indicated that time (i.e. layers of the circuit) will run upwards.
With this notation established, we now return to our objective of computing \eqref{eq:rhoAtdef} for the rooted tree (with $r=3$, for specificity, corresponding to the region $A$ in Fig.~\ref{fig:ee_regions}).
At $t=2$, the reduced density matrix on the rooted tree can be represented in the folded notation as
\begin{equation}\label{eq:rho2}
\rho_A(t=2)=\vcenter{\hbox{\includegraphics[height=0.4\columnwidth]{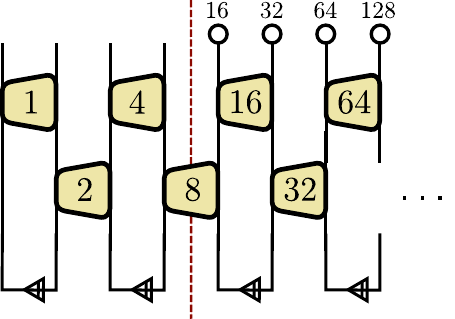}}}
\end{equation}
with the red dashed line separating $A$ from its complement $\bar{A}$, which is traced over.
Following the steps presented above, we use unitarity to remove as many gates as possible outside $A$
\begin{align}
    \rho_A(t=2)&=\vcenter{\hbox{\includegraphics[height=0.4\columnwidth]{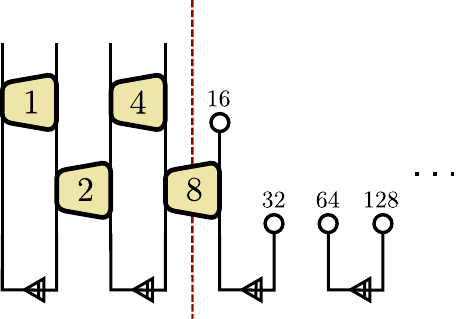}}} \nonumber\\
    &=
    \vcenter{\hbox{\includegraphics[height=0.4\columnwidth]{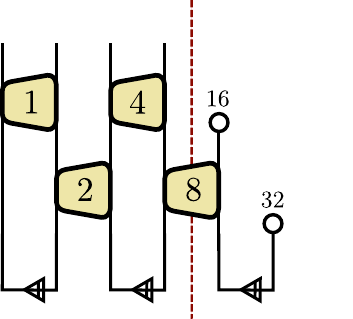}}}\,\,.
\end{align}
We see that unitarity has allowed us to disconnect the tensor diagram beyond the contraction front.
We can now apply the GHZ contraction property \eqref{eq:ghz_tr_folded}, then (folded) tree-unitarity  \eqref{eq:tu_folded}, to move the contraction front inwards:
\begin{align}
\rho_A(t=2)\,&=\, \vcenter{\hbox{\includegraphics[height=0.34\columnwidth]{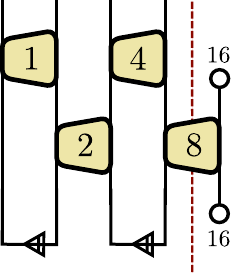}}}\,=\,\vcenter{\hbox{\includegraphics[height=0.34\columnwidth]{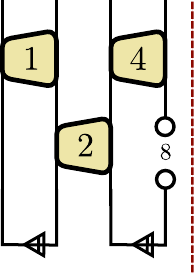}}} \nonumber\\\, &=\,\vcenter{\hbox{\includegraphics[height=0.34\columnwidth]{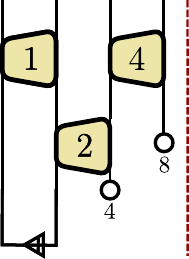}}}\,.
\end{align}
This final expression shows that $\rho_A(t=2)$ is a unitary transformation of $\ket{\mathrm{GHZ}}\bra{\mathrm{GHZ}}\otimes \frac{1}{q^{z(z-1)^2}}\mathbb{1}^{\otimes z(z-1)^2}$, where the original GHZ states closest to the boundary of $A$ have fully thermalized.
Since the unitary transformation will not affect the entanglement spectrum, the entanglement entropy is $S_A(t=2)=z(z-1)^2\ln{q}$. 
We also see that the entanglement spectrum is flat, such that all entanglement entropies are identical; the entanglement dynamics consists solely of growing the rank of the entanglement spectrum while maintaining a uniform distribution over the nontrivial eigenvalues, similar to the case of dual-unitary circuits.

At $t=4$, after using unitarity in step 1, the initial position of the contraction front has moved outwards by two steps,
\begin{equation}
\vcenter{\hbox{\includegraphics[height=0.5\columnwidth]{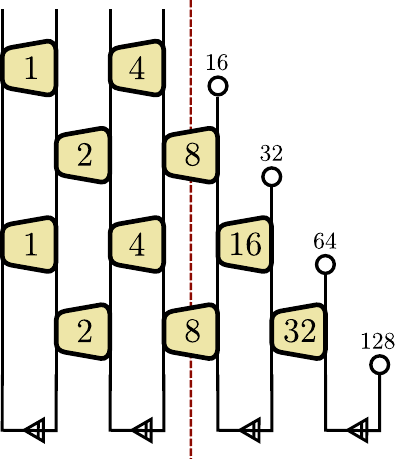}}} \,=\,\vcenter{\hbox{\includegraphics[height=0.45\columnwidth]{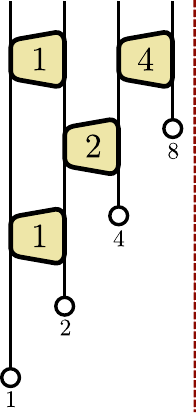}}}\,\,.
\end{equation}
In this equality we have used the GHZ contraction and tree-unitarity to move the contraction front inside region $A$, until we can go no further.
Observe that, although after step 1 the contraction front is two steps further outside $A$ than for $t=2$, the additional two layers of gates for $t=4$ allows us to use the GHZ/tree-unitarity conditions to move the front two steps further {\it inside} $A$ than in the $t=2$ case. 
For our choice of region with $r=4$, we explicitly see from the above graphical equation that at all sites are now thermalized, such that $\rho_A(t=4)$ is a unitary transformation of $\frac{1}{q^n}I^{\otimes n}$, where $n=\left[(z-1)^4-1\right]/(z-2)$ is the number of sites in $A$.

The extension to a larger region $A$ of $r$ generations on the rooted tree directly follows:  the initial reduced density matrix will be 
\begin{equation}
    \rho_A(0)=\bigotimes_{j=0}^{(r-1)/2} (\ket{\textrm{GHZ}}\bra{\textrm{GHZ}})^{\otimes(z-1)^{2j}},
\end{equation}
and for every two layers of the circuit applied, successive sets of GHZ states (moving inwards from the boundary of $A$ towards the root) will be transformed as
\begin{align}
    (\ket{\textrm{GHZ}}\bra{\textrm{GHZ}})^{\otimes (z-1)^{2j}} \to \frac{1}{q^{z\cdot z^{2j}}}\mathbb{1}^{\otimes z\cdot (z-1)^{2j}}
\end{align}
up to a unitary transformation --- the process that we termed `thermalization'.  
This therefore will increase the entanglement entropy by $z\cdot (z-1)^{2j}\ln{q}$, for every two time steps, until saturation when $t=r$.

Translating this into entanglement entropy, and keeping in mind the fact that the number of GHZ states is bigger for increasing $r$, we find that for even times $t\leq r$,
\begin{align}
\label{eq:eer_1}
    S_A^e(t)&=\frac{(z-1)^{r+1}}{z-2}(1-(z-1)^{-t})\ln{q},
\end{align}
and for odd times $t\leq r$
\begin{align}
\label{eq:eer_2}
    S_A^o(t)=S_A^e(t+1).
\end{align}
For $t\geq r$, the entanglement entropy saturates to its maximum value
\begin{align}
\label{eq:eer_3}
    S_{\infty}=\frac{(z-1)^{r+1}-1}{z-2}\ln{q}.
\end{align}

We emphasize that this result applies independent of the specific choice of tree-unitary gates and furthermore, that there is no requirement that all gates be the same: the only properties used in the calculation are tree-unitarity and the GHZ contraction.

The rooted tree allows for straightforward calculation of the entanglement entropy through the folding trick, but remains slightly unsatisfactory since the root is ascribed a special role. 
However, the calculation can be adapted to the unrooted tree, for `solvable regions' where $A$ is taken to be the interior of a light cone starting from any point. 

The key idea is that when we choose such a region, the same $3$-step process to find $\rho_A(t)$ applies, such that the outermost GHZ states thermalize successively in intervals $\Delta t=2$, just as for the rooted tree.  We will again illustrate this for a region $A$ of radius $r=3$, as in Fig.~\ref{fig:ee_regions}.
First, find the single `anomalous' gate whose root is at the boundary of $A$ (in Fig.~\ref{fig:ee_regions}, this is the green gate connected to the red site). Take the leaf of this gate (i.e. the red site), and fold all other gates from this point just as for the rooted tree. 
Although not all of these folded legs will have the same `radius', the calculation applies in the same way to all of them --- that is, we use unitarity to disconnect the region $t$ steps beyond the boundary, then use tree-unitarity to move this contraction front $t$ steps within the boundary (for even $t$). 

As long as we choose the region $A$ to be enclosed by a light cone of the unrooted tree, the same principle applies to the anomalous gate too. For even $t$, unitarity initially places the contraction front $t$ steps from the boundary, with tree-unitarity allowing us to move the contraction front $t$ steps inside.
To make this a little more concrete, consider again $\rho_A(t=2)$. After using unitarity, the part of the tree near the anomalous gate will be 
\begin{align}
\label{eq:ee_unrooted_unitarity}
\vcenter{\hbox{\includegraphics[height=0.4\columnwidth]{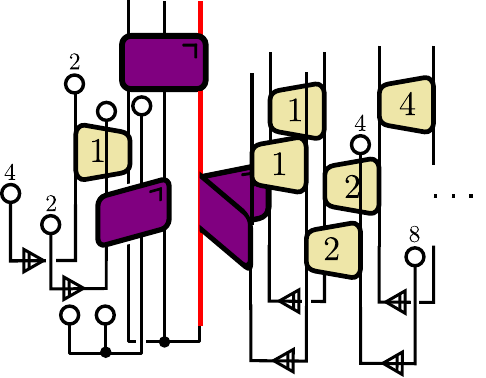}}}
\end{align}
Here, the thickened red line corresponds to the red site in Fig.~\ref{fig:ee_regions} (note that the first layer corresponds to orange bonds and the second layer to green), and the ellipsis indicates that the folded pattern is continued up to the edges of the region. The calculation there will be identical to the rooted tree. The shortened section in front corresponds to the top right of Fig.~\ref{fig:ee_regions}.

Now we can use GHZ contractions and tree-unitarity to move the contraction front inwards, to $2$ steps inside the region $A$ (here, moving inwards by a step means reversing the growth of the light cone by one time step)
\begin{align}
\label{eq:ee_unrooted_tu}
\vcenter{\hbox{\includegraphics[height=0.4\columnwidth]{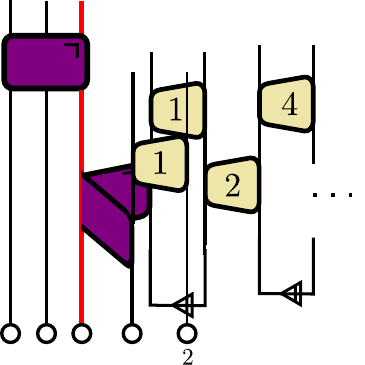}}}
\end{align}
The light cone structure allows this process to be continued with further time steps, with the outermost GHZ states thermalizing on alternating time steps as in the case of the rooted tree.

For a light cone of radius $r$ (i.e. generated by $r$ time steps), the number of sites inside this region is
\begin{equation}
    n_{LC}(r)=\frac{z}{z-2}\left[(z-1)^{r}-1 \right].
\end{equation}
From this, the entanglement entropy after an even time $t\leq r$ will be given by $n_{LC}(r)-n_{LC}(r-t)$, giving
\begin{align}
\label{eq:ee_1}
    S_A^e(t)=z\cdot \frac{(z-1)^r}{z-2}(1-(z-1)^{-t})\ln{q},
\end{align}
whereas for odd time steps, $t\leq r$,
\begin{align}
\label{eq:ee_2}
    S_A^o(t)=S_A^e(t+1).
\end{align} 
For $t\geq r$, $S(t)$ saturates to its maximum value 
\begin{align}
\label{eq:ee_3}
    S_{\infty}=z\frac{(z-1)^{r}-1}{z-2}\ln{q}.
\end{align}

The above results apply when the initial entanglement is $0$, with the boundary of $A$ not cutting any GHZ states. If such a boundary is shifted by one, the above formulae are simply shifted by a time step. On the unrooted tree, for \textit{odd} times  $t\leq r$,
\begin{equation}
    S_A^o(t)=z\cdot \frac{(z-1)^r}{z-2}(1-(z-1)^{-t})\ln{q}
\end{equation}
and for even times $t\leq r$,
\begin{equation}
    S_A^e(t)=S_A^o(t+1),
\end{equation}
 reaching the same $S_{\infty}$ for $t\geq r$.

 \begin{figure}[t]
\centering
\includegraphics[width=8cm]{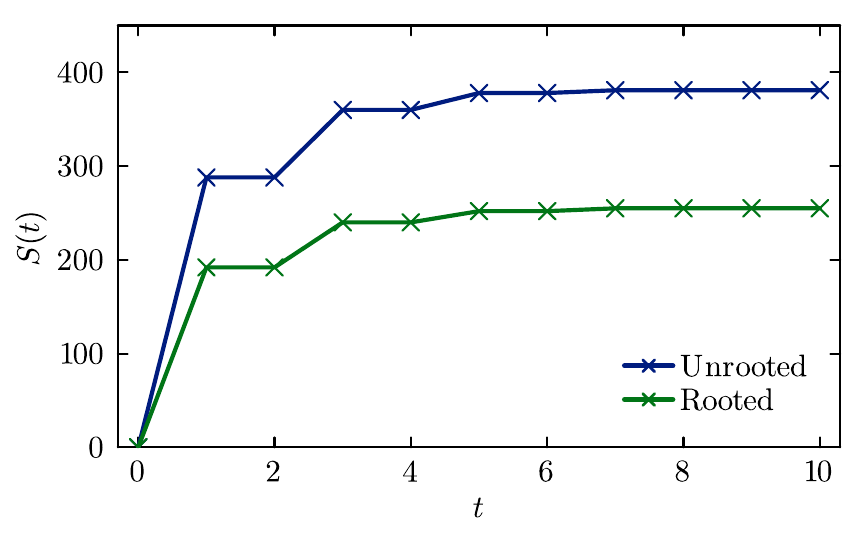}
\caption{Entanglement entropy $S(t)$ (in units of $\ln{2}$) starting from an exactly solvable GHZ state for a subsystem consisting of the first $r=7$ generations of the unrooted and rooted trees. Crosses are numerical data from simulating a tree-unitary Clifford circuit with $z=3$, and the solid lines correspond to Eqs.~(\ref{eq:eer_1}-\ref{eq:eer_3}) for the unrooted tree and Eqs.~(\ref{eq:ee_1}-\ref{eq:ee_3}) for the rooted tree.}
\label{fig:ee_tu}
\end{figure}

Had we not chosen a light cone region, then after using unitarity, there would be insufficiently many contractions to use tree-unitarity in going from \eqref{eq:ee_unrooted_unitarity} to \eqref{eq:ee_unrooted_tu}. To understand this, note that in the case of a light cone region, the final contraction fronts from different branches will reach the common node of those branches at the same time. This ensures that as we contract inwards, we are always able to trace over $z-1$ pairs of legs. However, with a region that is not a light cone, the final contraction fronts of two branches will reach their common node at different times; in the next time step, we will therefore not be able to contract over enough legs to use tree-unitarity to continue propagating the contraction front inwards.  
It is worth noting that the idea that the solvable regions for entanglement growth correspond to a light cone region applies also to the rooted tree with the light cone starting from the root, and can be seen as extending the notion of a contiguous region from (1+1)d circuits, in which the light cone region is just a line. 

We verify the formulas presented above for the entanglement growth for the rooted and unrooted tree by comparing against classical simulations of Clifford tree-unitary $z=3$ circuits, as shown in Fig.~\ref{fig:ee_tu}. Note that although Clifford circuits can be efficiently simulated in polynomial time in the number of sites $N$, the challenge in tree simulations is that $N$ itself grows exponentially with depth $r$, $N\sim z(z-1)^{r-1}$.

\section{\added{Non-Euclidean geometry, solvability, and maximum velocity}}
\label{subsec:tree_solvability}

\added{%To close this section on the dynamics of tree-unitary circuits, 
We now clarify the importance of the tree geometry in obtaining the above results. One of our striking results is the construction of a class of solvable models with a non-maximal butterfly velocity $v_B<1$. For models where the solvability arises because the correlation functions are constrained to be non-vanishing only on the edge of causal light cones (see Sec. \ref{subsec:correlations}), the butterfly velocity coincides with the light-cone velocity in all known constructions: in (1+1)d dual-unitarity~\cite{bertini_exact_2019,claeys_maximum_2020}, dual-unitarity in higher dimensions~\cite{suzuki_computational_2022,milbradt_ternary_2023}, triunitarity~\cite{jonay_triunitary_2021}, and hierarchical dual-unitarity~\cite{yu_hierarchical_2024,foligno_quantum_2024,rampp_entanglement_2023,rampp_geometric_2025,sommers_zero-temperature_2024}. Note that such light cones do not necessarily need to correspond to the geometric light cones, as e.g. in special classes of hierarchical dual-unitary gates~\cite{sommers_zero-temperature_2024,rampp_geometric_2025}.}

\added{We consider such a solvable circuit to be one for which
(a)  correlation functions are constrained to be non-vanishing only on isolated light cones; and (b) all light cone correlations at time $t$ can be obtained by acting with a sequence of $t$ quantum channels whose dimensions remain bounded as $t\to \infty$. 
We here first argue that for a circuit satisfying criterion (a) to have non-maximal butterfly velocity, the number of sites on the light cone must generally grow in time. 
Second, we argue that on Euclidean lattices, a light cone that grows in time implies a violation of criterion (b) above. In contrast, we show that even if the light cone grows with time, a {\it sufficient} condition for criterion (b) is for the lattice to possess a property known as $\delta$-hyperbolicity. The tree geometry is the simplest example with this property. We leave open the question of categorizing the full class of solvable lattice models that satisfy both the above criteria.}

\subsection{Growing Light Cone Requirement}

\added{As also discussed in Sec.~\ref{subsec:otoc}, the restriction that correlations are non-vanishing only on light cones strongly constrains operator spreading, which is in turn reflected by the OTOC.
Under unitary dynamics an initially localized Pauli operator $\sigma_{\beta}(j,t=0)$ will grow in time, as can be quantified by expanding the time-evolved operator in a basis of (products of) generalized Pauli operators}
\begin{align}
\label{eq:pauli_decomp}
\sigma_{\beta}(j,t) = \sum_{\mathcal{S}} c_{\mathcal{S}}(t)\, \mathcal{S},
\end{align}
\added{where $\mathcal{S}$ denotes a string of Pauli operators and $|c_{\mathcal{S}}(t)|^2$ is the operator weight on this Pauli string. (Recall that Pauli strings furnish a basis for Hermitian operators.) Under unitary dynamics the total operator weight is conserved: for an initially normalized operator we have that}
\begin{align}
\sum_{\mathcal{S}}|c_{\mathcal{S}}(t)|^2 = 1.
\end{align}
\added{%In the previously mentioned solvable models, 
In addition to the above constraint,  we now argue that in models satisfying criterion (a), the total weight on Pauli strings that act nontrivially on the outermost light cones is constant:}
\begin{align}\label{eq:wconst}
w=\sum_{\substack{\mathcal{S}, \mathcal{S}_{\rm LC} \neq \mathbb{1}}} |c_{\mathcal{S}}(t)|^2 = \textrm{const.},
\end{align}
\added{where $\mathcal{S}_{\rm LC}$ denotes the part of $\mathcal{S}$ which acts on the outermost light cones. 
To this end, we first show that \eqref{eq:wconst} is exactly fulfilled in a specific but large class of circuit geometries (which includes all examples of solvable dynamics mentioned above). These are geometries where each $z$-site unitary gate that acts on the light cone is constrained as follows: it has a single ingoing leg from the light cone at time $t$, with all the other $z-1$ ingoing legs from outside the light cone. As a consequence, at time $t+1$, this gate advances the light cone forward to the $z-1$ sites corresponding to the latter, leaving a single leg inside the light cone, corresponding to the site that was on the light cone at time $t$.
We then argue that even beyond this setting, violating \eqref{eq:wconst}, while possible in principle, would require excessive fine-tuning and hence can be ruled out by `naturalness'.}

\added{For the specific class of circuit geometries outlined above, the proof of \eqref{eq:wconst} is relatively straightforward. First, under unitary evolution every single-site Pauli operator on a light cone at time $t$ necessarily gets mapped to a linear combination of Pauli operators (single- or multi-site) that \emph{all} act nontrivially on the light cone at time $t+1$. Evolution to a single-site Pauli operator \textit{inside} the light cone would lead to a nontrivial `subluminal' correlation function inside the light cone, contradicting solvability condition (a). The only remaining possibilities are that this single-site Pauli operator gets mapped to single-site Pauli operators on the light cone or to multi-site Pauli operators. The latter necessarily act nontrivially on the light cone due to our constraint on the geometry. 
This argument immediately extends to multi-site Pauli operators which act nontrivially on the light cone, as these can always be factorized into a single-site Pauli matrix on the light cone and a Pauli operator acting nontrivially only inside the light cone. In all cases an operator with support on the light cone evolves into another operator with support on the light cone: the total operator weight on the light cone is conserved.}

\added{Finally, in the absence of the constraint that only one leg acts on the light cone, it is in principle possible that a single-site Pauli operator gets mapped to a multi-site Pauli operator inside the light cone. However, such multi-site Pauli operators inside the light cone could subsequently be mapped to a single-site Pauli operator and result in non-vanishing subluminal correlations; eliminating this possibility for more general geometries would require further constraints on the dynamics, and hence additional fine-tuning beyond `just' generalizations of dual-unitarity.}

\added{Assuming \eqref{eq:wconst}, we can now show that a non-maximal butterfly velocity requires that the size of the light cone grows with time. To see this, note that if the butterfly velocity is non-maximal, then the operator weight on any single site on the light cone vanishes as $t\to \infty$. Thus, if the total operator weight on the light cone is conserved, a non-maximal butterfly velocity requires that the number of sites on the light cone grows in time. Conversely, if in addition to the total operator weight the number of sites on the light cone is also conserved, then the butterfly velocity must be maximal. 
In Appendix~\ref{app:bound} we derive a precise bound relating the light cone weight to the light cone OTOCs. 
In particular, we introduce $\bar{O}$, which is the OTOC of a time-evolved operator $\sigma_{\beta}(j,t)$ with a single-site Pauli $\sigma_{\alpha}(i,0)$, averaged over all sites $i$ on the light cone of the time-evolved operator and all non-trivial Pauli operators $\sigma_{\beta}$,}
\begin{equation}
\label{eq:o_def}
    \bar{O}=\frac{1}{|R|} \sum_{i \in R} \frac{1}{q^2-1} \sum_{\beta} \langle \sigma_{\alpha}(i,0)\sigma_{\beta}(j,t)\sigma_{\alpha}(i,0)\sigma_{\alpha}(j,t)\rangle,
\end{equation}
\added{where $R$ denotes the light cone consisting of $|R|$ sites. For circuits where the butterfly velocity is non-maximal, $v_B<1$, $O$ approaches $1$ in the late-time limit. However, if the total operator weight on the light cone is fixed at $w$, then}
\begin{equation}
\label{eq:o_bound}
1-\bar{O} \addedtwo{\geq} \frac{2w}{|R|} \frac{q^2-2}{q^2-1}.
\end{equation}
\added{A non-maximal butterfly velocity with fixed light cone weight $w$ then \deletedtwo{clearly} requires that $|R|$ grows in time.}

\added{Note that, as demonstrated in the previous section, the tree-unitary circuits we construct exhibit maximal growth of entanglement entropy independent of whether they show the maximum butterfly velocity (see also the following section). This should be contrasted with (1+1)d, where the entanglement velocity lower bounds the butterfly velocity, and maximum entanglement velocity implies dual-unitarity \cite{zhou_maximal_2022}. Maximum-rate entanglement growth without maximum butterfly velocity is made possible by the fact that the weight on the light cone remains fixed in tree-unitary circuits.}

\subsection{Non-Euclidean Requirement}
\label{subsec:noneuclidean}

\added{We now turn to the constraint that correlation functions on the light cone be efficiently computable in the sense of solvability condition (b). We show that this cannot be true on Euclidean lattices unless correlation functions are constrained to be non-vanishing only along a line, which is ruled out by requiring them to be non-vanishing on a region growing with time. However, it is possible on graphs which are $\delta$-hyperbolic with finite $\delta \geq 0$, a term defined below.
The essential intuition behind this result is as follows: the computational hardness of computing correlation functions between two arbitrary sites is directly related to the intersection of their forward and backward light-cones. When considering light-like correlations (which are the only non-vanishing correlations in circuits satisfying criterion (a)), the size of such intersections are in turn are related to how geodesics between the sites ``spread'' with the separation between sites. The spreading of geodesics is fundamentally different in the Euclidean and $\delta$-hyperbolic settings, so that while the size of light-cone intersections is unbounded (for large time and distances) in the former case, it is bounded by a constant in the latter.}

\added{To make the above intuition concrete, we first set up the problem of computing light cone correlations more formally. To this end, we build an auxiliary graph $G$ whose vertices $V(G)$ correspond to the physical sites on the lattice (i.e., to the qubits), and whose
edges $E(G)$ are 
defined by the unitaries acting on this graph: if a given pair of sites $x,y \in V(G)$ are acted on together by a given unitary gate, then an (undirected) edge exists between them\footnote{To more faithfully account for the brickwork structure, we can increase the length of certain edges to account for delays in the light cone moving because a given unitary does not act in a given layer (as in Sec.~\ref{sec:2site}). However, for sensible gate choices (Sec.~\ref{subsec:isotropic_zsite}), only a small number of edges near the origin need to be adjusted, and this does not affect the conclusions below.}, i.e. $\{x,y\}\in E(G)$. Note that these are not necessarily the `natural' edges for the original lattice geometry.
We equip such a graph with a distance function between any two vertices $d(i,j)$, defined as the shortest number of edges that must be traversed to pass from $i$ to $j$, i.e. we use the `graph distance'.
$(V(G),d)$ then forms a metric space. 
}

\begin{figure}[t!]
\centering
\includegraphics[width=8cm]{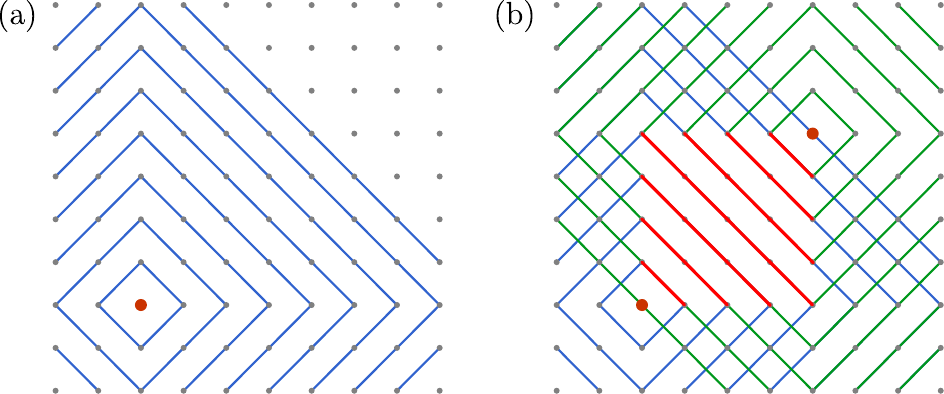}
\caption{\added{In (a) we plot the level sets for a square lattice with `graph distances'. These level sets correspond to the light cones when gates are allowed to act between nearest-neighbours on the lattice. In (b), we consider the intersections of level sets between two points. Intersections of two level sets containing more than one point are colored in red. The maximum size of any of these intersections here scales linearly with the separation between the two points. Only sites within these intersections contribute to the light-cone correlation function between the two points colored red.} }
\label{fig:levelsets}
\end{figure}

\added{The light cone of a given site $i$ after a time $t$ is equivalent to the \textit{level set} of $d(i)$ in $V(G)$, $L_i(t)=\{ x\in V(G) : d(i,x)=t\}$. This is illustrated for the square lattice in Fig.~\ref{fig:levelsets}(a).
Recall that from our criterion  (a) for solvability, the only nontrivial correlations are between sites $i$ and $j$ that are light-like separated. Thus, without loss of generality, let $j$ lie on the forward light cone of $i$, such that $t=d(i,j)$.
The correlations are then  determined only by gates that lie on the intersection of the forward light cone of $i$ and the backward light cone of $j$, since all other gates in the circuit can be contracted using unitarity. (Note that the preceding statement is true of light-cone correlations in any unitary circuit, though these will in general have nonzero correlations everywhere inside the light cone as well.)}

\added{The complexity of computing correlations between $i$ and $j$ is controlled by the number of gates in this intersection. To determine this, it suffices to characterize paths in $G$  that lie on the light cone intersections, which depend crucially on the graph geometry.
Given our definition of the light cones as level sets, it follows that such paths must pass 
through sites in $\intsec(i,j,s)=L_i(s) \cap L_j (t-s)=\{ x \in V(G): d(i,x) =s,~ d(j,x) = t-s\}$, with $s\leq t$. Notice that this implies $d(j,x)+d(i,x)=t$, i.e. all such paths from $i$ to $j$ are geodesics\footnote{Recall that in contrast to continuum Euclidean space with the $\ell_2$-norm where geodesics are unique, graph distance admits multiple  geodesics on graphs, a fact doubtless familiar to anyone who has visited  Manhattan.}. 
Crucially, if $|\intsec(i,j,s)|$ remains bounded\footnote{
Note the subtlety that  this does not require the {\it number} of geodesics to remain bounded. This is because even if   this number becomes unbounded due to a combinatorial blow-up in choosing paths through the intersections $\intsec(i,j,s)$, the resulting light cone dynamics can still be captured by the action of a channel of bounded size as long as $\intsec(i,j,s)$ remains bounded for all $s\leq t$ as $t\to\infty$.}
for all $s\leq t$ as $t \to \infty$, then this light cone correlation function is efficiently computable as demanded by solvability condition (b).
The two-point correlation function can only depend on those Pauli strings $\mathcal{S}$ in the expansion Eq.~\eqref{eq:pauli_decomp} which at time $s$ have 
weight \addedtwo{only} on $\intsec(i,j,s)$, and if $|\intsec(i,j,s)|$ remains bounded the total number of such Pauli strings also remains bounded. The former follows essentially from the fact that the dynamics traces light-like geodesics: causality  forbids contributions to  light-cone correlators from Pauli strings with support outside the intersection of forward and backward light cones of $i$ and $j$.
Computing the correlation function thus only requires us to compute, at each $s$, the amplitude  $\intsec(i,j,s)$ to $\intsec(i,j,s+1)$, which requires quantum channels  acting on $|\intsec(i,j,s)|$ ingoing sites and $|\intsec(i,j,s+1)|$ outgoing sites, whose size is $\exp(\max(|\intsec(i,j,s)|, |\intsec(i,j,s+1)|)$. The overall computational complexity is  $O(t\exp{\mathcal{D}_t})$, where 
\begin{equation}
\mathcal{D}_t = \max_{s\in[0,t]}|\intsec(i,j,s)|
\end{equation} is the size of the maximal intersection of the light cones.  Observe that if $\mathcal{D}_t$ grows polynomially or worse with $t$, then the problem is in principle exponentially difficult, absent some other physical simplification in the channel structure.}

\added{We now argue that, for any graph obtained from a Euclidean lattice, $\mathcal{D}_t = \Omega (t)$, i.e. is at least linear, meaning that computing correlations is exponentially difficult. 
It is straightforward to prove this for graphs that {are} $n$-dimensional lattices (i.e., isomorphic to $\mathbb{Z}^n$) with a set of lattice \addedtwo{basis} vectors $\{\hat{e}_k \}$ for $k=1,\dots n$. 
Let $i$ be the origin, and $j=a (\hat{e}_1+\hat{e}_2)$ for some $a>0$, such that $d(i,j)=2a$. Then the intersection of the level sets $L_i(a) \cap L_j (a)$ consists of all points $ m \hat{e}_1 + (a-m) \hat{e}_2$ for $0 \leq m \leq a$, so $|L_i(a) \cap L_j (a)|=a+1$, which scales linearly with $t=d(i,j)$; it follows that $\mathcal{D}_t>O(t)$ and is therefore unbounded. This is illustrated for the square lattice in Fig.~\ref{fig:levelsets}(b).
By further considering $j=a\sum_k \hat{e}_k$ one can show that $\mathcal{D}_t = \Omega(t^{n-1})$, but the simple example above with two lattice vectors is sufficient to show unboundedness.}

\begin{figure}[t!]
\centering
\includegraphics[width=5cm]{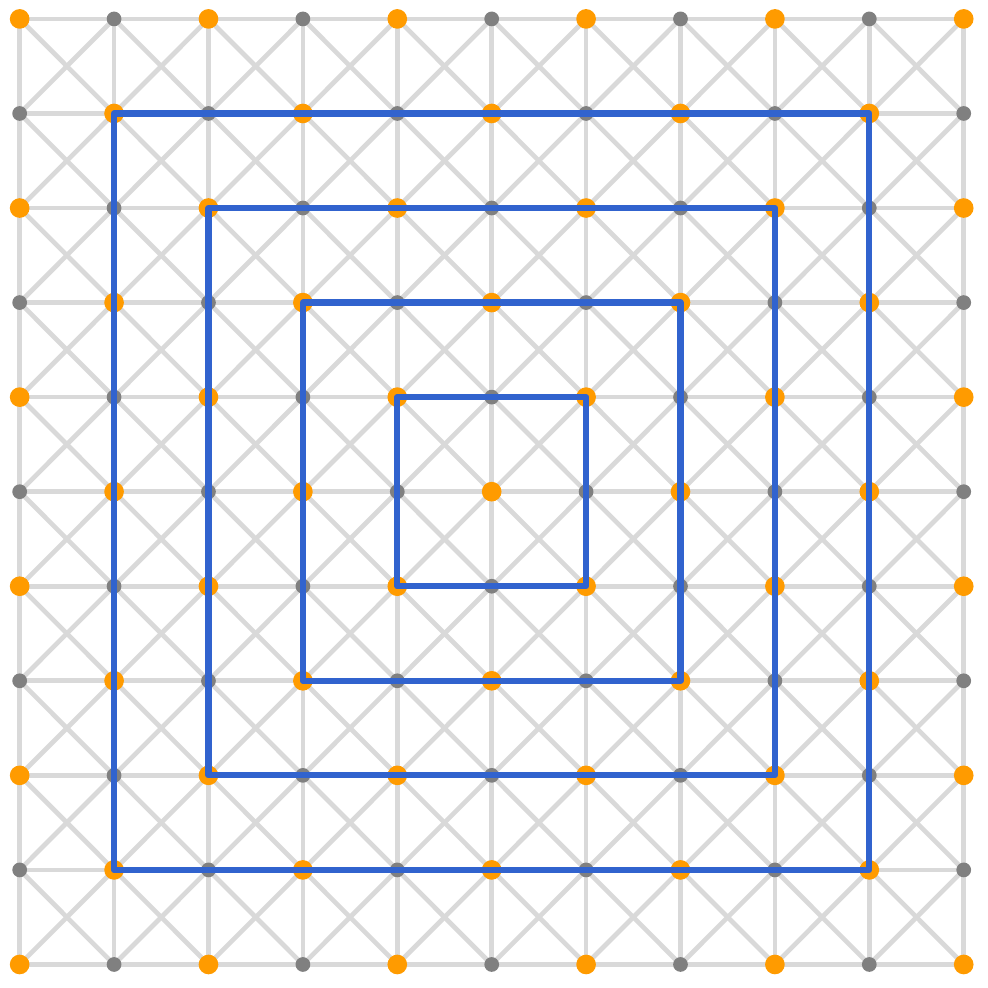}
\caption{\added{The structure of gates can modify the way distances are measured on the lattice. For example, if we allow 4-site gates to act on squares in the square lattice, we should add diagonal edges which allow one to cross the diagonal of the square with a distance $1$. This changes the level sets from diamonds (as in Fig.~\ref{fig:levelsets}) to squares. However, these can still be considered the level sets of the rotated square lattice shown in orange, and thus level set intersections are unbounded by the argument in the main text.}}
\label{fig:levelsetsrotated}
\end{figure}

\added{The alert reader might worry that some choices of unitary gate \addedtwo{structures} could lead to edges in $G$ such that its level sets are distinct from those of the underlying lattice of qubits.
However, for sensible gate structures for which the light cone grows `smoothly', we expect that the light cones (level sets of the graph) will remain the levels sets of \textit{some} Euclidean sublattice of the original sites, as demonstrated in Fig.~\ref{fig:levelsetsrotated} for the square lattice, and therefore level set intersections will still be unbounded. Indeed, one could simply {\it define} a well-behaved circuit to be one for which the level sets of the auxiliary graph $G$ satisfy this property.}

\added{It is of course possible to choose special pairs $i,j$ for which $\intsec(i,j,s)$ remains bounded in size. This is leveraged in existing higher-dimensional extensions of dual-unitarity, such as ternary-unitary circuits \cite{milbradt_ternary_2023}, where correlation functions are constrained to be non-vanishing only along such special directions that admit efficiently computations of light-cone correlations. However, since we require that correlations are non-vanishing on a growing number of sites, such correlations will not be efficiently computable in Euclidean geometries except between special such pairs of sites.}

\added{For what graphs $G$ \textit{does} this property hold for any pair of sites? One natural answer is in terms of a property called Gromov $\delta$-hyperbolicity \cite{bridson_metric_1999}. Consider a point $y_s \in \intsec(i,j,s)$. Then we can form a geodesic triangle consisting of the three points $\{i,j,y_s\}$ and three geodesics $[i,y_s], [y_s,j],[i,j]$ joining them. 
We say that this triangle is $\delta$-thin if each side lies within a distance $\delta$ of some point on the other two sides. In other words, for any point $p$ on one side (say $[i,j]$) there is some point $q$ on the union $[i,y_s]\cup [y_s,j]$ with $d(p,q)\leq \delta$.
If there exists a fixed finite $\delta \geq 0$ such that every geodesic triangle in $G$ is $\delta$-thin, then the metric space $(V(G),d)$ is called $\delta$-hyperbolic. 
In App.~\ref{app:delta}, we prove that a sufficient condition for $|\intsec(i,j,s)|$ to remain bounded for all $s\leq t$ is that $(V(G),d)$ is $\delta$-hyperbolic.
Intuitively, this is because in $\delta$-hyperbolic lattices any two geodesics between a pair sites $i,j$ must remain within a bounded distance of each other, and therefore the intersection of the level sets (through which all geodesics pass) cannot grow without bound. }

\added{This result establishes that on Euclidean geometries, quantum circuits with $v_B<1$ , i.e. non-maximal butterfly velocity, cannot be exactly solvable in the sense of our definition. In contrast $\delta$-hyperbolicity is a sufficient condition for efficiently computable light-cone correlations [criterion (b)], in the sense of linear-in-$t$ scaling of the relevant complexity. The simplest example satisfying this criterion is the Cayley tree (with $\delta =0$); here the light-cone correlations are even analytically tractable --- and not just computationally `simple' --- since they can be computed by repeated applications of one of a finite set of channels. However, $\delta$-hyperbolicity alone does not guarantee the latter: in principle each time slice from $s$ to $s+1$ could require a different channel (although each is of bounded size), so while the computation scales linearly it is still potentially  challenging to perform {\it analytically}. Whether simplifications similar to those on the tree emerge will require a more careful analysis, potentially on a case-by-case basis. Examples of $\delta$-hyperbolic graphs with non-zero $\delta$ include the hyperbolic tilings of the plane (with Schl{\"a}fli symbols $\{p,q\}$ where $(p-2)(q-2) > 4$ \cite{coxeter_regular_1973}). It remains to identify brickwork structures that realize solvable dynamics  (i.e., that also satisfy criterion (a), triviality of correlations off the light cone) on hyperbolic lattices and other $\delta$ hyperbolic graphs; we defer this to future work \cite{breach_inprep_2025}. }

\section{Maximum Velocity Tree-Unitary Circuits}
\label{sec:maxvel}

Above, we found that tree-unitarity alone does not imply operator spreading with a maximum butterfly velocity, in contrast with the situation for dual-unitary circuits. Fundamentally, this difference is due to the fact that the tree geometry allows the operators to move in multiple directions at a given time step, as opposed to a single direction in (1+1)d. However, by imposing further constraints beyond just tree-unitarity, it is possible to obtain tree-unitary circuits in which operators spread with maximal velocity in either specific directions or {\it all} directions.

\begin{figure}[t]
\centering
\includegraphics[width=8cm]{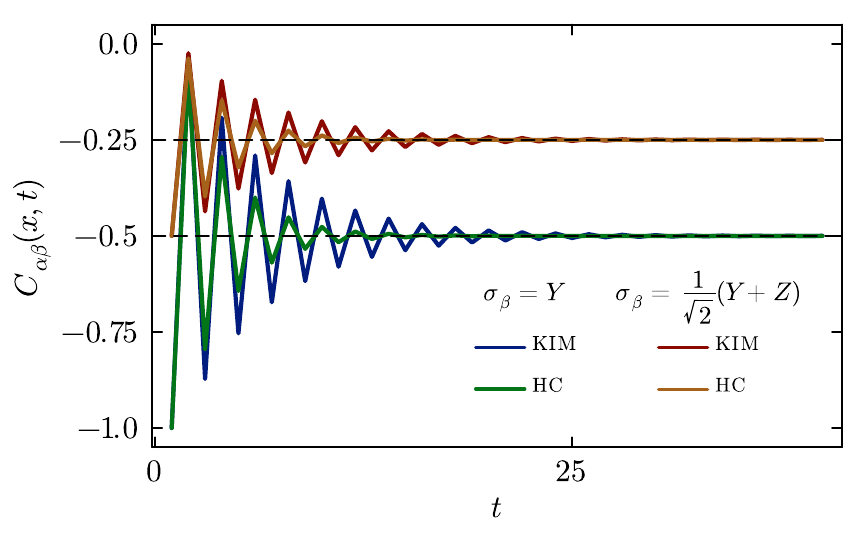}
\caption{OTOC $C_{\alpha \beta}(x,t)$ on the light cone direction $2\to 3$ with $\sigma_{\alpha} = \frac{1}{\sqrt{2}} (X+Z)$ and $\sigma_{\beta} = Y$ or $\sigma_{\beta} = (Y+Z)/\sqrt{2}$, for the kicked-Ising model (KIM) and a Hadamard-based construction (HC) discussed in App.~\ref{app:constructions}. The OTOC approaches the values of $-0.5$ or $-0.25$, in accordance with the calculation in App.~\ref{app:identities}, confirming the operators spread at maximum velocity.}
\label{fig:otocplot_kim}
\end{figure}

In fact, the latter scenario pertains to the kicked Ising dynamics of Sec.~\ref{subsec:kim}.
In Fig.~\ref{fig:otocplot_kim}, we show that for the KIM the OTOC approaches a non-trivial value on the light cone (dependent on the choice of operators), indicating a maximal butterfly velocity. This maximal butterfly velocity in all directions can be understood by observing that the KIM satisfies a further set of relations, which we term the {\it maximum-velocity conditions}. 
In order for a gate to be maximal velocity from ingoing leg $i$ to outgoing leg $j\neq i$, it needs to be unitary when 
these gates are swapped, i.e. when it is viewed as a transformation from the $z$ legs consisting of the $z-1$ incoming legs different from $i$ and the outgoing leg $j$, to the remaining set of $z$ legs consisting of $i$ grouped with the $z-1$ outgoing legs distinct from $j$. In the folded graphical notation, focusing on $z=3$ and $i=2,j=1$ for concreteness, this condition reads
\begin{equation}
\label{eq:maxvel}
    \vcenter{\hbox{\includegraphics[height=0.2\columnwidth]{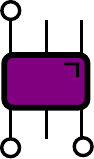}}}\,=\,
    \vcenter{\hbox{\includegraphics[height=0.16\columnwidth]{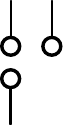}}}\,,
\end{equation}
which is equivalent to tracing over the other set of legs
\begin{equation}
\label{eq:maxvel_2}
    \vcenter{\hbox{\includegraphics[height=0.2\columnwidth]{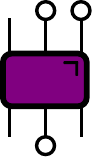}}}\,=\,
    \vcenter{\hbox{\includegraphics[height=0.16\columnwidth]{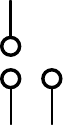}}}\,.
\end{equation}
\replacedtwo{T}{Clearly, t}here are $z(z-1)$ such maximum-velocity conditions corresponding to the possible choices for the ingoing and outgoing legs. When a gate satisfies the maximum-velocity condition for at least one  pair of ingoing and outgoing legs,  we shall term it a maximum-velocity gate.

The maximum-velocity conditions are stronger than tree-unitarity, since unitarity is stronger than isometry. Put differently, the maximum-velocity condition ~\eqref{eq:maxvel} automatically implies the corresponding tree-unitarity condition \eqref{eq:tree_unitarity_conditions}. 
In the same way that dual-unitarity implies a maximal butterfly velocity, imposing the maximal velocity condition %\eqref{eq:maxvel} 
between sites $i$ and $j$ implies maximal operator spreading from site $i$ to site $j$. Considering Eq.~\eqref{eq:maxvel}  (where $i=2$, $j=1$) for concreteness, this identity implies that $\mathrm{tr}_1[U^{\dagger}\sigma(2)U]=0$. 
As discussed in the context of correlation functions, tree-unitarity implies that $U^{\dagger}(\mathbb{1} \otimes \sigma \otimes \mathbb{1})U$ can only have single-site Pauli contributions where the single-site Pauli matrix `hops' to another site. This additional condition, however,  implies that only terms in which there is a Pauli matrix on site $1$ can appear. As such, the operator front necessarily moves to site $1$. More generally, if tracing over all ingoing legs except for $i$ and tracing over the single outgoing leg $j$ yields the identity (i.e. the maximum-velocity condition between $i$ and $j$ is satisfied), then the operator front initially localized at site $i$ necessarily grows to site $j$. 
This property manifests itself as a non-trivial light cone OTOC in the $i\to j$ direction. 

To see this more formally, consider the channel $\mathcal{T}_{e\tilde{e}}$ introduced in Sec.~\ref{subsec:otoc} [Eq.~\eqref{eq:Tee_otoc}]. If the circuit satisfies the maximum-velocity condition for $e\to\tilde{e}$, then we can construct an additional (left and right) eigenoperator of $\mathcal{T}_{e\tilde{e}}$ with unit eigenvalue:
\begin{equation}
    |R')=\,\vcenter{\hbox{\includegraphics[height=0.125\columnwidth]{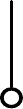}}}\,, \qquad (L'|=\,\vcenter{\hbox{\includegraphics[height=0.125\columnwidth]{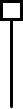}}}
    \,,
\end{equation}
such that $\mathcal{T}_{e\tilde{e}} |R') = |R')$ and $(L'|\mathcal{T}_{e\tilde{e}} = (L'|$.
To understand this, suppose that Eq.~\eqref{eq:maxvel} is satisfied, such that the circuit is maximum velocity in the $2\to 1$ direction. In the four-folded notation, this becomes
\begin{align}
    \vcenter{\hbox{\includegraphics[height=0.18\columnwidth]{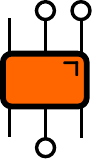}}}\,=\,
    \vcenter{\hbox{\includegraphics[height=0.16\columnwidth]{figures_final/2d1u.pdf}}}
\end{align}
and therefore
\begin{align}
    \mathcal{T}_{21}|R')=\vcenter{\hbox{\includegraphics[height=0.18\columnwidth]{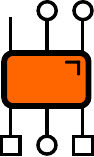}}}=\vcenter{\hbox{\includegraphics[height=0.13\columnwidth]{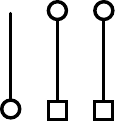}}}=\vcenter{\hbox{\includegraphics[height=0.13\columnwidth]{figures_final/Rprime_circle.pdf}}} = |R').
\end{align}

\begin{figure}[t]
\centering
\includegraphics[width=8cm]{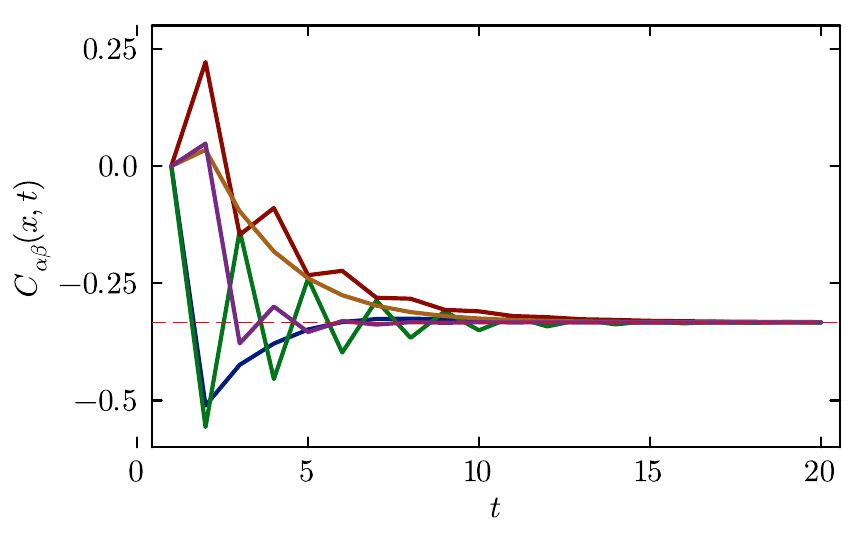}
\caption{OTOC $C_{\alpha \beta}(x,t)$ on the light cone direction $2\to 3$ with $\sigma_{\alpha} = \frac{1}{\sqrt{2}} (X+Z)$ and $\sigma_{\beta} = Z$, for randomly generated maximum-velocity tree-unitary gates (as described in App.~\ref{app:constructions}). The OTOC approaches $-1/(q^2-1)=-1/3$ on the light cone, as expected from the calculation in the main text.}
\label{fig:otocplot_mvgen}
\end{figure}

The steady-state value of the OTOC on the light cone is set by these non-trivial eigenvalues.
From the right eigenstates $|R)$ and $|R')$ and the left eigenstates $(L|$ and $(L'|$ we construct a biorthonormal basis
\begin{align}
    |v)&=|R),\quad |v')=\frac{|R')-q|R)}{\sqrt{q^2-1}},\\
    (v|&=(L|,\quad (v'|=\frac{q(L|-(L'|}{\sqrt{q^2-1}}.
\end{align}
satisfying $(v|v)= (v'|v')= 1$ and $(v'|v) = (v|v') = 0$. 
At late times we can replace the repeated action of the channel by a projection on these states, i.e.
\begin{align}
    \lim_{t \to \infty} \mathcal{T}_{e \tilde{e}}^t = |v)(v| + |v')(v'|
\end{align}
The OTOC will hence approach a non-trivial value on the light cone
\begin{align}
    \lim_{t \to \infty} C_{\alpha \beta}(i,j;t) &= (\sigma_{\alpha}|v)(v|\sigma_{\beta})+(\sigma_{\alpha}|v')(v'|\sigma_{\beta}) \nonumber\\
    &=1-\frac{q^2}{q^2-1}=\frac{-1}{q^2-1},
\end{align}
where we have evaluated the overlaps as 
\begin{align}
    &(\sigma_{\alpha}|v)=1, \quad (\sigma_{\alpha}|v')=-\frac{q}{\sqrt{q^2-1}}, \\
    &(v|\sigma_{\beta})=1, \quad (v'|\sigma_{\beta})=\frac{q}{\sqrt{q^2-1}}.
\end{align}
This value indicates a maximal butterfly velocity as well as maximal scrambling on the edge of the causal light cone. In Fig.~\ref{fig:otocplot_mvgen}, we show this for 5 randomly generated maximum-velocity tree-unitary gates (constructed as described in App.~\ref{app:constructions}).

Despite their appearance, these additional maximum-velocity conditions \eqref{eq:maxvel} are quite natural, since various explicit constructions of tree-unitary gates satisfy one or more maximum-velocity conditions. The KIM of Sec.~\ref{subsec:kim}, for example, is maximum velocity between neighbouring legs. It can be directly checked that, for example,
\begin{align}
\label{eqn:kim_maxvel}
    \addedtwo{16}\vcenter{\hbox{\includegraphics[height=0.22\columnwidth]{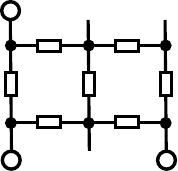}}}&= 
    \addedtwo{8}\vcenter{\hbox{\includegraphics[height=0.2\columnwidth]{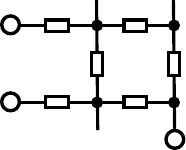}}}=
    \addedtwo{4}\vcenter{\hbox{\includegraphics[height=0.2\columnwidth]{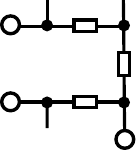}}} \nonumber\\
    &=\addedtwo{2}\vcenter{\hbox{\includegraphics[height=0.2\columnwidth]{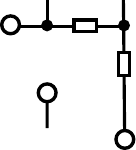}}}=
    % \addedtwo{2}\vcenter{\hbox{\includegraphics[height=0.18\columnwidth]{figures_final/kim/kim_mv5.pdf}}}=
    \vcenter{\hbox{\includegraphics[height=0.18\columnwidth]{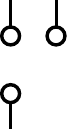}}},
\end{align}
which constrains $U^{\dagger}\sigma_{\alpha}(2)U$ to only generate terms of the form $\sigma\otimes \sigma\otimes \sigma$ or $\sigma\otimes \mathbb{1} \otimes \sigma$.
In fact, for ingoing site $2$ the KIM satisfies an additional identity, as proven in App.~\ref{app:identities},
\begin{equation}
    \addedtwo{16}\label{eq:kim_identity_main}\vcenter{\hbox{\includegraphics[height=0.22\columnwidth]{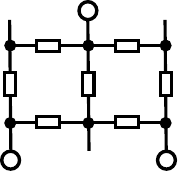}}}= 
    \vcenter{\hbox{\includegraphics[height=0.15\columnwidth]{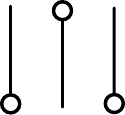}}}
\end{equation}
which additionally removes the possibility of $\sigma\otimes \mathbb{1}\otimes \sigma$ terms. 

The prediction $-1/(q^2-1)$ assumes that there are no other nontrivial unit-modulus eigenoperators; for the KIM, more eigenoperators can be constructed, giving rise to a light cone OTOC value that depends on the choice of operators Fig.~\ref{fig:otocplot_kim} (see App.~\ref{app:identities}). Taking the KIM again, for an operator initially at the root of a gate in the first layer, all correlation functions vanish, indicating maximum ergodicity. The only non-vanishing correlation functions are for an operator passing from a leaf to the root of the gate always.
For higher-$z$ KIM, an operator initially at the root of the gate will spread with maximum velocity to all leaf sites (but unlike $z=3$, need not remain on the root). An operator initially at a leaf will spread with maximum velocity to the root of the gate, without spreading to the other leaves. 

A different class of tree-unitary gates that are maximum velocity in {\it all} directions are perfect tensors. Perfect tensors (for an even number of legs) encode a unitary transformation between any equal-size bipartitions of their legs, i.e. for $z=3$ these satisfy
\begin{gather}
    \vcenter{\hbox{\includegraphics[height=0.1\columnwidth]{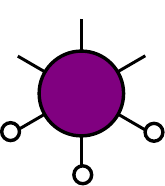}}}= 
    \vcenter{\hbox{\includegraphics[height=0.1\columnwidth]{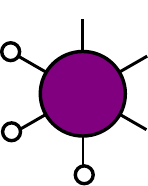}}}= 
    \vcenter{\hbox{\includegraphics[height=0.1\columnwidth]{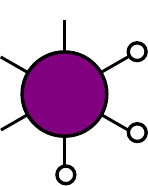}}}= 
    \vcenter{\hbox{\includegraphics[height=0.1\columnwidth]{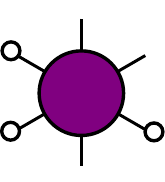}}}= 
    \vcenter{\hbox{\includegraphics[height=0.1\columnwidth]{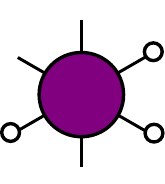}}}= \nonumber\\
    \vcenter{\hbox{\includegraphics[height=0.1\columnwidth]{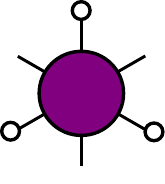}}}= 
    \vcenter{\hbox{\includegraphics[height=0.1\columnwidth]{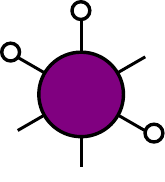}}}= 
    \vcenter{\hbox{\includegraphics[height=0.1\columnwidth]{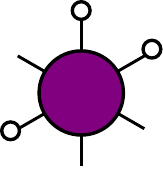}}}= 
    \vcenter{\hbox{\includegraphics[height=0.1\columnwidth]{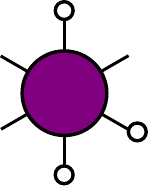}}}= 
    \vcenter{\hbox{\includegraphics[height=0.1\columnwidth]{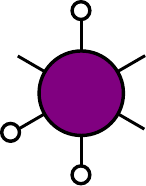}}}= 
    \vcenter{\hbox{\includegraphics[height=0.08\columnwidth]{figures_final/triple_identity.pdf}}}.
\end{gather}
The maximum-velocity conditions correspond to a subset of these conditions. For $z=2$, perfect tensors return dual-unitary gates that are additionally maximally ergodic, i.e. all correlation functions vanish identically after a single time step~\cite{aravinda_dual-unitary_2021}. Remarkably, for tree-unitary gates this is also the case, but with a much less stringent condition: all correlation functions vanish identically after a single time step provided the gate satisfies at least two maximum-velocity conditions. In the language of Ref.~\onlinecite{aravinda_dual-unitary_2021}, these tree-unitary gates result in {\it Bernoulli dynamics}. Interestingly, while Bernoulli dynamics cannot be realized for $q=2$ in dual-unitary gates, moving to tree-unitarity with $z >2$ directly returns a large class of such models.

This result can be understood by noting that the maximum-velocity condition strongly constrains the quantum channel~\eqref{eq:channelcorrelator} in the calculation of the correlation functions. If Eq.~\eqref{eq:maxvel} is satisfied, then $M_{23}(\sigma)=0$, so correlation functions in the $2\to 3$ direction vanish. More generally, if the gate satisfies maximum velocity for $i\to j$, then $M_{i\tilde{e}}(\sigma)=0$ for $\tilde{e} \neq j$. A particular direction being maximum velocity means that no other direction can have non-vanishing correlation functions. Accordingly, if a gate is maximum velocity in two directions (from a given ingoing leg), then all correlation functions stemming from this ingoing site must vanish. 
Both perfect tensors and the KIM, along with further constructions which we consider in App.~\ref{app:constructions}, are not just maximum velocity but are maximal velocity in more than one direction. All correlation functions therefore vanish, indicating maximum ergodicity on the level of the correlation functions.

\section{Concluding Remarks}
\label{sec:conclusion}

We began this paper with two central objectives. Our first and more conceptual goal was to understand the basic properties of local unitary dynamics on trees. Our second was more technical: namely, to identify a notion of solvability akin to dual-unitarity in one dimension in the resulting ``tree+1-dimensional'' unitary circuits. The obstacle to the former is the desire to preserve (to the greatest extent possible) the extensively large set of tree symmetries that arise from the freedom to permute the branches at any given node. For the latter, the challenge is that a spatial structure that is a tree clearly cannot admit a  space-time duality of the type central to the original formulation of dual-unitarity. 
\deletedtwo{(This is in contrast, for example, to problems where the entire dynamical evolution in spacetime is described by a tree tensor network.)}

In spite of these challenges, as reported in the preceding sections, we have made substantial progress on both of our main objectives. First, we demonstrated that a seemingly-natural generalization of local unitary dynamics to trees, based on applying 2-site gates according to a $z$-coloring of the bonds, fails to preserve the tree symmetries --- most obviously reflected in the existence of slow and fast portions of the light cone (despite such pathologies, the resulting circuits allow exact solvability using dual-unitary techniques, summarized for the interested reader in App.~\ref{app:2site_dynamics}). Instead, we showed that a distinct unitary circuit construction, involving $z$-site gates applied alternately to sets of sites dictated by a certain bond 2-coloring, results in an isotropic light cone. 

For these $z$-site-gate circuits we then defined a notion of ``tree-unitarity'', that trades the unitary space-time duality for a requirement of space-time isometry between the space-like (tree-like) and time-like directions. 
Notably, a physically-motivated example of dual-unitarity, the self-dual kicked Ising model, exhibits tree-unitarity when appropriately generalized to the tree setting. 
Imposing tree-unitarity in turn allowed us to lift much of the solvability of dual-unitary circuits to trees. For example, in tree-unitary circuits,  operator correlations vanish except on the light cone, where they can be exactly computed using quantum channels \added{despite the exponential growth of the number of sites}; the ``out-of-time order correlator'' (OTOC), a key diagnostic of operator scrambling and quantum chaos,  can be computed on the light cone using similar techniques;  and  exact calculations of entanglement growth are possible for a set of special initial conditions, essentially by `folding' the problem into a one-dimensional system whose Hilbert space dimension grows exponentially with distance from the origin.

However, the move from unitarity to isometry also revealed that tree-unitary quantum circuits are richer than their dual-unitary counterparts. This is especially evident in the computation of OTOC: whereas for dual-unitary circuits the OTOC spreads at the maximum possible ``butterfly velocity'', manifesting in nontrivial operator scrambling already {\it at} the causal light cone, this is not automatically the case for a generic tree-unitary circuit. Instead, requiring nontrivial OTOCs on the light cone stems from an additional set of ``maximum velocity'' conditions, which are not satisfied by generic tree-unitaries. We discuss examples of such maximum-velocity tree-unitary circuits in App.~\ref{app:constructions}, but note that the KIM is one such example. This weakening of the connection between solvability and ``maximum velocity'' means that, in a certain sense, tree-unitary circuits are more `generic' than dual unitaries --- at least in their operator scrambling properties. 
\added{We argue that, subject to certain mild assumptions on the solvability constraints and geometry, this feature can \textit{never} occur in Euclidean geometries, but is possible on graphs which are $\delta$-hyperbolic, of which the Cayley tree studied here is the simplest example.}

Having established these new ideas and results, in the remainder of this section we discuss their possible extensions, links to other work, and questions they raise about dynamics more generally. 

First, we note that there are other interesting computations one can envisage on the tree, or even locally tree-like graphs. For instance, all our present results make no assumptions of regularity in either space or time (other than, of course, tree-unitarity). However, one can compute additional quantities when such structure is present --- as for instance is the case for {\it Floquet} circuits such as the KIM, that repeat periodically in time. One such quantity is the so-called spectral form factor (SFF) \cite{bertini_exact_2018}, defined as $K(t) = \text{Tr} [U^t]$, where $U$ is the unitary evolution operator for a single full time step (in our case, corresponding to a single application of gates on each of the disjoint sets of bonds). The SFF encodes correlations between the eigenphases of the Floquet operator $U$, and as such is a sensitive spectral probe of chaos (complementing the OTOC). One challenge in computing the SFF is that it is not self-averaging, necessitating the introduction of some sort of disorder averaging to obtain sensible results. While such an analytic computation may be feasible on a tree, a more subtle issue is to interpret the result, since the thermodynamic limit of locally tree-like problems is often subtle\footnote{For instance, the spectrum of the adjacency matrix on a tree has distinct thermodynamic limit depending on whether one takes this limit through a sequence of large Cayley trees, or as (a suitable average over) an ensemble of regular random graphs.}. Numerical computations are challenged both by the exponential growth of the system size with depth, and, more seriously, by the finite bulk-boundary ratio; it is hence crucial to identify ways to consider ``closed'' trees by suitably rewiring sites on the boundary. Addressing these issues, both for the computation of the SFF and more generally, is an important direction for future work. 

This connects naturally to the broader question of extending tree-unitarity \added{and similar notions} to more general ``expander'' graphs, such as hyperbolic lattices, that share  a similar finite bulk-boundary ratio for subsystems. Dynamics is such systems is clearly important to address, given the relevance of expanders to quantum error correction and related ideas~\cite{breuckmann2016hyperbolic,breuckmann2017hyperbolic,fahimniya2023hyperbolic_floquet,higgott2024hyperbolic_floquet,breuckmann2021qldpc_review}.
The crucial new feature in such settings, in common with the closed trees mentioned above, is the presence of loops (which are usually short in the former case, and long in the latter). \added{From our discussion in Sec.~\ref{subsec:tree_solvability}, it follows that our definition of exact solvability could extend to the ($\delta$-)hyperbolic case.
We hope to present a construction of brickwork structures that realize such dynamics in future work.
% Constructing approximations to} tree-unitary-type solvability in the presence of loops is a question to which we have no clear answer at the present time -- yet one that is clearly important to address, given the relevance of expanders to quantum error correction and related ideas~\cite{breuckmann2016hyperbolic,breuckmann2017hyperbolic,fahimniya2023hyperbolic_floquet,higgott2024hyperbolic_floquet,breuckmann2021qldpc_review}. 
For more general graphs which do not admit such solvability,} one promising direction is to utilize techniques such as the cavity method~\cite{mezard_spin_1986,mezard_bethe_2001,mezard_cavity_2003,laumann_cavity_2008,laumann_aklt_2010} and belief propagation~\cite{alkabetz_tensor_2021,tindall_efficient_2024,mezard_information_2009} that have proven to be powerful tools in more conventional statistical mechanics problems on trees. 
\added{Along these lines, Ref.~\cite{park_simulating_2025} developed influence-functional belief propagation to simulate local observable dynamics on arbitrary graphs, with `tree+1' forming a starting point.}
In a different direction, multi-unitary gates arranged on expander graphs have gained interest as holographic codes, and it is natural to explore how tree-unitarity modifies e.g. the coding properties and correlation structure within such models~\cite{pastawski_holographic_2015,evenbly_hyper-invariant_2017,harlow_tasi_2018,bistron_bulk-boundary_2025}.

Even within the tree+1 dimensional unitary setting, there are several open problems that we briefly flag. We have already mentioned the use of Clifford tree unitaries as a means of simulating entanglement growth (albeit for small system sizes due to the exponential-growth problem). Of course, Clifford dynamics is more general and would allow a deeper understanding of non-tree-unitary circuits on trees~\cite{aaronson_improved_2004}.  A second class of solvable quantum circuits are built from so-called ``matchgate'' tensors, which can often be mapped to free fermions. Studying such matchgate circuits on trees  potentially  offers a distinct tractable limit to tree-unitarity~\cite{jozsa_matchgates_2008,valiant_quantum_2002,projansky_entanglement_2024,richelli_brick_2024}. A rather different question is whether tree-unitarity can be extended in a hierarchical manner, as in the case of dual-unitary circuits, where such constructions can lead to circuits with multiple light cones and submaximal entanglement growth~\cite{yu_hierarchical_2024,rampp_entanglement_2023,rampp_geometric_2025,sommers_zero-temperature_2024,foligno_quantum_2024,liu_solvable_2025}.
Moving away from tree-unitary problems and instead considering Haar-random unitary gates offers an entirely new line of investigation \cite{von_keyserlingk_operator_2018,nahum_operator_2018}. Such random unitary circuits are usually analytically tractable in the limit of infinite Hilbert space dimension of each individual gate \cite{chan_solution_2018,chan_spectral_2018,jian_measurement_2020}. Since for the $z$-site construction, each gate is a $q^z\times q^z$ unitary, this limit can be accessed by taking {\it either} $q$ or $z$ to infinity while holding the other fixed. Understanding the differences, if any, between these limits is an intriguing question, as is the development of an `entanglement membrane'-like picture for trees \cite{nahum_quantum_2017,zhou_entanglement_2020,rampp_entanglement_2023,sommers_zero-temperature_2024,foligno_quantum_2024}. 
For structured models, these different limits should be reflected in the eigenstates of the unitary evolution operator. Eigenstates of systems with local (one-to-one, $z=2$) and long-range (all-to-all, similar to $z\to \infty$) interactions have different entanglement structure~\cite{dalessio_quantum_2016,defenu_outequilibrium_2024}, and it is natural to wonder where tree-like interactions fall in between these limits using e.g. probes of the eigenstate thermalization hypothesis.
Still further extensions of both random and tree-unitary circuits are possible by modifying them to include conservation laws \cite{khemani_operator_2018,richter_transport_2023,foligno_nonequilibrium_2025}, and exploring the resulting dynamics. 

Breaking unitarity by introducing measurements is yet another possible extension of our work \cite{li_quantum_2018,skinner_measurement-induced_2019,chan_unitary-projective_2019,gullans_dynamical_2020,jian_measurement_2020,yao_temporal_2024}. Here, the relevant question is whether there are new insights to be gleaned into the measurement-induced phase transition (MIPT) by moving to the tree setting. Since the boundary of a region now scales in proportion to its volume, we expect scrambling to have a stronger effect than in finite dimension. However, given that MIPTs (characterised by information retained from an initial state) are observed even in all-to-all models \cite{nahum_measurement_2021,gullans_dynamical_2020,vijay_measurement-driven_2020}, an MIPT in tree+1 dimensions seems likely.
We note that although dynamical quantum trees have allowed for analytic insight into the MIPT beyond what is achievable in (1+1)d \cite{nahum_measurement_2021,feng_measurement-induced_2023}, these simplifications are unlikely to hold in the tree+1 setting. 
Within the tree-unitary setting, one pertinent consideration is if the restricted choice of projective measurements and associated circuit modifications that are necessary in order to preserve solvability \cite{claeys_exact_2022} are richer than their counterparts in finite dimensional circuits.

Finally, one can consider the $z$-site tree-unitary gates, but now acting in one dimension. The isometry conditions now offer a different class of dynamics beyond dual-unitarity; \added{although the dynamics here would not necessarily be solvable}, it would be interesting to explore the attendant physical consequences of the isometries. 

We could continue to enumerate further new directions, but close here with the hope that answers to those raised above might appear in the not-too-distant future.

\section*{Code Availability}

\added{We provide implementations of the algorithm to generate tree-unitary gates, \addedtwo{estimation of their dimensionality,} calculation of correlation functions and OTOCs, and Clifford simulations of entanglement entropy, in a GitHub repository~\cite{code}.}

\begin{acknowledgements}
We thank Vedika Khemani, Yaodong Li, Frank Pollman, Bal\'azs Pozsgay, Suhail Rather, Charles Stahl and \addedtwo{Fredy Yip} for useful discussions. \added{We are especially grateful to Dawid Kielak for discussions on $\delta$-hyperbolicity and related graph theoretic issues.  
We also thank the anonymous referees for helpful comments on the manuscript.}
O.B. and S.A.P. are grateful for the hospitality of the Max Planck Institute for the Physics of Complex Systems in Dresden, where this work was initiated, during an extended visit supported through a 2024 Gutzwiller Award to S.A.P. We  acknowledge support from the UKRI Horizon Europe Guarantee Grant No. EP/Z002419/1 (to S.A.P.). B.P. acknowledges funding through a Leverhulme-Peierls Fellowship at the
University of Oxford and the Alexander von Humboldt Foundation through a Feodor-Lynen fellowship. O.B. acknowledges support from a Leverhulme Trust International Professorship [Grant Number LIP-202-014], Merton College, and the Clarendon Fund.
\end{acknowledgements}

\appendix

\section{Numerical constructions of general Tree-Unitary Gates}
\label{app:gen_tu}

The KIM is an example of a tree-unitary gate which in many ways is non-generic: it is an example of a `maximum velocity' tree-unitary gate, as we discuss in Sec.~\ref{sec:maxvel}. In this section we describe how to construct more generic tree-unitary gates, and thereby numerically estimate the dimensionality of the set of such gates.

To generate ensembles of tree-unitary gates, we extend the method of Ref.~\onlinecite{rather_creating_2020}, which generates dual-unitary gates by iteratively shuffling indices and performing a polar decomposition to find the closest unitary. For clarity, we focus on the $z=3$ case; the generalization for $z>3$ is straightforward.

Given a matrix $M$, with matrix elements $\langle abc|M|def\rangle$, we define `reshuffled' matrices as
\begin{align}
\langle abde|M_{R_1}|cf \rangle=\langle abc|M|def\rangle, \\
\langle acdf|M_{R_2}|be \rangle=\langle abc|M|def\rangle, \\
\langle bcef|M_{R_3}|ad \rangle=\langle abc|M|def\rangle.
\end{align}
Tree-unitarity of $M$ is the statement that $M_{R_i}$ is an isometry $\forall i$, i.e. $M_{R_i}^{\dagger}M_{R_i}^{\phantom{\dagger}} =\mathbb{1}_{q^2}$.

We now define the map $T_{R_i}:\mathbb{C}^{q^3\times q^3}\to\mathbb{C}^{q^3\times q^3}$ to consist of three stages. First, shuffle the indices to take $M\to M_{R_i}$. Second, map $M_{R_i}\to N_{R_i}$ where $N_{R_i}$ is the nearest isometry to $M_{R_i}$. Finally, unshuffle the indices $N_{R_i}\to N$.
The second step, finding the nearest isometry to $M_{R_i}$, is achieved by performing a polar decomposition 
\begin{align}
M_{R_i}^{\phantom{\dagger}}=N_{R_i}^{\phantom{\dagger}}\sqrt{M_{R_i}^{\dagger} M_{R_i}^{\phantom{\dagger}}},
\end{align}
where $N_{R_i}$ is a $q^4\times q^2$ matrix satisfying $N_{R_i}^{\dagger}N_{R_i}=\mathbb{1}_{q^2\times q^2}$. We shall also need the analogous result for a square matrix: the closest unitary matrix to $M$ is $N$ where $M=N\sqrt{M^{\dagger} M}$ is the polar decomposition. We shall denote this map $M\to N$ by $T$.

Define the composite map $T_{c}: \mathbb{C}^{q^3\times q^3}\to\mathbb{C}^{q^3\times q^3}$ to act as $T_c(M)=T_{R_3}(T_{R_2}(T_{R_1}(T(M))))$.
Given a random $q^3 \times q^3$ `seed' matrix $M_0$, we act with $T_c$ a large number of times $n$ to produce a matrix $M_n=T_c^n(M_0)$.
If this procedure converges, as numerically seems to be the case for generic initial matrices, it converges to a matrix satisfying all the tree-unitary conditions.

One can generate random tree-unitary gates that satisfy one or more of the maximum-velocity conditions by including this unitarity constraint (in place of the tree-unitarity condition that follows from it) in the above algorithm. \added{We provide an implementation of this algorithm \cite{code}.}

Since we do not have a general parametrization of tree-unitary gates, we numerically estimate the dimensionality of the set of tree-unitary gates. To do this we follow Ref.~\onlinecite{prosen_many-body_2021}, writing the tree-unitarity constraints as a vector function $\vec{f}(\vec{a})=0$ of all real parameters $\vec{a}$ in our matrix. Finding the rank of the Jacobian matrix $\frac{\partial f_i}{\partial a_j}$ gives the (local) dimension of the manifold of solutions to this equation. For 3-site gates with $q=2$, for each of $100$ samples this approach returns that tree-unitary operators form a $37$-dimensional space.

\section{Constructions of Tree-Unitary Gates}
\label{app:constructions}

We now give more explicit examples of tree-unitary gates beyond the KIM. All examples will turn out to be `maximum velocity' as defined in Sec.~\ref{sec:maxvel} in at least one direction. While we do not have any explicit constructions of tree-unitary gates which do not satisfy maximum-velocity conditions, such gates can be obtained numerically using the algorithm described in App.~\ref{app:gen_tu}.

\subsection{Two Dual-Unitary Gates}
\label{app:2du}

The first explicit construction consists of two dual-unitary gates applied on adjacent bonds, 
\begin{equation}
\label{eq:2du}
    U=\vcenter{\hbox{\includegraphics[height=0.26\columnwidth]{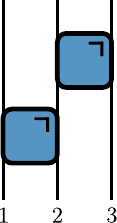}}}.
\end{equation}
The entire circuit from this construction could in fact be viewed as a 4-layer circuit of 2-site gates. Tree-unitarity of this composite gate follows directly from dual-unitary of the 2-site gate. 

The maps to calculate light cone correlation functions [Eq.~\eqref{eq:M_def} of the main text] here obey $M_{12}(\sigma)=M_{23}(\sigma)=M_{31}(\sigma)=0$, while $M_{32}(\sigma),~M_{13}(\sigma),~M_{21}(\sigma) \neq 0$ in general. The indices in Eq.~\eqref{eq:2du} correspond to our labeling convention of the nodes, with $M_{ij}$ dictating the correlation functions from ingoing leg $i$ to outgoing leg $j$.
Consistent with this, these gates satisfy the maximum-velocity condition for $2\to 1$, $1\to 3$, and $3\to 2$. As a result, for an operator initially on ingoing leg $2$ of a gate in the first layer, correlations are only non-vanishing along the single ray in the $2\to 1$ direction, along which operator spreading is maximum velocity. For an operator initially on ingoing leg $1$, after one time step the operator will have spread to site $3$. This site will then be ingoing leg $2$ for the next layer, leading to the single ray $2\to 1$. Similarly, an operator initially on ingoing leg $3$ will spread to site $1$, at which point it will move along the $2\to 1$ direction. 
We see that the maximum-velocity condition here constrains the non-vanishing correlations to be only along that direction.
This construction directly extends to higher $z$, for which $z-1$ dual unitaries applied to different bonds in any order also yields a tree-unitary.

\subsection{Hadamard Construction}

A more symmetric example can be constructed from Hadamard matrices and $\delta$-tensors. This 2-site `Hadamard construction' is physically obtained through Ising terms and kicks, like the KIM, but we include Ising couplings between \textit{all} pairs of spins (not just nearest neighbours according to the tree geometry) 
\begin{align}\label{eq:had_gate}
    U = \prod_{( i,j)}^z\mathcal{I}_{ij} \prod_{j=1}^z \mathcal{K}_j \prod_{( i,j)}^z\mathcal{I}_{ij} ,
\end{align}
where $I_{ij}$ are the Ising terms and $K_j$ are the kick terms [Eq.~\eqref{eq:ising_kick}], with $(i,j)$ indicating all pairs $i\neq j$.
For $z=3$, this gate is represented as 
\begin{align}
    \addedtwo{8}\vcenter{\hbox{\includegraphics[height=1.8cm]{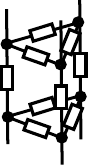}}}
\end{align}

In this construction, the circuit is maximum velocity in all directions, and therefore all two-point correlation functions vanish, $M_{ij}(\sigma)=0$, as discussed in Sec.~\ref{sec:maxvel} of the main text.
The light-cone OTOC is plotted in Fig.~\ref{fig:otocplot_kim} of the main text, and tends to the same non-trivial values as the KIM, since it possesses the same additional eigenoperators constructed in App.~\ref{app:identities}.

As with all tree-unitary gates, dressing each outgoing leg with a single-site unitary gate does not alter the tree-unitarity conditions. Furthermore, if we dress this Hadamard construction with a single 2-site unitary, 
\begin{equation}
    \addedtwo{8}\vcenter{\hbox{\includegraphics[height=2.8cm]{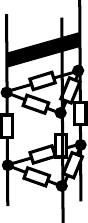}}},
\end{equation}
the maximum-velocity conditions of the original construction imply that we still obey the tree-unitary conditions. Suppose that the 2-site unitary is applied between sites 1 and 3. Then the circuit is no longer maximum velocity in the $2\to 3$ or $2\to 1$ directions, but maintains the maximum velocity along $3 \to 2$ and $1 \to 2$. However, the 2-point correlation functions still vanish, because of a further identity satisfied by the gate (see App.~\ref{app:identities})
\begin{equation}
    \addedtwo{8^2}\vcenter{\hbox{\includegraphics[height=1.8cm]{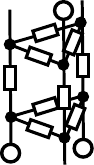}}}=
    \vcenter{\hbox{\includegraphics[height=1.2cm]{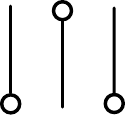}}},
\end{equation}
which can be interpreted as saying that for the undressed construction,
\begin{equation}
\label{had_hops}
    U^{\dagger}(\mathbb{1} \otimes \sigma \otimes \mathbb{1})U \longrightarrow  \sigma \otimes \sigma \otimes \sigma, 
\end{equation}
while for the dressed construction
\begin{equation}
\label{haddressed_hops}
    U^{\dagger}( \mathbb{1} \otimes \sigma \otimes \mathbb{1})U \longrightarrow \begin{cases}
    \sigma \otimes \sigma \otimes \sigma  \\
     \sigma \otimes \sigma \otimes \mathbb{1} \\
      \mathbb{1} \otimes \sigma \otimes \sigma \\
\end{cases}.
\end{equation}

\subsection{Controlled Gates and Tri-Unitary Gates}
\label{app:triunitary}

Tree-unitary gates can also be constructed from a gate with two control qubits and one target qubit, along with SWAP gates
\begin{align}
    U=\vcenter{\hbox{\includegraphics[height=1.8cm]{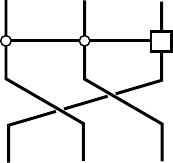}}}\,,
\end{align}
with open circles representing the controls. This gate is maximum velocity in $1\to 3$, $2\to 1$, and $3\to 2$ directions. The extension to higher $z$ is to have $z-1$ controls, and a cyclic SWAP on all sites. 
One can further dress this construction with a further SWAP and a dual-unitary gate to obtain
\begin{align}
    U=\vcenter{\hbox{\includegraphics[height=2.3cm]{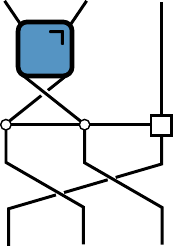}}}
\end{align}
which is also tree-unitary.

Tree-unitary gates acting on three sites can also be obtained from a family of triunitary gates~\cite{jonay_triunitary_2021}, which are 3-site gates satisfying unitarity under $\pi/3$ rotations
\begin{align}
\label{eq:triunitary}
    \vcenter{\hbox{\includegraphics[height=1.8cm]{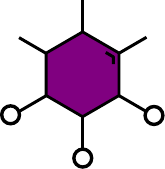}}}=
    \vcenter{\hbox{\includegraphics[height=1.8cm]{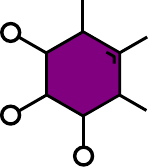}}}=
    \vcenter{\hbox{\includegraphics[height=1.8cm]{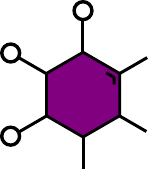}}}=
    \vcenter{\hbox{\includegraphics[height=1.3cm]{figures_final/triple_identity.pdf}}}.
\end{align}
Given a triunitary gate, we can obtain a tree-unitary gate by applying a SWAP to sites $2$ and $3$ or sites $1$ and $2$
\begin{align}
    U=\vcenter{\hbox{\includegraphics[height=1.7cm]{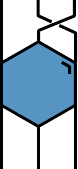}}}\,.
\end{align}
This gate is maximum velocity in the $3\to 1$ and $1\to 2$  directions. 
It also satisfies 
\begin{align}
    \vcenter{\hbox{\includegraphics[height=2.0cm]{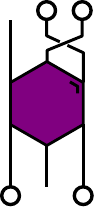}}}=\vcenter{\hbox{\includegraphics[width=0.025\columnwidth]{figures_final/double_identity.pdf}}}\,,
\end{align}
which is neither a tree-unitary condition nor a maximum-velocity condition, but enforces $M_{21}(\sigma)=M_{22}(\sigma)=0$. In words, the gate is not maximum velocity from $2\to 3$, but for operators starting on site $2$ of the gate, they can only have non-zero correlation functions in the $2\to 3$ direction. \added{More explicitly, an operator initially on site $2$ can evolve into terms of the form}
\begin{equation}
\label{triu_2}
    U^{\dagger}( \mathbb{1} \otimes \sigma \otimes \mathbb{1})U \longrightarrow \begin{cases}
    \sigma \otimes \sigma \otimes \sigma  \\
     \sigma \otimes \sigma \otimes \mathbb{1} \\
      \mathbb{1} \otimes \sigma \otimes \sigma \\
      \sigma \otimes \mathbb{1} \otimes \sigma \\
      \mathbb{1} \otimes \mathbb{1} \otimes \sigma
      
\end{cases}.
\end{equation}
This, along with the maximum-velocity conditions, is what allows for the solvability of the circuits in Ref.~\onlinecite{jonay_triunitary_2021}.
Ref.~\onlinecite{jonay_triunitary_2021} provides a partial parametrization of tri-unitary gates.
Using controlled phases $CP_{ij}(\phi)=e^{-i\frac{\phi}{4}(Z_i-1)(Z_j-1)}$ (indicated with blue circles) and single site unitaries (white boxes) we construct the tree-unitary gate
\begin{align}
    U=\vcenter{\hbox{\includegraphics[height=2.3cm]{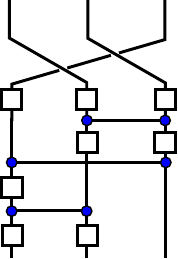}}}
\end{align}
This gives a $31$-parameter space of circuits, but is maximum velocity in the $2\to 1$, $3\to 2$ and $1\to 3$ directions, and therefore is highly non-generic.
Notice that the dual-unitary construction in Eq.~\eqref{app:2du} is contained in this construction as a special case.

\section{\added{Bound on the OTOC}}
\label{app:bound}

\added{In this section we derive a bound on the average light-cone OTOC, given that an operator has a total weight $w$ on the light cone. 
As argued in the main text (Sec.~\ref{subsec:tree_solvability}), a solvable model will generically conserve operator weight on the light cone and the following bound will apply.}

\added{Under unitary dynamics a Pauli operator, initially located at site $j$, will grow into a superposition of Pauli strings}
\begin{equation}
\label{eq:pauli_decomp_app}
\sigma_{\alpha}(j,t) = \sum_{\mathcal{S}} c_{\mathcal{S}} (t) {\mathcal{S}},
\end{equation}
\added{where unitarity conserves the total operator weight $\sum_{\mathcal{S}} c_{\mathcal{S}}(t)^2 = 1$ (from here on, we suppress the time for notational convenience $c_{\mathcal{S}}(t)=c_{\mathcal{S}}$). Note that since unitary evolution of a Hermitian operator preserves hermiticity, we take all $c_{\mathcal{S}}$ to be real.}

\added{Denote the region of sites on the light cone $R$, with number of sites $|R|$. The support of ${\mathcal{S}}$ on $R$ is defined as $\mathrm{supp}_R({\mathcal{S}}) = \{i \in R : {\mathcal{S}}_i \neq I\}$, where ${\mathcal{S}}_i$ is the value of the Pauli string on site $i$. $|\mathrm{supp}_R({\mathcal{S}})|$ therefore counts the number of sites in $R$ on which ${\mathcal{S}}$ does not act as the identity.
For example, if we consider the Pauli string $\mathcal{S}=XIY$ and the region $R=\{1,2\}$, then $S_1=X$ and $|\mathrm{supp}_R({\mathcal{S}})|=1$.
The total weight $w$ of $\sigma_{\alpha}(j,t)$ on $R$ is defined as }
\begin{equation}
w=\sum_{{\mathcal{S}}:|\mathrm{supp}_R({\mathcal{S}})|\neq 0} c_{\mathcal{S}}^2.
\end{equation}
\added{For solvable circuits, as argued in the main text, the weight $w$ on the light cone is constant. For tree-unitary circuits in particular, the weight $w=1$, i.e. every Pauli string with non-zero weight has support on the light cone.}

\added{We now seek to relate this weight $w$ to the OTOCs on any given site in $R$. To this end, consider taking the OTOC [Eq.~(\ref{eq:otoc_def})] between the growing operator $\sigma_{\alpha}(j,t)$, and a local Pauli operator on a site in $R$, $\sigma_{\beta}(i,0)$. We average over all sites $i \in R$, and over all non-trivial Pauli operators $\beta=1,\dots,q^2-1$, yielding }
\begin{equation}
\label{eq:o_def_app}
    \bar{O}=\frac{1}{|R|} \sum_{i \in R} \frac{1}{q^2-1} \sum_{\beta} \langle \sigma_{\beta}(i,0)\sigma_{\alpha}(j,t)\sigma_{\beta}(i,0)\sigma_{\alpha}(j,t) \rangle.
\end{equation}
\added{If $\sigma_{\alpha}(j,t)$ has no support on $R$, i.e. $w=0$, then $\bar{O}=1$. When $w\neq 0$, $\bar{O}\neq 1$, and we now find a bound on the difference $1-\bar{O}$ given $w$.}

\added{Using the decomposition Eq.~(\ref{eq:pauli_decomp_app}), we can write the OTOC in terms of the coefficients $c_{\mathcal{S}}$}
\begin{align}
&\langle \sigma_{\beta}(i,0)\sigma_{\alpha}(j,t)\sigma_{\beta}(i,0)\sigma_{\alpha}(j,t) \rangle \nonumber \\
&\quad=\sum_{{\mathcal{S}},{\mathcal{S}}'} c_{\mathcal{S}} c_{\mathcal{S}'} \langle \sigma_{\beta}(i,0) {\mathcal{S}} \sigma_{\beta}(i,0){\mathcal{S}}'\rangle \nonumber \\
&\quad= \sum_{{\mathcal{S}}:{\mathcal{S}}_i \in \{\sigma_{\beta} , I\}} \kern-6pt c_{\mathcal{S}}^2 \,\, - \kern-4pt \sum_{{\mathcal{S}}:{\mathcal{S}}_i \notin \{\sigma_{\beta} , I\}} \kern-6pt c_{\mathcal{S}}^2,
\end{align}
\added{since if ${\mathcal{S}}_i \in \{\sigma_{\beta}, I\}$, then ${\mathcal{S}}$ commutes with $\sigma_{\beta}(i,0)$, otherwise ${\mathcal{S}}$ anticommutes with $\sigma_{\beta}(i,0)$. }
\added{Returning now to the average Eq.~\eqref{eq:o_def_app},}
\begin{equation}
\bar{O}=\frac{1}{|R|} \sum_{i \in R} \frac{1}{q^2-1} \sum_{\beta}  \left[ \sum_{{\mathcal{S}}:{\mathcal{S}}_i \in \{\sigma_{\beta} , I\}} \kern-6pt c_{\mathcal{S}}^2 \,\,- \kern-4pt \sum_{{\mathcal{S}}:{\mathcal{S}}_i \notin \{\sigma_{\beta} , I\}} \kern-6pt c_{\mathcal{S}}^2 \right],
\end{equation} 
\added{we seek to rewrite the sums over $\beta$ and ${\mathcal{S}}$ into a more unified form.}
\added{First note that }
\begin{equation}
\sum_{\beta} \sum_{{\mathcal{S}}:{\mathcal{S}}_i =I}c_{\mathcal{S}}^2 = (q^2-1)\sum_{{\mathcal{S}}:{\mathcal{S}}_i=I} c_{\mathcal{S}}^2
\end{equation}
\added{because the inner sum is independent of $\beta$.
Second, we have }
\begin{align}
\sum_{\beta} \sum_{{\mathcal{S}}:{\mathcal{S}}_i=\sigma_{\beta}} c_{\mathcal{S}}^2 &= \sum_{\mathcal{S}}c_{\mathcal{S}}^2 \,\,- \sum_{{\mathcal{S}}:{\mathcal{S}}_i=I}c_{\mathcal{S}}^2 \nonumber \\
&=1-\sum_{{\mathcal{S}}:{\mathcal{S}}_i=I}c_{\mathcal{S}}^2.
\end{align}
\added{Finally,}
\begin{align}
\sum_{\beta} \sum_{{\mathcal{S}}:{\mathcal{S}}_i \neq \{I,\sigma_{\beta}\}} c_{\mathcal{S}}^2 
&= \sum_{\beta} \sum_{\gamma \neq \beta} \sum_{{\mathcal{S}}:{\mathcal{S}}_i=\sigma_{\gamma}} c_{\mathcal{S}}^2 \nonumber  \\ 
&= (q^2-2) \sum_{\beta} \sum_{{\mathcal{S}}:{\mathcal{S}}_i =\sigma_{\beta}}c_{\mathcal{S}}^2 \nonumber \\
&= (q^2-2)\left[ 1-\sum_{{\mathcal{S}}:{\mathcal{S}}_i=I}c_{\mathcal{S}}^2\right].
\end{align}
\added{Collecting these, we obtain}
\begin{equation}
\label{eq:o_intermediate}
\bar{O}=\frac{1}{|R|} \sum_{i\in R} \left[ -\frac{q^2-3}{q^2-1} + 2 \frac{q^2-2}{q^2-1} \sum_{{\mathcal{S}}:{\mathcal{S}}_i=I}c_{\mathcal{S}}^2 \right].
\end{equation}

\added{So far we have been exact, but now place a bound on $\sum_{i\in R} \sum_{{\mathcal{S}}:{\mathcal{S}}_i=I}c_{\mathcal{S}}^2$ given a weight on $R$ of $w$. To do so, let $w_n$ be the total weight of $\sigma_{\alpha}(j,t)$ for which $|\mathrm{supp}_R({\mathcal{S}})|=n$, i.e.}
\begin{equation}
w_n=\sum_{{\mathcal{S}}:|\mathrm{supp}_R({\mathcal{S}})|=n}c_{\mathcal{S}}^2.
\end{equation}
\added{These $w_n$ satisfy}
\begin{align}
\sum_{n=0}^{|R|} w_n=1, ~\sum_{n=1}^{|R|} w_n = w.
\end{align}
\added{Now, restricting the sum to include only strings $S$ with $|\mathrm{supp}_R({\mathcal{S}})|=n$,}
\begin{align}
\sum_{i\in R} \sum_{\substack{{\mathcal{S}} : {\mathcal{S}}_i=I\\ |\mathrm{supp}_R({\mathcal{S}})|=n}} \kern-8pt c_{\mathcal{S}}^2 
&=(|R|-n)\sum_{i\in R} \sum_{\substack{{\mathcal{S}}\\ |\mathrm{supp}_R({\mathcal{S}})|=n}}  \kern-8pt c_{\mathcal{S}}^2 \nonumber \\
&= w_n(|R|-n),
\end{align}
\added{because a Pauli string with ${\mathcal{S}}_k=I$ will be counted whenever $i=k$, and by definition this occurs on $|R|-n$ sites $k$.
This means that }
\begin{align}
\sum_{i\in R} \sum_{{\mathcal{S}}:{\mathcal{S}}_i=I}c_{\mathcal{S}}^2 &= \sum_{n=0}^{|R|} w_n(|R|-n) \nonumber \\
&= (1-w) |R| + \sum_{n=1}^{|R|} w_n (|R|-n) \nonumber \\
&\leq (1-w) |R| + \sum_{n=1}^{|R|} w_n (|R|-1) ,
\end{align}
\added{so we have}
\begin{equation}
\sum_{i\in R} \sum_{{\mathcal{S}}:{\mathcal{S}}_i=I}c_{\mathcal{S}}^2 \leq (1-w)|R| + w (|R|-1).
\end{equation}
\added{Putting this into Eq.~(\ref{eq:o_intermediate}) yields, after some simple algebra,}
\begin{equation}
1-\bar{O} \geq \frac{2w}{|R|} \frac{q^2-2}{q^2-1}.
\end{equation}

\section{Sufficiency of $\delta$-hyperbolicity for computable light-cone correlations}
\label{app:delta}

\added{Here we prove that $|\intsec(i,j,s)|$ remains bounded if the metric space $(V(G),d)$ is $\delta$-hyperbolic (see Sec.~\ref{subsec:noneuclidean} for definitions). This means that $\delta$-hyperbolicity is a sufficient condition for efficiently-computable light cone correlation functions for any circuit. }

\added{Assume that $(V(G),d)$ is $\delta$-hyperbolic. Recall from the main text that this means that any geodesic triangle $\{i,j,y\}$ is $\delta$-thin, namely that for any point $p$ on one side there exists a point $q$ on the union of the other two sides with $d(p,q)
\leq \delta$.
We now show that for any two points $x_s,x'_s \in \intsec(i,j,s)$, $d(x_s,x'_s) \leq 2\delta$,
which in turn implies that the number of sites in $\intsec(i,j,s)$ is bounded. 
To see this, consider two geodesics $\gamma$ and $\gamma'$ which go from $i \to j$. Consider the `degenerate' triangle which consists of $\{i,j,j\}$, with sides being geodesics $\gamma$, $\gamma'$, and a geodesic of $0$ length. Then by the assumption this triangle is $\delta$-thin, which implies that any $p$ point on $\gamma$ is within $\delta$ of a point on $\gamma'$ (more precisely, the union of $\gamma'$ with the geodesic of $0$ length, but this is just $\gamma'$). Consider now two points $x_s \in \gamma$, and $x_s'\in \gamma'$, for which $x_s, x_s' \in \intsec(i,j,s)$.  Again, by the previous argument, there exists a point $u' \in \gamma'$ such that $d(x_s,u') \leq \delta$. By the triangle inequality, $|d(i,u')-d(i,x_s)| \leq d(u',x_s) \leq \delta$. But since $d(u',x_s')=|d(x_s',i)-d(u',i)|$, and $d(i,x_s) =s=d(i,x'_s)$, we have $d(u',x_s') \leq \delta$. Therefore, using the triangle inequality once more, $d(x_s,x_s') \leq d(u',x_s) + d(u',x_s') \leq 2\delta$. 
Reiterating, we have shown that for any pair of points $x_s, x_s' \in \intsec(i,j,s)$, $d(x_s,x_s') \leq 2\delta$. Since every vertex has a finite number of edges, a bounded distance between any pair of points in $|\intsec(i,j,s)|$ implies a bounded size $|\intsec(i,j,s)|$.}

\section{Identities for KIM}
\label{app:identities}

\subsection{Trace Identity}
Here we prove an identity for the 3-site KIM used in the main text (Eq.~\ref{eq:kim_identity_main}) to constrain the form of Pauli operators under conjugation. We show that 
\begin{equation}
\label{eq:target}
    \addedtwo{4^2}\vcenter{\hbox{\includegraphics[height=0.22\columnwidth]{figures_final/kim/kim_wi1.pdf}}}= 
    \vcenter{\hbox{\includegraphics[height=0.15\columnwidth]{figures_final/kim_wi2.pdf}}}.
\end{equation}
This equality cannot be shown simply using the identities \eqref{eq:had_identities}, but still relies on the properties of complex Hadamard matrices. 
Consider separating out the LHS of \eqref{eq:target} into the following parts
\begin{align}
    \label{eq:parts}
    \vcenter{\hbox{\includegraphics[height=0.32\columnwidth]{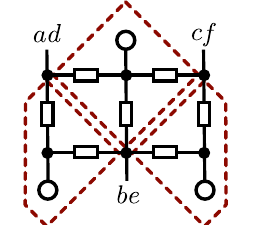}}} 
\end{align}

Each of these parts have the form of a $\delta$-tensor surrounded by an equal number of complex Hadamard matrices and their Hermitian conjugates. Our goal will be to show that these parts are themselves (essentially) $\delta$-tensors, following similar arguments as in Ref.~\onlinecite{claeys_operator_2024}, from which the identity will follow. 
The complex Hadamard matrix $H$ [Eq.~\eqref{eq:hadamard_gate}], making up the kick and the Ising terms, can be written as
\begin{equation}
    H=DFD,
\end{equation}
where $F$ is the $q=2$ Fourier matrix 
\begin{equation}
    F=\begin{pmatrix}
    1 & 1 \\
    1 & -1
    \end{pmatrix}\,,
\end{equation}
and $D$ is a diagonal matrix with matrix elements $D_{aa}=e^{i\phi_a}$.

Since each of the parts above consists of a complex Hadamard and its Hermitian conjugate, the phases on and next to the $\delta$-tensor cancel out. For the parts consisting of two folded gates, this leaves us with 
\begin{align}
    \vcenter{\hbox{\includegraphics[height=0.15\columnwidth]{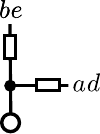}}} &= \frac{1}{2}e^{i(\phi_d+\phi_e-\phi_a-\phi_b)}\sum_f (-1)^{(f-1)(a+b+d+e-4)} \nonumber\\
    &=e^{i(\phi_d+\phi_e-\phi_a-\phi_b)} \delta_{(a+b+c+d)}
\end{align}
where $\delta_{(i)} \equiv \delta_{i~\mathrm{mod}~2,0}$. 
Similarly, with three folded gates we have
\begin{align}
    \vcenter{\hbox{\includegraphics[height=0.15\columnwidth]{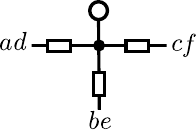}}}  = e^{i(\phi_d+\phi_e+\phi_f-\phi_a-\phi_b-\phi_c)} \delta_{(a+b+c+d+e+f)}.
\end{align}

Piecing these together in \eqref{eq:parts} and
cancelling common phase factors, we obtain
\begin{align}
    \delta_{(abde)}\delta_{(abcdef)}\delta_{(bcef)} =\delta_{a,d}\delta_{b,e}\delta_{c,f}
\end{align}
where, e.g. $\delta_{(abde)}\equiv \delta_{(a+b+d+e)}$, and so on.
This returns the RHS of the identity \eqref{eq:target} that we set out to show.

\subsection{OTOC Channel Eigenoperators}

In Sec.~\ref{sec:maxvel} of the main text, we showed that satisfying a maximum-velocity condition led to a non-trivial eigenoperator with unit-eigenvalue of the OTOC channel. Constructing this eigenoperator explicitly, we showed that the light cone OTOC approached the value of $-1/(q^2-1)$. 
The KIM, however, has additional symmetries that allow for the construction of further eigenoperators with eigenvalue $1$, changing this non-trivial light cone OTOC value in a way which depends explicitly on the choice of operators, as for the 2-site KIM in Refs.~\onlinecite{bertini_operator_2020,claeys_maximum_2020}. 

Let's briefly recall the status in the main text. The light-cone OTOC is computed using the channel $\mathcal{T}_{e\tilde{e}}$ defined by, for example,
\begin{align}%\label{eq:Tee_otoc}
    \mathcal{T}_{23}|\sigma)=\vcenter{\hbox{\includegraphics[height=0.2\columnwidth]{figures_final/t23_circ.pdf}}}\,.
\end{align}

Unitarity allows the construction of a single left/right eigenoperator with eigenvalue $1$,
\begin{equation}
    |R)=\,\vcenter{\hbox{\includegraphics[height=0.125\columnwidth]{figures_final/squaretop.pdf}}}\,, \qquad (L|=\,\vcenter{\hbox{\includegraphics[height=0.125\columnwidth]{figures_final/circletop.pdf}}}\,.
\end{equation}

Maximum velocity in the direction $e\to \tilde{e}$ allows the construction of a further left/right eigenoperator with eigenvalue $1$, 
\begin{equation}
    |R')=\,\vcenter{\hbox{\includegraphics[height=0.125\columnwidth]{figures_final/Rprime_circle.pdf}}}\,, \qquad (L'|=\,\vcenter{\hbox{\includegraphics[height=0.125\columnwidth]{figures_final/Lprime.pdf}}}
    \,.
\end{equation}

Without any further structure, these are the only unit-eigenvalue eigenoperators, and from these one can show that the light cone OTOC approaches $-1/(q^2-1)$, as done in the main text. 
For the KIM, however, we can construct one additional left/right eigenoperator with eigenvalue $1$~\cite{bertini_operator_2020},
\begin{equation}
    |R'')=\,\vcenter{\hbox{\includegraphics[height=0.15\columnwidth]{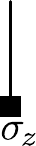}}}\,, \qquad (L''|=\,\vcenter{\hbox{\includegraphics[height=0.15\columnwidth]{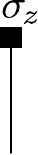}}}
    \,.
\end{equation}
We now orthonormalize the first pair as before (with $q=2$ for the KIM)
\begin{align}
    |v)&=|R),\quad |v')=\frac{|R')-2|R)}{\sqrt{3}},\\
    (v|&=(L|,\quad (v'|=\frac{2(L|-(L'|}{\sqrt{3}},
\end{align}
with the new eigenoperators orthonormalized to
\begin{align}
    |v'')&=\frac{1}{\sqrt{2}}\left(\frac{1}{\sqrt{6}}|R)-\sqrt{\frac{2}{3}}|R') + \sqrt{\frac{3}{2}}|R'')\right),\\
    (v''|&=\frac{1}{\sqrt{2}}\left(-\sqrt{\frac{2}{3}}(L|+\frac{1}{\sqrt{6}}(L'| + \sqrt{\frac{3}{2}}(L''|\right).
\end{align}

The overlaps required for the OTOC now explicitly depend on the operator. We take $\sigma_{\alpha}=\alpha_x \sigma_x + \alpha_y \sigma_y + \alpha_z \sigma_z$ and $\sigma_{\beta} = \beta_x \sigma_x + \beta_y \sigma_y + \beta_z \sigma_z$, with $\alpha_x^2 +\alpha_y^2+\alpha_z^2=\beta_x^2+\beta_y^2+\beta_z^2=1$. Then
\begin{align}
(\sigma_{\alpha}|R'')&=(2\alpha_z^2-1), \qquad
(L''|\sigma_{\beta})=2\beta_z^2,
\end{align}
such that
\begin{align}
(\sigma_{\alpha}|v'')&=\frac{1}{\sqrt{2}}\left[\sqrt{\frac{3}{2}}(2\alpha_z^2-1) + \sqrt{\frac{1}{6}}\right], \\
(\sigma_{\alpha}|v'')&=\frac{1}{\sqrt{2}}\left[\sqrt{\frac{3}{2}}(2\beta_z^2) - \sqrt{\frac{2}{3}}\right].
\end{align}
From the projection on the resulting eigenspace, we find that the light cone OTOC approaches 
\begin{align}
    &\lim_{t \to \infty} C_{\alpha \beta}(i,j;t) = \sum_{i=\{v,v',v''\}}(\sigma_{\alpha}|i)(i|\sigma_{\beta}) \nonumber\\
    &\qquad=1-\frac{4}{3}+\frac{1}{2}(\sqrt{6}\alpha_z^2-\sqrt{\frac{2}{3}})(\sqrt{6}\beta_z^2-\sqrt{\frac{2}{3}}) \nonumber \\
    &\qquad= 3\alpha_z^2\beta_z^2-\alpha_z^2-\beta_z^2.
\end{align}
For Fig.~\ref{fig:otocplot_kim}, $\alpha_z=1/\sqrt{2}$ and $\beta_z=0$ when $\sigma_{\beta}=Y$, so the OTOC approaches $-1/2$. When $\sigma_{\beta}=(Y+Z)/\sqrt{2}$, $\beta_z=1/\sqrt{2}$ and the OTOC approaches $-1/4$.

\section{Dynamics of 2-site Gate Brickwork Circuits}
\label{app:2site_dynamics}

In the main text we briefly explored the construction of brickwork circuits on the tree using 2-site gates, finding that they suffer from anisotropy. Nevertheless, it is noteworthy that many of the techniques used in (1+1)d circuits can be applied in this setting to calculate properties such as light-cone correlation functions, light-cone OTOCs, and entanglement growth following a quench. To illustrate how these calculations proceed, and to contrast them with the isotropic, tree-unitary case, we summarize them in this Appendix.

\subsection{Correlation Functions}

\subsubsection{On the light cone}

In the Sec.~\ref{sec:2site} of the main text we discussed the anisotropy of the light cone in brickwork circuits constructed from 2-site gates. Despite this anisotropy, it is still possible to efficiently calculate light cone correlation functions for any choice of unitary gate, along any light cone path. 

Let us describe a light cone path by its sequence of fast steps (F) and slow steps (S). To calculate the correlation function along such a path, we introduce the channels 
\begin{equation}
\label{m1}
    M_1(\sigma)=\mathrm{tr}_1\left[U^{\dagger} (\sigma \otimes I)U \right]/q=\vcenter{\hbox{\includegraphics[height=0.25\columnwidth]{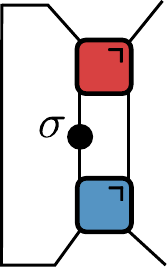}}}
\end{equation}
and 
\begin{equation}
\label{m2}
    M_2(\sigma)=\mathrm{tr}_2\left[U^{\dagger} (\sigma \otimes I)U \right]/q=\vcenter{\hbox{\includegraphics[height=0.25\columnwidth]{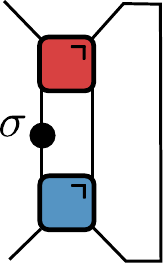}}}.
\end{equation}
For each fast step we apply $M_1$, and for each slow step we apply $M_1 M_2$ (for $z=3$). For example, a path consisting of steps FFS has correlation function
\begin{equation}
    \langle \sigma_{\beta} (x_{LC},t) \sigma_{\alpha} \rangle = \mathrm{tr}\left[\sigma_{\beta}  M_1M_2M_1M_1(\sigma_{\alpha}) \right].
\end{equation}
Physically, the $M_2$ channel arises due to the unitary gate applied to the site which is not along the path of interest. For $z>3$, we have a hierachy of steps from fast (advance immediately) to slow (wait $z-1$ time steps before advancing). In a step which waits $m$ time steps before advancing, one applies the channel $M_1M_2^m$.

\subsubsection{Inside the light cone}

The calculation of correlation functions inside the light cone is complicated because a large number of unitary gates can contribute (those within the intersection of light cones from the two operators). 
In (1+1)d, imposing dual-unitarity on the gates leads to correlation functions vanishing inside the light cone~\cite{bertini_exact_2018}, so all correlation functions can be calculated efficiently. 

On the Cayley tree, dual-unitary gates have the same consequence --- correlation functions inside the light cone vanish. 
This result follows from the identity (in the folded notation introduced in the main text)
\begin{equation}
\label{eq:dualwayshop}
    \mathrm{tr}_2 \left[ U^{\dagger} (\sigma \otimes \mathbb{1})U \right] = \vcenter{\hbox{\includegraphics[height=0.15\columnwidth]{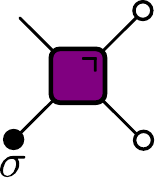}}} =\vcenter{\hbox{\includegraphics[height=0.07\columnwidth]{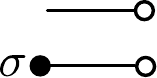}}}\propto \mathrm{tr}(\sigma)=0,
\end{equation}
which implies that $U^{\dagger}(\sigma_{\alpha}\otimes\mathbb{1})U$ cannot generate terms of the form $\sigma_{\beta}\otimes\mathbb{1}$. Physically, the operator has amplitude $1$ to spread. Applying layers of the circuit, this means that only terms which have non-zero weight on the \textit{fastest path} can be generated. Single-site correlation functions hence vanish identically (by trace orthonormality of Pauli operators) everywhere except for the fastest path.

In passing we note that alternating layers of dual-unitary and unitary gates also leads to correlation functions vanishing inside the light cone, but correlation functions can now be non-vanishing on a range of paths around the light cone.

\subsection{Out-of-time-order Correlators}

\begin{figure}[t]
\centering
\includegraphics[width=\columnwidth]{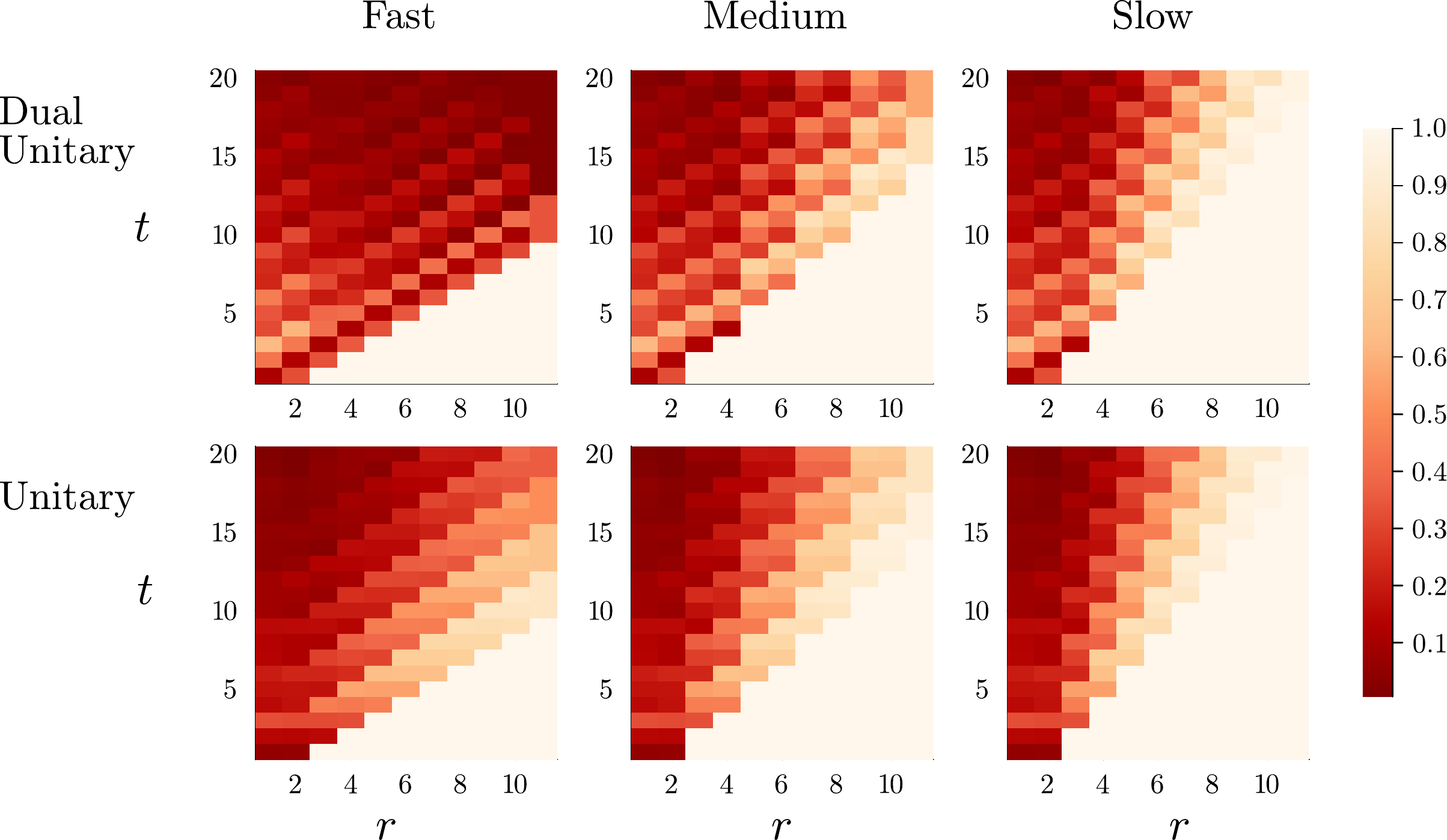}
\caption{
OTOCs for dual-unitary and unitary circuits on the $z=3$ tree along fast, medium (alternating fast and slow steps), and slow directions. For dual-unitary circuits in the fastest direction, the OTOC converges to a non-trivial value on the light cone, corresponding to a butterfly velocity $v_B=1$. In all other directions, and for unitary-only circuits, the typical rate at which operators grow is $v_B<1$, with a front that broadens in time. The simulations were performed using the QuantumClifford.jl package~\cite{quantumclifford}, with the random unitary gates generated using the algorithm in Ref.~\onlinecite{bravyi_hadamard_2021}, and random dual-unitary gates from the SDKI-r class introduced in Ref.~\onlinecite{yao_temporal_2024}.}
\label{fig:2site_otocs}
\end{figure}

\begin{figure}[t]
\centering
\includegraphics[width=\columnwidth]{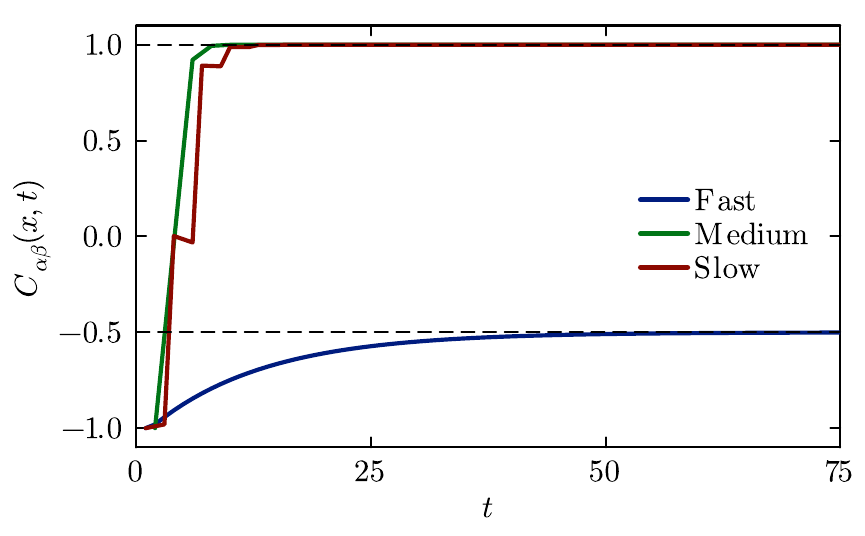}
\caption{
OTOC for the 2-site brickwork circuit on the tree, where each 2-site gate is the kicked Ising model at its dual-unitary point. Here, $\sigma_{\alpha}=(X+Z)/\sqrt{2}$ and $\sigma_{\beta}=Y$. Along the fastest direction the OTOC converges to the non-trivial value of $-0.5$. In the slowest direction and a direction consisting of alternating fast/slow steps, the OTOC approaches the trivial value of $1$.}
\label{fig:2site_kim_otocs}
\end{figure}

In Fig.~\ref{fig:2site_otocs}, we plot the OTOC for random Clifford simulations of both dual-unitary and unitary circuits in fast, intermediate, and slow directions. The first aspect to notice is that the asymmetry of the light cone is nicely visible in these plots. Second, on the dual-unitary fast path, the OTOC converges to a non-trivial value , and propagates with a butterfly velocity $v_B=1$. However, away from the fast direction, we observe a butterfly velocity that is slower than the light velocity, with spreading of the OTOC inside the light-cone. In Fig.~\ref{fig:2site_kim_otocs} we plot the OTOC value on the light cone only in these directions, this time where the dual-unitary gate is the kicked Ising model at the dual-unitary point.

To understand these properties, we follow Ref.~\onlinecite{claeys_maximum_2020} and diagrammatically write the OTOC in the form of a quantum channel. We use the same notation developed in the main text in Sec.~\ref{subsec:otoc}.
As for the correlation functions, we define two maps (using the four-folded notation of the main text)
\begin{equation}
\label{t1t2}
    T_1=\vcenter{\hbox{\includegraphics[height=1.8cm]{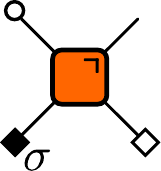}}}, \quad T_2=\vcenter{\hbox{\includegraphics[height=1.8cm]{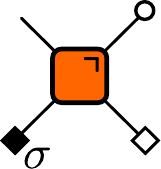}}},
\end{equation}
where each fast step corresponds to $T_1$ and each slow step for $z=3$ to $T_1T_2$.

For unitary gates, right/left eigenoperators of both $T_2$ and $T_1$ with eigenvalue $1$ are given by
\begin{equation}    |R)=\vcenter{\hbox{\includegraphics[width=0.03\columnwidth]{figures_final/squaretop.pdf}}}, \qquad (L|=\vcenter{\hbox{\includegraphics[width=0.03\columnwidth]{figures_final/circletop.pdf}}}.
\end{equation}Assuming only unitarity, we expect no other unit eigenvalues, and the light cone OTOC approaches the trivial value 
\begin{align}
    &\lim_{t \to \infty} C_{\alpha \beta}(i,j;t) = (\sigma_{\alpha}|R)(L|\sigma_{\beta})\nonumber\\
    &\quad=\vcenter{\hbox{\includegraphics[height=0.15\columnwidth]{figures_final/sigmaRLsigma.pdf}}}=\frac{1}{q^2}\mathrm{tr}[\sigma_{\alpha}^2]~\mathrm{tr}[\sigma_{\beta}^2]=1.
\end{align}

Imposing dual-unitarity on the gates allows one to construct a further left/right eigenoperator pair of $T_1$, 
\begin{equation}
    |R')=\vcenter{\hbox{\includegraphics[height=0.125\columnwidth]{figures_final/Rprime_circle.pdf}}}, \qquad (L'|=\vcenter{\hbox{\includegraphics[height=0.125\columnwidth]{figures_final/Lprime.pdf}}}.
\end{equation}
As a result, the OTOC on the fastest path (which involves only $T_1$) approaches a non-trivial light cone value~\cite{claeys_maximum_2020}, indicating butterfly velocity $v_B=1$.
Notably, however, $|R')$ and $(L'|$ are \textit{not} eigenoperators of $T_2$. As a result, paths involving a finite fraction of slow steps have a light cone OTOC which approaches the trivial value of $1$, consistent with the observations made in Figs.~\ref{fig:2site_otocs}~and~\ref{fig:2site_kim_otocs}. 

This result is perhaps unsurprising, because the `maximum velocity' condition follows from the `always-hop' property of dual-unitary gates in Eq.~\eqref{eq:dualwayshop}. There is no guarantee that operator weight is left behind on the original site, and therefore no guarantee that weight will be moved forwards in a slow step. 

\subsection{Entanglement Spreading}

\begin{figure}[t!]
\centering
\includegraphics[width=\columnwidth]{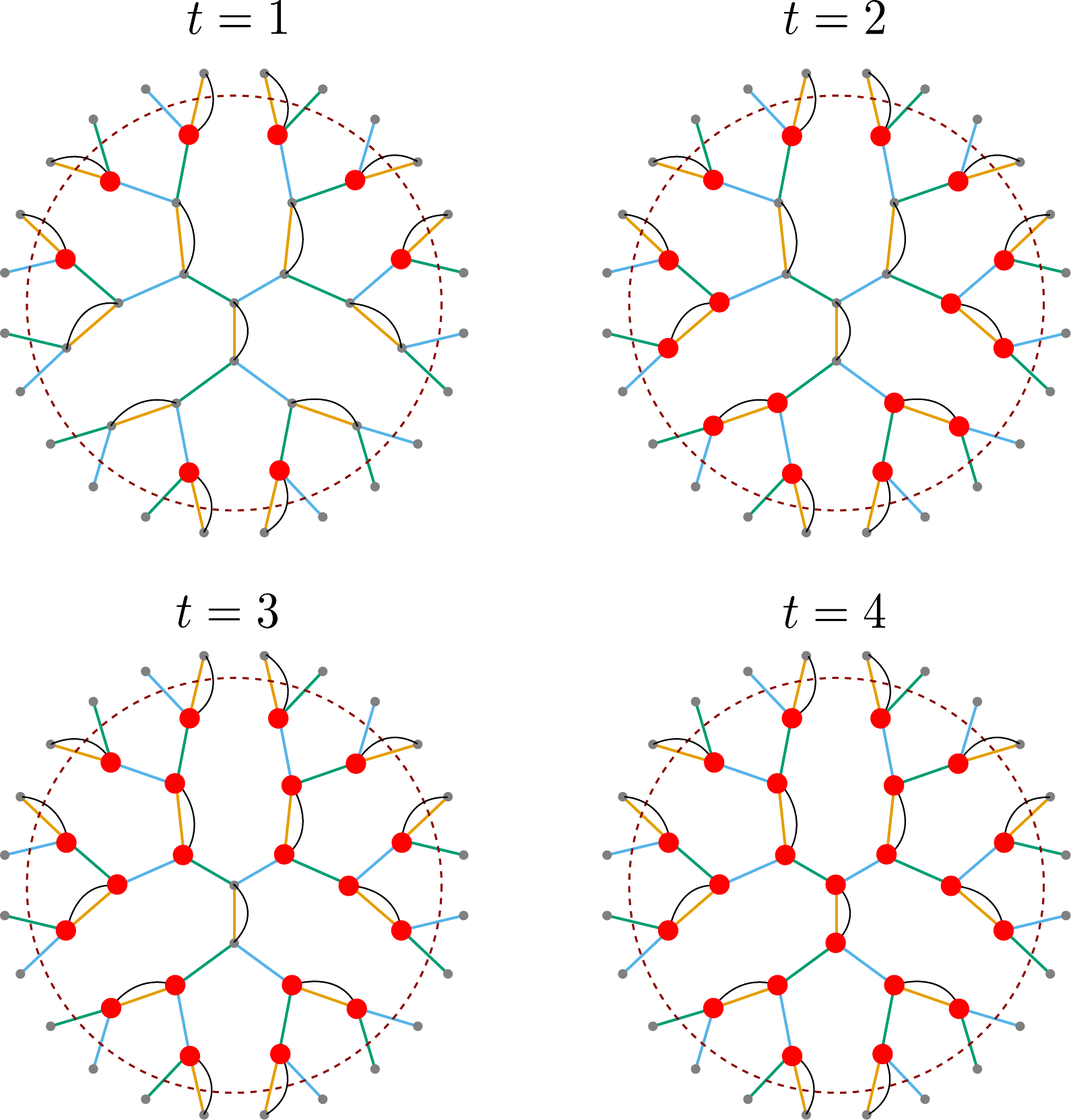}
\caption{Here we demonstrate how sites in region $A$ thermalize under dual-unitary dynamics into maximally mixed states. The initial state consists of Bell pairs on orange bonds, shown as the black arcs. We apply the unitary gates in the order orange, green, blue. At each successive time step, Bell pairs across another radial layer thermalize to $\frac{1}{2}\mathbb{1}$ (up to a unitary transformation), shown in red.}
\label{thermalisation}
\end{figure}

As in (1+1)d dual-unitary circuits, the entanglement dynamics can be exactly calculated for specific classes of initial states when the gates are dual-unitary. In particular, we start from a product state, and consider the entanglement entropy (EE) of a region A of $r$ generations from the center (shown in Fig.~\ref{thermalisation}). One solvable class of initial states consists of Bell pairs $\frac{1}{\sqrt{2}}\left(\ket{\uparrow\uparrow}+\ket{\downarrow\downarrow}\right)$ on the bonds corresponding to layer $z$ of the circuit. 
In Fig.~\ref{thermalisation} with layers applied in the order orange, blue, green, this corresponds to Bell pairs on the green bonds.

To calculate the EE, we construct the reduced density matrix $\rho_A(t)=\mathrm{tr}_{\bar{A}}\left(U(t)\ket{\psi_0}\bra{\psi_0}U^{\dagger}(t)\right)$. Analogously to the (1+1)d calculation, we use unitarity and dual-unitarity to write $\rho_A(t)$ as a unitary transformation of a simple state in region $A$. 
As an example, we will consider $z=3$ and a region of size $r=3$. The number of spins in the layer at radius $r$ is $n(r)=3\cdot2^{r-1}$. 

At $t=0$, all the entanglement comes from the Bell pairs which span the boundary; looking at Fig.~\ref{thermalisation} we see that there are $\frac{2}{3}n(r)=2^r$ of these (on the orange bonds). Tracing over $\bar{A}$, the residual spin of any Bell pair which crosses the boundary is in a maximally mixed state. The reduced density matrix is then $\rho_A(0)=(\ket{\textrm{Bell}}\bra{\textrm{Bell}})^{\otimes (2^r-1)} \otimes (\mathbb{1})^{\otimes 2^r}$, so $S(0)=2^r$. We show this in Fig.~\ref{thermalisation} at $t=0$ by highlighting the sites which contribute a $\left(\frac{1}{2}\mathbb{1}\right)$ to the reduced density matrix. 

\begin{figure}[t!]
\centering
\includegraphics[width=\columnwidth]{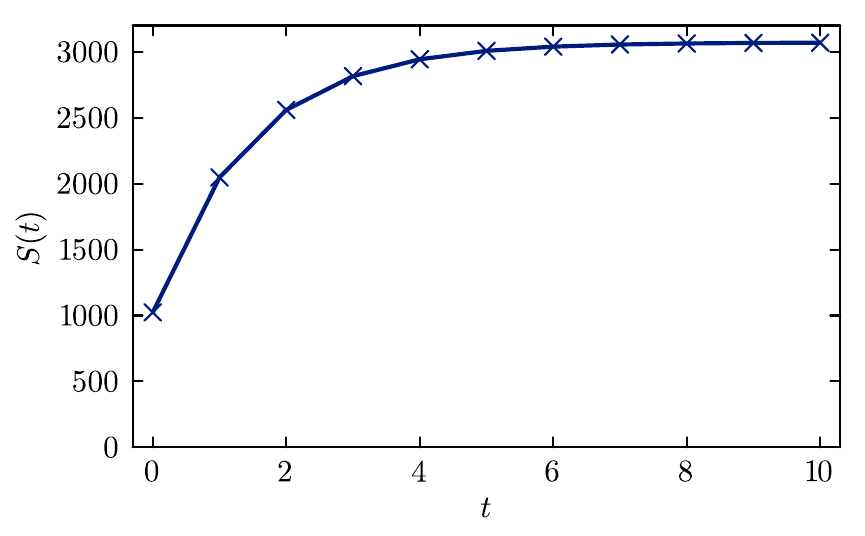}
\caption{Entanglement entropy growth for a $z=3$ tree, under dual-unitary dynamics starting from an exactly solvable Bell pair state for a subsystem $A$ consisting of the first $r=10$ generations of the tree.}
\label{2site_ee}
\end{figure}

For each layer of the circuit applied, the next outermost Bell pairs thermalize. The argument is fully analogous to the (1+1)d case, so we do not repeat it here. 
This pattern of thermalization leads to an exponentially saturating entanglement entropy, as shown in Fig.~\ref{2site_ee} for $r=10$. Since the number of initial Bell pairs between layer $r$ and layer $r+1$ is $2^r$, and noting that Bell pairs on the edge only contribute one site, the entanglement entropy in units of $\ln{2}$ is
\begin{equation}
    S(t)=2^r+(2^{r-1}+...+2^{r-t})\cdot2 = 2^r+2^{r+1}(1-2^{-t}).
\end{equation}
For general $z$,
\begin{equation}
S(t)=(z-1)^r+\frac{2(z-1)^r}{z-2}\left[1-(z-1)^{-t}\right].
\end{equation}

\bibliography{references_trees}

\end{document}